\documentclass[twocolumn]{aastex63}

\usepackage{epstopdf}
\usepackage{multirow}
\usepackage{float}
\usepackage{amsmath}
\usepackage{graphicx}
\usepackage[T1]{fontenc}
\usepackage{apjfonts}

\newcommand{\editone}[1]{{#1}}
\newcommand{\edittwo}[1]{{#1}}
\newcommand{\editthree}[1]{{#1}}

\shorttitle{On the Lack of Evolution of Stellar Mass Function of Massive Galaxies}
\shortauthors{Kawinwanichakij et al.}
\begin{document}

\title{\large \textbf{On the (Lack of) Evolution of the Stellar Mass Function of Massive Galaxies from $\mathbf{z}$=1.5 to 0.4 \footnote{Released on \today}}}


\correspondingauthor{Lalitwadee Kawinwanichakij}
\email{lalitwadee.kawinwanichakij@ipmu.jp}
\author[0000-0003-4032-2445]{Lalitwadee Kawinwanichakij}
\affiliation{Kavli Institute for the Physics and Mathematics of the Universe, The University of Tokyo, Kashiwa, Japan 277-8583 (Kavli IPMU, WPI)}
\affiliation{Department of Physics and Astronomy, Texas A\&M University, College
Station, TX, 77843-4242 USA}
\affiliation{George P.\ and Cynthia Woods Mitchell Institute for
  Fundamental Physics and Astronomy, Texas A\&M University, College
  Station, TX, 77843-4242}
\affiliation{LSSTC Data Science Fellow}  
\author[0000-0001-7503-8482]{Casey Papovich}
\affiliation{Department of Physics and Astronomy, Texas A\&M University, College
Station, TX, 77843-4242 USA}
\affiliation{George P.\ and Cynthia Woods Mitchell Institute for
  Fundamental Physics and Astronomy, Texas A\&M University, College
  Station, TX, 77843-4242}
\author[0000-0002-1328-0211]{Robin Ciardullo}
\affiliation{Department of Astronomy \& Astrophysics, The Pennsylvania State University, University Park, PA 16802, USA}
\affiliation{Institute for Gravitation and the Cosmos, The Pennsylvania State University, University Park, PA 16802, USA}

\author[0000-0001-8519-1130]{Steven L. Finkelstein}
\affiliation{Department of Astronomy, The University of Texas at Austin, 2515 Speedway, Stop C1400, Austin, Texas 78712, USA}

\author[0000-0001-8379-7606]{Matthew L. Stevans}
\affiliation{Department of Astronomy, The University of Texas at Austin, 2515 Speedway, Stop C1400, Austin, Texas 78712, USA}

\author[0000-0002-0784-1852]{Isak G. B. Wold}
\affiliation{NASA Goddard Space Flight Center, Greenbelt, MD, 20771, USA}

\author[0000-0002-1590-0568]{Shardha Jogee}
\affiliation{Department of Astronomy, The University of Texas at Austin, 2515 Speedway, Stop C1400, Austin, Texas 78712, USA}

\author[0000-0003-1885-2490]{Sydney Sherman}
\affiliation{Department of Astronomy, The University of Texas at Austin, 2515 Speedway, Stop C1400, Austin, Texas 78712, USA}

\author{Jonathan Florez}
\affiliation{Department of Astronomy, The University of Texas at Austin, 2515 Speedway, Stop C1400, Austin, Texas 78712, USA}

\author{Caryl Gronwall}
\affiliation{Department of Astronomy \& Astrophysics, The Pennsylvania State University, University Park, PA 16802, USA}
\affiliation{Institute for Gravitation and the Cosmos, The Pennsylvania State University, University Park, PA 16802, USA}


\begin{abstract}
We study the evolution in the number density of the highest mass galaxies over $0.4<z<1.5$ (covering 9 Gyr). We use the \emph{Spitzer}/HETDEX Exploratory Large-Area (SHELA) Survey, which covers 17.5 $\mathrm{deg}^2$ with eight photometric bands spanning 0.3--4.5~\micron\ within the SDSS Stripe 82 field.  \editthree{This size produces the lowest counting uncertainties and cosmic variance yet for massive galaxies at $z\sim1.0$}.  We study the stellar mass function (SMF) for galaxies with $\log(M_\ast/M_\odot)>10.3$ using a forward-modeling method that fully accounts for statistical and systematic uncertainties on stellar mass. From $z$=0.4 to 1.5 the massive end of the SMF shows minimal evolution in its shape: the characteristic mass ($M^\ast$) evolves by less than 0.1~dex ($\pm$0.05~dex);  the number density of galaxies with $\log M_\ast/M_\odot >11$ stays roughly constant at $\log (n/\mathrm{Mpc}^{-3})$ $\simeq$ $-$3.4 ($\pm$0.05), then declines to $\log n/\mathrm{Mpc}^{-3}$=$-$3.7 ($\pm$0.05) at $z$=1.5. We discuss the uncertainties in the SMF, which are dominated by assumptions in the star formation history and details of stellar population synthesis models for stellar mass estimations. \editthree{For quiescent galaxies, the data are consistent with no (or slight) evolution ($\lesssim0.1$~dex) in the characteristic mass nor number density from $z\sim 1.5$ to the present.} This implies that any mass growth (presumably through ``dry' mergers) of the quiescent massive galaxy population must balance the rate of mass losses from late-stage stellar evolution and the formation of quenching galaxies from the star-forming population. We provide a limit on this mass growth from $z=1.0$ to 0.4 of $\Delta M_\ast/M_\ast\leq$~45\% (i.e., $\simeq0.16$~dex) for quiescent galaxies more massive than $10^{11}$~$M_\odot$.
\end{abstract}

\keywords{galaxies: evolution -- galaxies: abundance}

\section{Introduction} \label{sec:intro}
\par One of the major features of the cold dark matter dominated universe is the hierarchical structure-formation, by which increasingly larger dark matter halos are formed through the assembly of smaller ones. As galaxies reside in these halos, they trace the underlying dark matter distribution, and therefore we expect these galaxies to undergo hierarchical growth as well \citep[e.g.,][]{White1978,Blumenthal1984, White1991,Lacey1993,Springel2005}. 

\par Within this hierarchical growth, it is generally argued that the formation of massive galaxies follows a ``two-phase'' formation scenario \cite[e.g.,][]{Oser2010,Oser2012,Wellons2015,Belli2019}. According to this scenario, galaxies form compact cores through an early rapid phase of dissipational in situ star formation  at $2 \lesssim z  \lesssim 6$  \citep{Keres2005,Dekel2009apj,Oser2010} and/or gas-rich mergers \citep{Weinzirl2011,Wellons2015}. The subsequent evolution is dominated by assembly of stellar halos by dissipationless minor (dry) mergers \citep[e.g.,][]{Khochfar2006,Naab2006,Oser2010,Oser2012,Johansson2012,Hilz2013}. This two-phase formation scenario provides an explanation for the observed growth in the effective radii \cite[e.g.,][]{Newman2012,VanderWel2014} and stellar halos of massive galaxies \citep[e.g.,][]{Szomoru2012,Patel2013,Huang2018}.

\par Numerical simulations \citep[e.g.,][]{Oser2010,Wellons2015}, semi-analytic models \citep[SAM; e.g.,][]{Lee2013,Lee2017}, and stellar-mass---halo-mass (SHAM) analyses \citep[e.g.,][]{Moster2013,Moster2018,Behroozi2013,Rodriguez2017}  generally show that the fraction of accreted stars through mergers increase with total galaxy stellar mass or halo mass \citep[e.g.,][]{Lackner2012,Cooper2013,Rodriguez2016,Qu2017,Pillepich2018}.
%
%
For example, \citet{Qu2017} analyzed the mass assembly of central galaxies in the EAGLE cosmological simulation and found that $\sim20\%$ of the stellar mass of present day massive galaxies ($>10^{11}M_{\odot}$) is built up through mergers, and more massive galaxies have experienced more stellar-mass growth by mergers. The implied growth should be reflected in the evolution of the characteristic mass and number density in the galaxy stellar mass function (SMF), particularly during the past $\sim$10~Gyr from $z\sim$1.5 to the present \citep[e.g.,][]{Conroy2009,Mutch2013,Moster2013,Behroozi2013,Rodriguez2017}. These studies also consistently show that more massive galaxies have experienced more stellar mass growth by mergers.

\par 
%
Previous observational studies of the galaxy SMF have been based mostly on deep surveys \editthree{\citep[see, e.g.,][]{Conselice2007,Pozzetti2007,Drory2009,Brammer2011,Moustakas2013,Tomczak2014,Muzzin2013,Mortlock2015,Davidzon2017,Wright2018, Arcila-Osejo2019}}. These studies provide the global view on the evolution of stellar mass assembly over a wide range of redshift and mass.  These different studies have consistently demonstrated that since $z\sim1$ most massive galaxies have undergone less evolution than their less massive counterparts, revealing a faster stellar mass assembly for the most massive systems \citep[e.g.,][]{Fontana2006,Pozzetti2007,Moustakas2013,Beare2019}. However, given their small angular coverage, \edittwo{these surveys are subject to relatively large cosmic variance}, particularly at low redshift ($z \lesssim 1$). 
Cosmic variance uncertainties add noise and dilute the signal of the true evolution of galaxy number densities.  Imaging surveys that cover large cosmic volumes are crucial for probing the accurate number densities of galaxies, particularly at the high-mass end where the exponentially declining SMF makes them very rare.  

\par  
%
Previous attempts \edittwo{to measure the evolution of the SMF out to $z\lesssim 1$} have utilized surveys covering tens of square degrees, in order to minimize the contribution from cosmic variance and focus on the evolution of high-mass galaxies \editthree{\citep[e.g.,][]{Maraston2013,Bundy2017,Capozzi2017}.  However, these studies often lack coverage to the rest-frame near-IR, which is needed to derive accurate stellar masses,} nor the depth required to measure the evolution of the SMF to very higher redshift $(z \gtrsim 1)$.  \citet{Matsuoka2010} presented an analysis of massive ($M_{\ast} > 10^{11}M_{\odot}$) galaxies out to $z=1$ using $55\ \mathrm{deg}^{2}$ of the UKIDSS Large Area Survey $K$-band images on the SDSS southern equatorial stripe.  These authors found that massive galaxies with $M_{\ast} = 10^{11.0-11.5} M_{\odot}$ and $M_{\ast} = 10^{11.5-12.0} M_{\odot}$ have experienced rapid growth in the number density since $z=1$, by factors of 3 and 10, respectively. Similarly, \citet{Moutard2016} analyzed the evolution of the galaxy SMF from $0.2 < z < 1.5$ using a $K_{s} < 22-$selected, photometric redshift-based sample over $\sim22.4~\mathrm{deg}^2$ of the footprint of the VIPERS spectroscopic survey. The authors provided evidence \edittwo{for a factor of $\sim2$ increase in the number density of massive galaxies ($M_{\ast} > 10^{11.5}M_{\odot}$)} from $z\sim1$ to $z\sim0.3$. Even though both \citet{Matsuoka2010} and \citet{Moutard2016} detected growth in the number density of massive galaxies, there is an inconsistency concerning the amount of the evolution in number density

\par 
The discrepancy in the evolution of the number density of massive galaxies highlights the challenges of probing the high-mass end of the SMF and raises concerns about other systematic errors. One of the largest sources of uncertainty comes from assumptions in the modeling of the galaxy spectral energy distribution (SED), such as model templates, initial mass function, metallicity, and treatment of dust attenuation. This could significantly contribute to the total error budget, and consequently affect the robustness of the detected evolution of galaxy stellar mass function \citep{Marchesini2009}. \citet{Bundy2017} recently studied these issues and their effect on the evolution of the SMF by exploiting the Stripe 82 Massive Galaxy Catalog (S82-MGC) and constructing a mass-limited sample at $0.3 < z  < 0.65$ that is complete to $M_{\ast} > 10^{11.3}M_{\odot}$, over a large area of $140~\mathrm{deg}^2$. After accounting for both random and systematic uncertainties,  they reported no evolution in the characteristic stellar mass of the stellar mass function over the redshift range probed, \edittwo{contrasting with the findings of both \citet{Matsuoka2010} and \citet{Moutard2016}}. 


\par 
Here, we utilize the 17.5 $\mathrm{deg}^2$ \emph{Spitzer}/HETDEX Exploratory Large-Area Survey (SHELA) survey dataset to probe the stellar mass function, particularly for massive galaxies with $\log (M_{\ast} / M_{\odot}) >10.3$ over $0.4 < z < 1.5$.  At these redshifts, SHELA covers $\sim0.15~\mathrm{Gpc}^3$ in comoving volume.  This allows us to test the evolution of the SMF using a method similar to that of \citet[see above]{Bundy2017}, 
%
%
but using a sample of galaxies with a larger redshift range (out to $z < 1.5$) and extending to lower stellar masses (down to $\log M_\ast/M_\odot = 10.3$).   Motivated by \citeauthor{Bundy2017}, we consider the potential systematic uncertainties in the derivation of stellar masses in our sample, including the assumptions in modeling SED and random errors. After accounting for systematic uncertainties arising from differences in star-formation histories and stellar population synthesis (SPS) models, we find no redshift evolution ($\lesssim0.1$~dex) in the characteristic stellar mass ($M^{\ast}$) and cumulative number density of massive galaxies  ($>10^{11}~M_{\odot}$) from $z=1.0$ to $z=0.4$. In contrast, we find a 0.3 dex increase in the number density of massive galaxies from $z=1.5$ to $z=1.0$, where our sample is mass-complete.
%

\par The plan for this paper is as follows. Section~\ref{sec:dataset} begins by summarizing how we build our sample from eight photometric bands spanning optical to mid-infrared wavelength. In Section~\ref{sec:tractorphotoz}, we detail how we assign photometric redshifts to our galaxies and in Section~\ref{sec:massestimates}, we describe the various estimates of stellar mass. In Section~\ref{sec:method} we discuss potential biases in the derived SMF for large samples, including the impact of the stellar mass uncertainties on the SMF. Additionally, we discuss the effect of cosmic variance on the measured SMF.   We present our results in Section~\ref{sec:results}, and we discuss the significance of our results as well as comparisons to other works in Section~\ref{sec:discussion}.   Finally, we provide summary in Section~\ref{sec:conclusion}. Throughout this paper, we use the AB magnitude system and adopt a standard cosmology with $H_{0}=70~h_{70}~\mathrm{km}~\mathrm{s}^{-1}~\mathrm{Mpc}^{-1}$, $\Omega_{M} =  0.3$, and $\Omega_{\Lambda} =0.7$, consistent with Planck 2018 data \citep{Planck2018} and local measurements of $H_0$ \citep{Riess2019}.

\section{Data and sample selection}

\label{sec:dataset}
\subsection{NEWFIRM $K$ photometry}
For this study, we use a new $K$-band-selected catalog for SHELA.  This catalog is based on imaging from the NEWFIRM-HETDEX Survey (Stevans et al. 2019, \edittwo{submitted}), which covers $17.5~\mathrm{deg}^2$ to a median $5\sigma$ depth of $K_s$=22.7 AB mag (2\arcsec-diameter apertures).    The advantage of using the $K$--band catalog is that for our galaxies of interest, this filter samples the rest-frame near-IR ($\sim1$~\micron) and is therefore more sensitive to the older stellar populations which dominate the light of many systems. We have also verified that including the $K$-band photometry improves the quality of our photometric redshifts and stellar population parameter fits (including stellar masses).   
%
A full description of the NEWFIRM $K$-band imaging, catalog construction, and derived properties (photometric redshifts and stellar masses) is provided by Stevans et al. 2019 (submitted).  For reference, we use the NEWFIRM $K$-band photometry measured in the FLUX\_AUTO apertures from SExtractor (this is relevant to the next subsection).  

\subsection{Forced Photometry from DECam $ugriz$ and Spitzer/IRAC data}
\par  In addition to the NEWFIRM $K$-band data, the SHELA survey includes $ugriz$ imaging from the Dark Energy Camera over 17.5$\ \mathrm{deg}^2$ (DECam; \citet{Wold2019}) and deep 3.6 and 4.5 $\mu$m imaging from  \emph{Spitzer}/IRAC \citep{Papovich2016}. The $riz-$band selected DECam catalogs reach a $5\sigma$ depth of $\sim24.5$ AB mag. \edittwo{However, we do not use the DECam catalogs and the DECam-selected \emph{Spitzer}/IRAC forced photometry \citep{Wold2019}. In our analysis, we use our new $K$-band image for detection for the reason we have described in the previous section.} Following our work in \cite{Kawinwanichakij2018} and \cite{Wold2019},  we perform ``forced photometry'' to derive optimal fluxes in the $ugriz$ + [3.6] and [4.5] data for sources detected in the $K_s$ catalog. We use the code ``The Tractor''  \citep{Lang2016aj, Lang2016ascl} for this process.
%
%
This allows us to measure flux densities (or accurate limits) even for sources below the $5\sigma$ depth threshold of the original DECam or IRAC imaging or for sources blended at the resolution of {\it Spitzer}/IRAC.
%
%
We follow identical procedures as those described in \cite{Kawinwanichakij2018} and \cite{Wold2019} except we use the NEWFIRM $K$--band image for detection. \edittwo{With this technique, our forced photometric IRAC catalog is $80\%$ completeness to limiting magnitude of 22.6 AB mag in both 3.6 and 4.5 $\mu$m bands, in contrast to 22.3 AB mag for the original published {\it Spitzer}/IRAC--selected catalog \citep{Papovich2016}}.  

\section{Photometric redshift estimates}
\label{sec:tractorphotoz}
To estimate photometric redshifts of our eight-band photometric data set (NEWFIRM $K$, DECam $ugriz$, and Spitzer/IRAC 3.6 and 4.5 $\mu$m), we use \editthree{the publicly available software package EAZY-py\footnote{https://github.com/gbrammer/eazy-py} which is based on the EAZY code} \citep{Brammer2008}. We utilize EAZY-py's ability to apply a $K$-band magnitude prior\editthree{, and the software's ``template error function'' option to account for both random and systematic differences between observed photometry and the templates. This allows us to minimize systematic errors in the photometric redshift without the need to optimize either the templates or the photometry  \citep[see][]{Brammer2008}. We use the default template error function, set the amplitude of the template error function (TEMP\textunderscore ERR\textunderscore A2) to 0.20 and the minimal fractional error added to the uncertainties of every filter and at every redshift ($\sigma_{\mathrm{sys}}$) to 0.01. These parameters were chosen so that the median offset and the scatter when comparing the photometric redshift to the spectroscopic redshift are minimized (see below).}
%
%
\par
\begin{figure}
	\centering
	\includegraphics[width=0.5\textwidth]{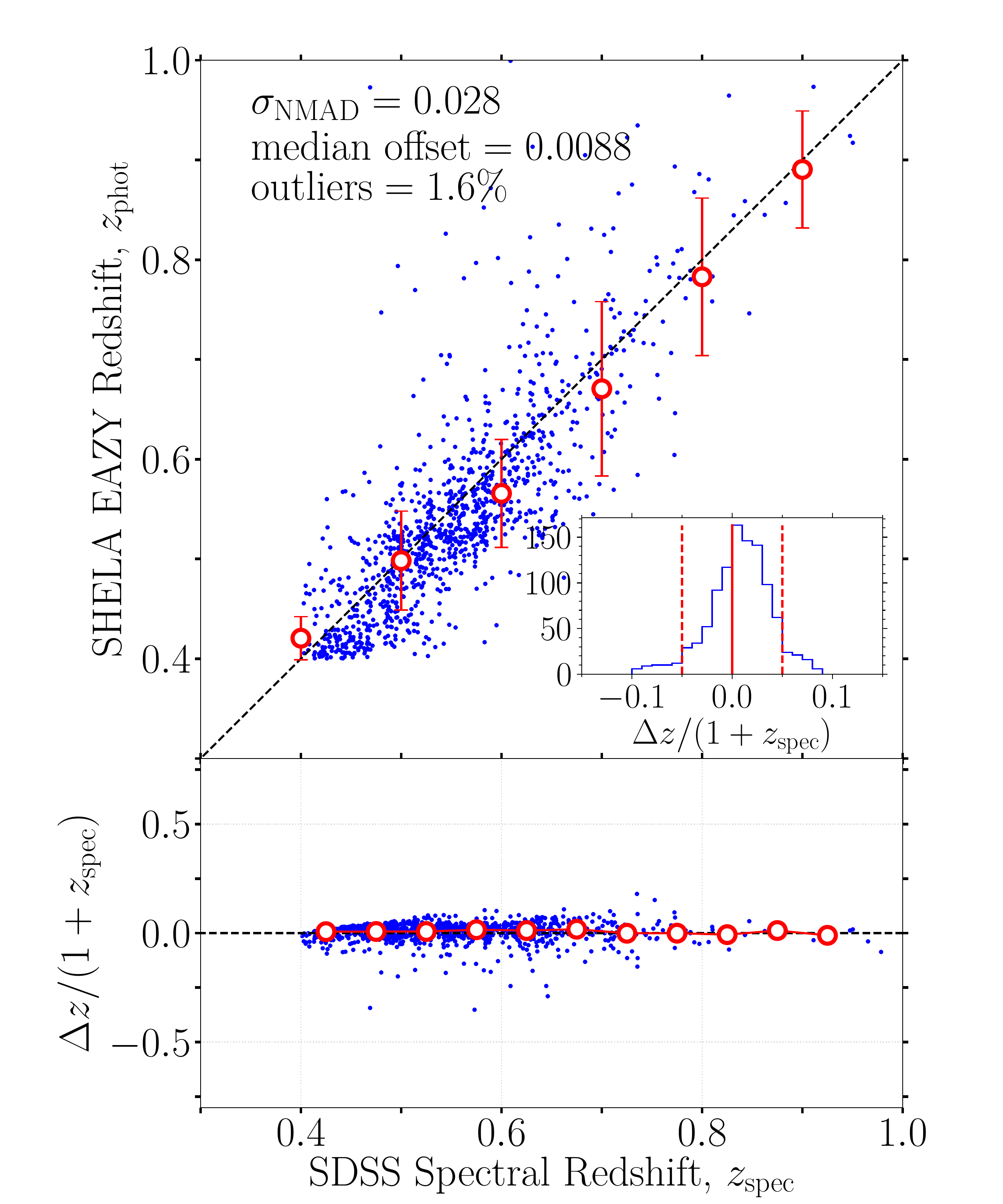}
	\caption{Comparison between our SHELA photometric redshifts ($z_{\mathrm{phot}}$) and spectroscopic redshift ($z_{\mathrm{spec}}$) from SDSS DR13 for our galaxy sample at $0.4 < z < 1.0$ with $\log(M_\ast/M_\odot) > 10$, our mass completeness at $z\sim1$. We derived the photometric redshifts using EAZY-py \citep{Brammer2008} with eight-band photometry: NEWFIRM $K$, DECam $ugriz$, and IRAC 3.6 and 4.5 $\mu$m. The $\sigma_{\mathrm{NMAD}}$ denotes 1.48 times the median absolute deviation of the difference between $z_{\mathrm{spec}}$ and $z_{\mathrm{phot}}$ ($\Delta z$), normalized to $1+z_{\mathrm{spec}}$. The percentage of outlier corresponds to the fraction of sources with $\Delta z/ (1+z_{\mathrm{spec}})$ exceeding $5\sigma$. }
	\label{fig:photoz}
\end{figure}

 We compared our photometric redshifts to the SDSS spectral catalog (DR13; \citealt{Albareti2017}), which includes optical spectra of galaxies and quasars from the Baryonic Oscillation Spectroscopic Survey (BOSS). For our purpose, we select galaxies from the SDSS spectral catalog within the SHELA footprint using CLASS=`GALAXY'.
Figure~\ref{fig:photoz} demonstrates the quality of our derived photometric redshifts (EAZY's $z_{{\rm {peak}}}$ parameter\footnote{This parameter corresponds to the peak probability of the P($z$) function, and is considered the best $z_{{\rm {photo}}}$ estimate \citep{Muzzin2013}.}) by comparing our dataset to the available SDSS spectroscopic sample at $z<1.0$. 
Focusing on the galaxy sample at $0.4 < z < 1.5$ with $\log(M_\ast/M_\odot) > 10$ (this is our $80\%$ stellar mass completeness limit at $z=1$; see Section~\ref{sec:masscomplete}), the median offset (bias) between the spectroscopic and photometric redshift $\Delta z=(z_{{\rm {photo}}}-z_{{\rm {spec}}})/(1+z_{\rm{spec}})$ is 0.0088.  Similarly, the normalized median absolute deviation (the scatter), defined as:
\begin{equation}
\sigma_{\mathrm{NMAD}}=1.48\times \mathrm{median}\left(\left|\frac{\Delta z-\mathrm{median}(\Delta z)}{1+z_{{\rm {spec}}}}\right|\right),
\end{equation}
is 0.028 with $\sim2\%$ of sources found to be $5\sigma$ outliers. 
%



\section{Stellar Masses Estimates}
\label{sec:massestimates}



\subsection{Derivation of Stellar Mass Estimates}
\par For all of the analyses here, we estimate the stellar mass, $M_{\ast,\mathrm{iSED}}$, using the Bayesian \emph{iSEDfit}\footnote{https://github.com/moustakas/iSEDfit} package \citep{Moustakas2017} presented in \citet{Moustakas2013}. The \emph{iSEDfit} code performs a refined grid search of the posterior distributions of stellar mass and enables priors with nonflat probability distributions.  The advantage of using \emph{iSEDfit} is that it allows the use of different assumptions in the star-formation histories (including ``bursts'' of star-formation) and in the stellar population synthesis (SPS) models, including those of the Flexible Stellar Population Synthesis\footnote{\editthree{As described in \citet{Moustakas2013} and \cite{Conroy2010}, we use the FSPS models to the Padova stellar evolutionary isochrones \citep{Girardi2000,Marigo2007,Marigo2008}. These evolutionary tracks have been supplemented with the post-AGB models of \cite{Vassiliadis1994}. The integrated spectra are generated using the empirical MILES stellar library \citep{Sanchez2006}. }} \citep[FSPS;][]{Conroy2010ascl,Conroy2010}, \cite{Bruzual2003}, and \cite{Maraston2005} models.  In this way we are able to test for systematic uncertainties in the stellar masses resulting from differences in the underlying assumptions, without systematic uncertainties resulting from different stellar-population fitting codes.  We summarize the key aspects of \emph{iSEDfit} code below.


\par  For \emph{iSEDfit}, we adopt fiducial prior parameters for $M_{\ast,\mathrm{iSED}}$ from  \citet{Moustakas2013}.   The basic set of  \emph{iSEDfit} priors is based only on a set (randomly generated for each run of \emph{iSEDfit}) of star-formation histories using 10,000 declining exponential models, where SFR $\propto$ $\exp(-t/\tau)$, for age $t$ and e-folding timescale $\tau$. The parameters for each \emph{iSEDfit} model vary independently, and it is therefore important to sample the entire the range of each prior.  We also allow the \emph{iSEDfit} age $t$ (time since the onset of star formation) of each model to have uniform probability from $0.1-13$~Gyr\editthree{; however, we disallow ages older than the age of the universe at the redshift of each galaxy.}   We draw the e-folding time ($\tau$) from the linear range $0.1-5$ Gyr. We assume a uniform prior on stellar metallicity, $Z$, in the range of $Z=0.004-0.03$ (roughly 20\%-150\% times the solar metallicity; \citealt{Asplund2009}).   For the stellar masses based on \cite{Bruzual2003} and FSPS \citep{Conroy2010ascl,Conroy2010} models, we assume the \cite{Chabrier2003} initial mass function (IMF).  For the stellar masses derived from \cite{Maraston2005} models, we assume the \cite{Kroupa2001} IMF (these choices of IMF produce systematic shifts in the derived stellar masses of $\simeq$0.04~dex).   Finally, we adopt the time-dependent dust--attenuation curve of \citet{Charlot2000}, in which stellar populations older than 10 Myr are attenuated by a factor of $\mu$ times less than younger stellar populations. Following \citet{Moustakas2013},  we draw $\mu$ from an order four Gamma distribution that  range from zero to unity centered on a typical value ($\left \langle  \mu \right \rangle=0.3$).

\par  We consider both smooth star-formation histories and superpositions of smooth star-formation histories with ``bursts'' of star-formation with varying strengths and duration.
%
%
Following \cite{Bundy2017} and \cite{Moustakas2013}, we add stochastic bursts randomly to the star formation histories (SFHs).   For every 2 Gyr interval over the lifetime of a given model, the cumulative probability that a burst occurs is $P_{\mathrm{burst}}= 0.2$.  The SFH of each burst is modeled as a Gaussian as a function of time, with an amplitude of $F_b$, defined as the total amount of stellar mass formed in the burst divided by the underlying mass of the smooth SFH at the burst's peak time. The values of $F_b$ are drawn from the range of $0.03-4.0$. The allowed burst duration (or the width of a Gaussian distribution characterizing the SF burst)  uniformly ranges from 0.03 to 0.3 Gyr. 

Additionally, we allow the time for the onset of star formation of each model to range with equal probability from 0.1 to 13$~\mathrm{Gyr}$ \citep{Salim2007,Wild2009}\editthree{, but as above, we restrict these times to be less than the age of the universe at the redshift of each galaxy.} 

We then apply \emph{iSEDfit} to the galaxies in our catalog with difference assumptions for the star-formation histories, stellar population libraries, and models.   Table~\ref{table:massestimates} describes the details of each set of runs. 
We adopt fiducial prior parameters for $M_{\ast,\mathrm{iSED}}$ from  \citet{Moustakas2013}, marginalizing over all stellar population parameters, to produce posterior probability distribution functions (PDFs). We then adopt the median PDF as the reported value for each quantity and derive 68\% confidence intervals by taking the values that correspond to the PDF integrated between 0.16 and 0.84.  We refer to the stellar mass derived for each of these runs using the ``name'' listed in the first column of Table~\ref{table:massestimates}. We discuss how the different model assumptions impact both the stellar mass estimates and the derived SMF in Section~\ref{sec:results}. 

\par
\begin{figure*}
	\centering
	\includegraphics[width=1.0\textwidth]{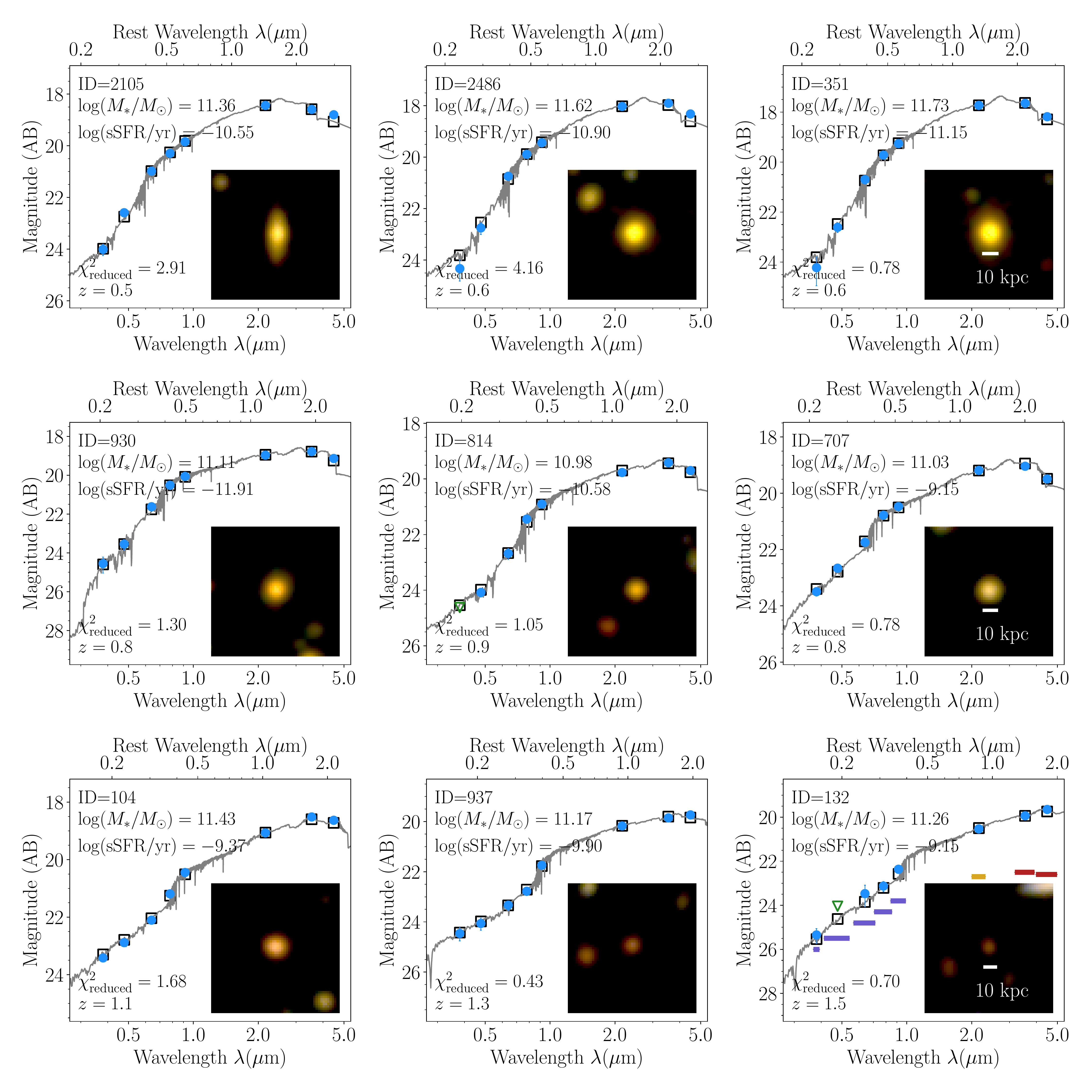}
	\caption{Representative examples of the spectral energy distributions of our SHELA galaxies with $\log(M_\ast/M_\odot) > 11$ ordered by redshift:  $0.5 < z < 0.75$ (top row), $0.75 < z < 1.0$ (middle row), and $1.0 < z < 1.5$ (bottom row). Observed datapoints (DECam $ugriz$, NEWFIRM $K$, Spitzer/IRAC 3.6 and 4.5 $\mu$m) are shown as blue circles with error bars. Upper limits are indicated with green triangles. The solid curves and squares are the \emph{iSEDfit} best-fit SEDs and photometry based on FSPS models without stochastic bursts. The inset is an 15\arcsec$\times$ 15\arcsec~false-color RGB image of the corresponding galaxy. In each RGB image, the red, green, and blue colors correspond to the image from the DECam $z$-, $i$-, and $g$-bands, respectively.
	 In the last panel (bottom right), the colored bars show the $80\%$ completeness limits for our SHELA/DECam $ugriz$ and Spitzer/IRAC 3.6 and 4.5 $\mu$m from the forced photometry of NEWFIRM $K$-band-selected sources, as well as the median 5$\sigma$ depth for the NEWFIRM $K$-band.}
	\label{fig:sedplotRGB}

\end{figure*}

 \edittwo{In Figure~\ref{fig:sedplotRGB} 
we show representative examples of SHELA galaxies with $\log(M_\ast/M_\odot) > 11$ in three redshift bins: $0.5 < z < 0.75$, $0.75 < z < 1.0$, and $1.0 < z < 1.5$. The best-fit SEDs and photometry are based on FSPS models without stochastic bursts. Figure~\ref{fig:sedplotRGB} also shows false-color images of the massive galaxies in the DECam $z$- (red color) combined with $i$- (green color) and $g$-band (blue color) images. By inspection, our massive galaxies are typically spheroidal, or reddened, bulge-dominated disks.}

\par \editthree{Finally, we note that our stellar mass estimates ($M_{\ast,\mathrm{iSED}}$) refer to the stellar mass implied via the visible flux from the living stellar population within a galaxy, and not the total living plus stellar remnants (i.e., white dwarf, neutron stars, black holes etc.). Stellar remnants can make an important contribution to the total stellar mass of a galaxy and the SMFs for massive galaxies at low-redshift \citep[e.g.,][]{Shimizu2013,Bernardi2016}. For example, \cite{Shimizu2013} found a weak correlation between the remnant mass fraction and the total stellar mass of galaxies, and the remnant fraction can be regarded as a redshift-dependent constant. Also, the shape of the SMF is almost unchanged, but simply shifts horizontally depending on the inclusion or omission of the remnant mass. This shift in the SMF is larger at lower redshift ($\sim0.05$ dex at $z=3$ and $\sim0.15$ dex at $z=0$). However, we find that this difference is small and comparable to the SHELA stellar mass uncertainty, which we already take into account using a forward-modeling method. Therefore, the choice of including or excluding stellar remnants to the ``stellar mass" does not significantly impact our inferred evolution of the SMFs. } 
 

\begin{deluxetable*}{cccccc}
\tablecaption{Stellar mass estimates \label{table:massestimates}}
\tablehead{ & & & \multicolumn{3}{c}{Prior} \\
\colhead{Name} & \colhead{Models} & \colhead{Star formation history} &  \colhead{Metallicity ($Z$)} & \colhead{Decay time scale ($\tau/\mathrm{Gyr}$)} & \colhead{Age ($t/\mathrm{Gyr}$)}} 
\colnumbers
\startdata
$M_{\ast,\mathrm{iSED}}^{\mathrm{FSPS,\ burst}}$  & FSPS & Exponentially 
&  [0.004,0.03] &[0.5-1] & [0.1-13]  \\
& \citep{Conroy2010ascl}  & declining with $P_{\mathrm{burst}} = 0.2$ &  &  \\ \\
$M_{\ast,\mathrm{iSED}}^{\mathrm{FSPS,\ no\ burst}}$  & FSPS   & Exponentially & [0.004,0.03] & [0.5-1] & [0.1-13]  \\
& \citep{Conroy2010ascl} & declining  &  & \\ \\
$M_{\ast,\mathrm{iSED}}^{\mathrm{BC03}}$  &  \cite{Bruzual2003}  & Exponentially & [0.004,0.03] & [0.5-1] &[0.1-13] \\
& & declining &   & \\ \\
$M_{\ast,\mathrm{iSED}}^{\mathrm{Ma05}}$  & \cite{Maraston2005}  & Exponentially & [0.004,0.03] & [0.5-1] & [0.1-13] \\ 
& & declining  &  &  \\ \\
\enddata
\tablecomments{(1) Name of the stellar mass estimate, (2) stellar population synthesis (SPS) model,  (4) the prior on metallicity, (5) the prior on exponentially decline star-formation time scale, $\tau$, in unit of Gyr, (6) the prior on time for the onset of star formation, $t$, in unit of Gyr. Priors of the form [A,B] are flat with minimum and maximum given by A and B. $P_{\mathrm{burst}}$ denotes the cumulative probability that a burst occurs. We assume \cite{Chabrier2003} initial mass function.}
\end{deluxetable*}

\subsection{Effects of Model Assumptions on Stellar Mass Estimates}

\begin{figure*}

\centering
\includegraphics[width=1.0\textwidth]{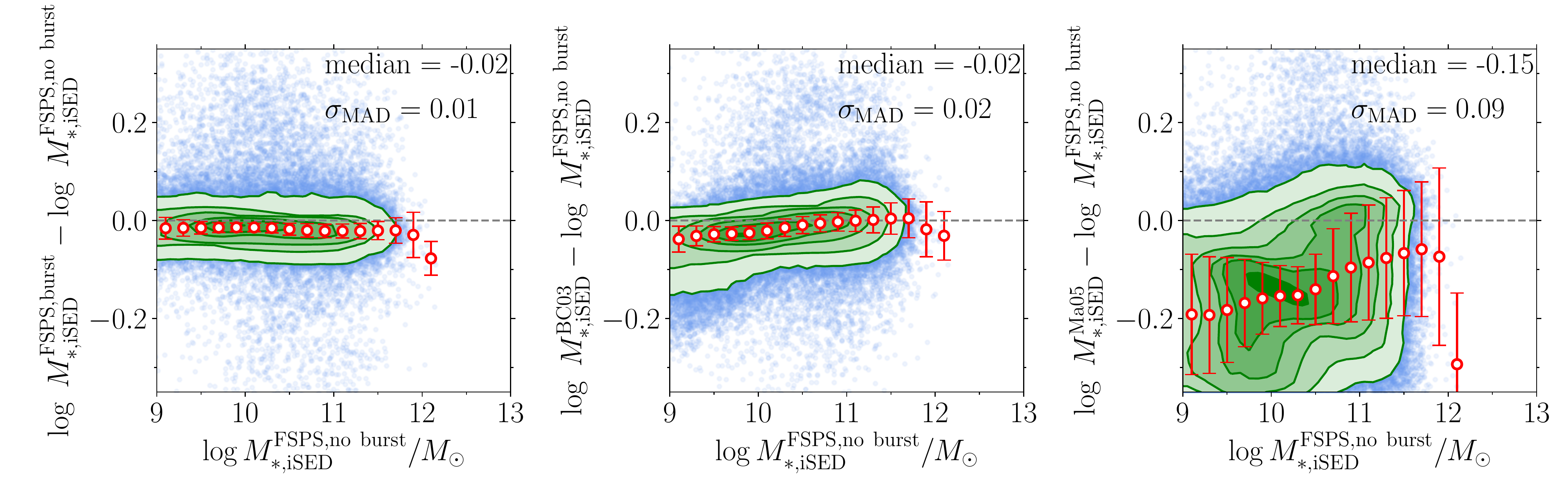}

\caption{Comparison of stellar masses derived from \emph{iSEDfit} ($M_{\ast,\mathrm{iSED}}$), based on FSPS models without stochastic bursts ($M_{\ast,\mathrm{iSED}}^{\mathrm{FSPS, no~burst}}$) with other \emph{iSEDfit} stellar masses: FSPS models with stochastic bursts ($M_{\ast,\mathrm{iSED}}^{\mathrm{FSPS, burst}}$; left), \cite{Bruzual2003} SPS models ($M_{\ast,\mathrm{iSED}}^{\mathrm{BC03}}$; middle), and \cite{Maraston2005} SPS models ($M_{\ast,\mathrm{iSED}}^{\mathrm{Ma05}}$; right). The  $\sigma_{\mathrm{MAD}}$ denotes 1.48 times the median absolute deviation of the difference between stellar masses. The large red open circles with error bars indicate the median and $\sigma_{\mathrm{MAD}}$ in each bin of $M_{\ast,\mathrm{iSED}}^{\mathrm{FSPS, no~burst}}$. The green contours indicates the distribution of the stellar mass differences from $0.5\sigma$ to $3\sigma$ in units of 0.5$\sigma$.}
\label{fig:comparefastised}
\end{figure*}

The left panel of Figure~\ref{fig:comparefastised} compares the stellar masses of FSPS, no burst star-formation histories ($M_{\ast,\mathrm{iSED}}^{\mathrm{FSPS, no~burst}}$) to models with bursts ($M_{\ast,\mathrm{iSED}}^{\mathrm{FSPS, burst}}$).  For these two cases, the  median offset is $\sim-0.02$ dex, and the scatter is tight ($0.01$~dex).  There is no measurable dependence on stellar mass for the range ($M_{\ast}=10^{9-12}~M_{\odot}$).

The middle of Figure~\ref{fig:comparefastised} shows the comparison between FSPS $M_{\ast,\mathrm{iSED}}^{\mathrm{FSPS, no~burst}}$ stellar masses and those based on the \cite{Bruzual2003} SPS model ($M_{\ast,\mathrm{iSED}}^{\mathrm{BC03}}$).  In both cases we compare models with smoothly varying SFHs.   For these two cases, the median offset is $-$0.02 dex, with a weak dependence on stellar mass.  The scatter is likewise small with  $\sigma$=0.02 dex. \editthree{This result confirms our expectations that adding bursts primarily modifies the bluer bandpasses of a galaxy, while leaving the redder wavelengths, which count the accumulated stellar mass of a galaxy, relatively unaffected.}

The right panel of Figure~\ref{fig:comparefastised} compares the stellar masses of FSPS $M_{\ast,\mathrm{iSED}}^{\mathrm{FSPS, no~burst}}$ models to those based on the \cite{Maraston2005} ($M_{\ast,\mathrm{iSED}}^{\mathrm{Ma05}}$) stellar population models.  These SPS models exhibit larger mass dependent offsets, with a median of $-$0.15 dex and a scatter substantially larger than in the previous comparisons, $\sigma$=0.09 dex. \edittwo{This is most likely due to the different treatment of the thermally pulsing asymptotic giant branch (TP-AGB) phase; the \cite{Maraston2005} models have a much larger flux at wavelengths longward of $\sim0.7-0.8\mu m$.} 


In summary, the stellar masses for the galaxies in our samples are fairly robust to changes in star-formation history or model stellar population (with the exception of the \citeauthor{Maraston2005} models). 
\edittwo{For this reason, we average the results from the SMFs derived from different mass estimates (defined as ``assumption-averaged'' SMF, see Section~\ref{sec:assumpavgsmf}) to infer biases associated with assumptions of star formation history and SPS models.} We adopt the assumption-averaged SMF as our fiducial measurement. However, in Section~\ref{sec:dependpop}, we further discuss how the differences in stellar masses from the different model assumptions impact the SMFs.


\subsection{Stellar mass completeness limit}
\label{sec:masscomplete}
\par
\begin{figure}
	\centering
	\includegraphics[width=0.48\textwidth]{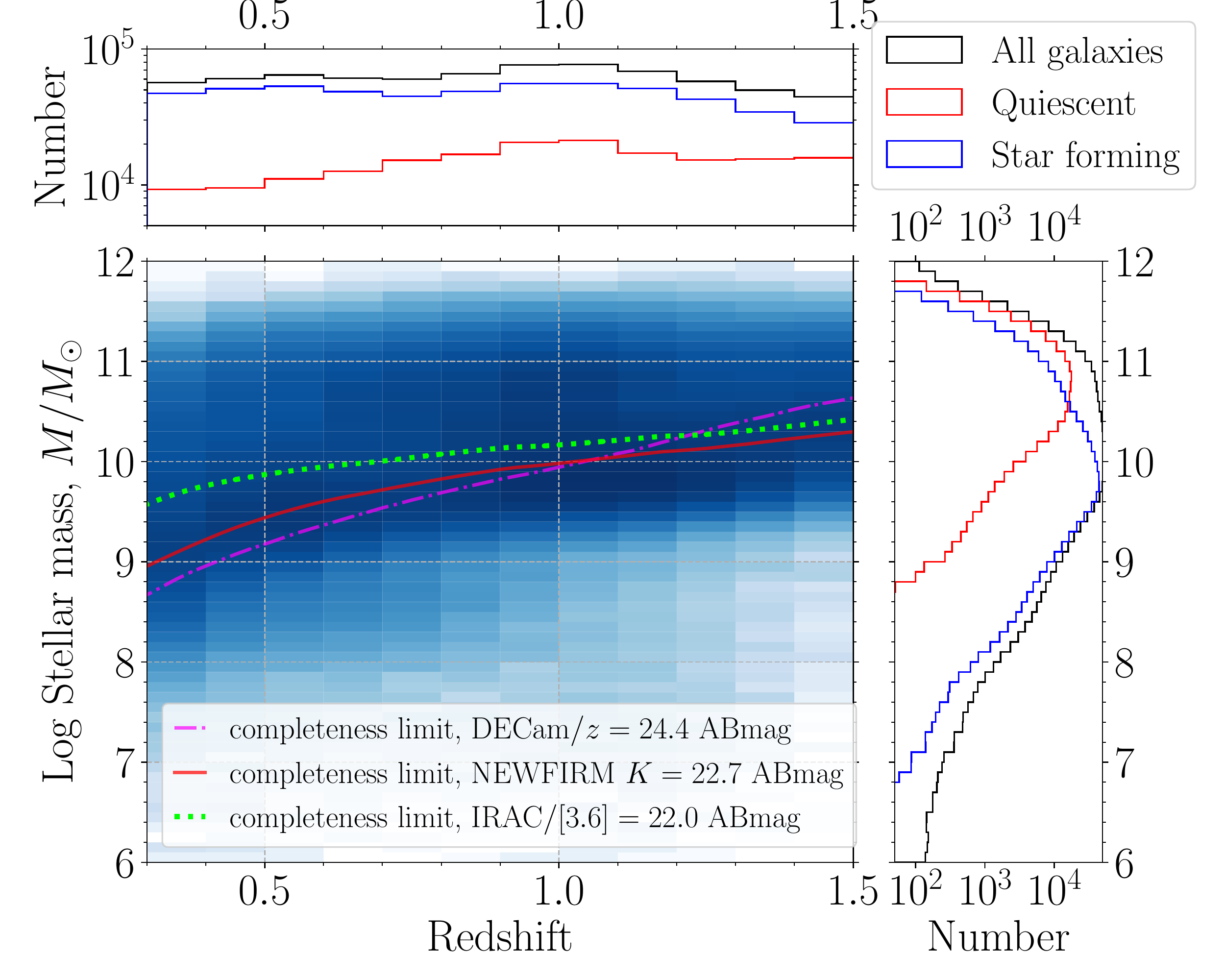}
	\caption{Distribution of stellar masses derived from \emph{iSEDfit}, based on \cite{Bruzual2003} models without burst ($M_{\ast,\mathrm{iSED}}^{\mathrm{BC03}}$) as function of redshift. The colorscale indicates the density in each bin of stellar mass (0.1 dex) and redshift ($\Delta z = 0.1$). The solid red line indicates the mass completeness limit determined from passively evolving an SSP with a formation redshift $z_f=4$, when using a limiting NEWFIRM $K$-band magnitude of 22.7 AB mag ($5\sigma$ depth in 2\arcsec-diameter apertures). The dotted dashed magenta line indicates the mass completeness limit determined from an SSP with $z_f = 4$ when using a limiting DECam/$z$-band magnitude of 24.4 AB mag (80\% complete). The dotted green line indicates the mass completeness limit determined from an SSP with $z_f = 4$ when using a limiting IRAC/[3.6] magnitude of 22.0 AB mag (80\% complete). \edittwo{The distributions of redshift and stellar mass for quiescent (red), star forming (blue), and all galaxies (black) are shown in the top and the right panels, respectively.}}
	\label{fig:massredshift}
\end{figure}

 We estimated our stellar mass completeness limits by using \cite{Bruzual2003} to generate a series of solar metallicity Simple Stellar Populations (SSPs) models with a formation redshift of $z_f = 4$.  We then and used EZGal \citep{Mancone2012} to infer the models’ observable and compared these values to our \cite{Bruzual2003} models without bursts.  The result is shown in Figure~\ref{fig:massredshift} as a function of redshift using the limits defined from our limiting NEWFIRM $K$-band magnitude of 22.7 AB mag ($5\sigma$ depth in 2\arcsec-diameter apertures).   We adopt this stellar mass completeness for the following analysis. We also computed the stellar mass completeness limits using other star formation histories, SPS models, metallicity ($Z$), and exponential decay timescales ($\tau$) spanning the parameters we used. We find less than $0.2$ dex change in the derived stellar mass completeness limits, consistent with the results from the previous section.

Figure~\ref{fig:massredshift} also shows that the $K$-band provides the deepest stellar mass limit at all redshifts we consider here i.e., $z < 1.5$.   For $z < 1.0$ the IRAC depth corresponds to slightly lower stellar mass limits.  However, at $1.0 < z < 1.5$ the $K$-band provides a deeper stellar mass limit, and is well matched to the DECam imaging.   For this reason we use the limit derived from the $K$--band data, which provides a galaxy sample ``complete'' to $\log M_\ast/M_\odot = 10.0$ at $z=1$ and $\log M_\ast/M_\odot$=10.3 at $z=1.5$.   



\subsection{Selecting Quiescent and Star-forming Galaxies}
\label{sec:selectquisf}
\begin{figure}
	\includegraphics[width=0.5\textwidth]{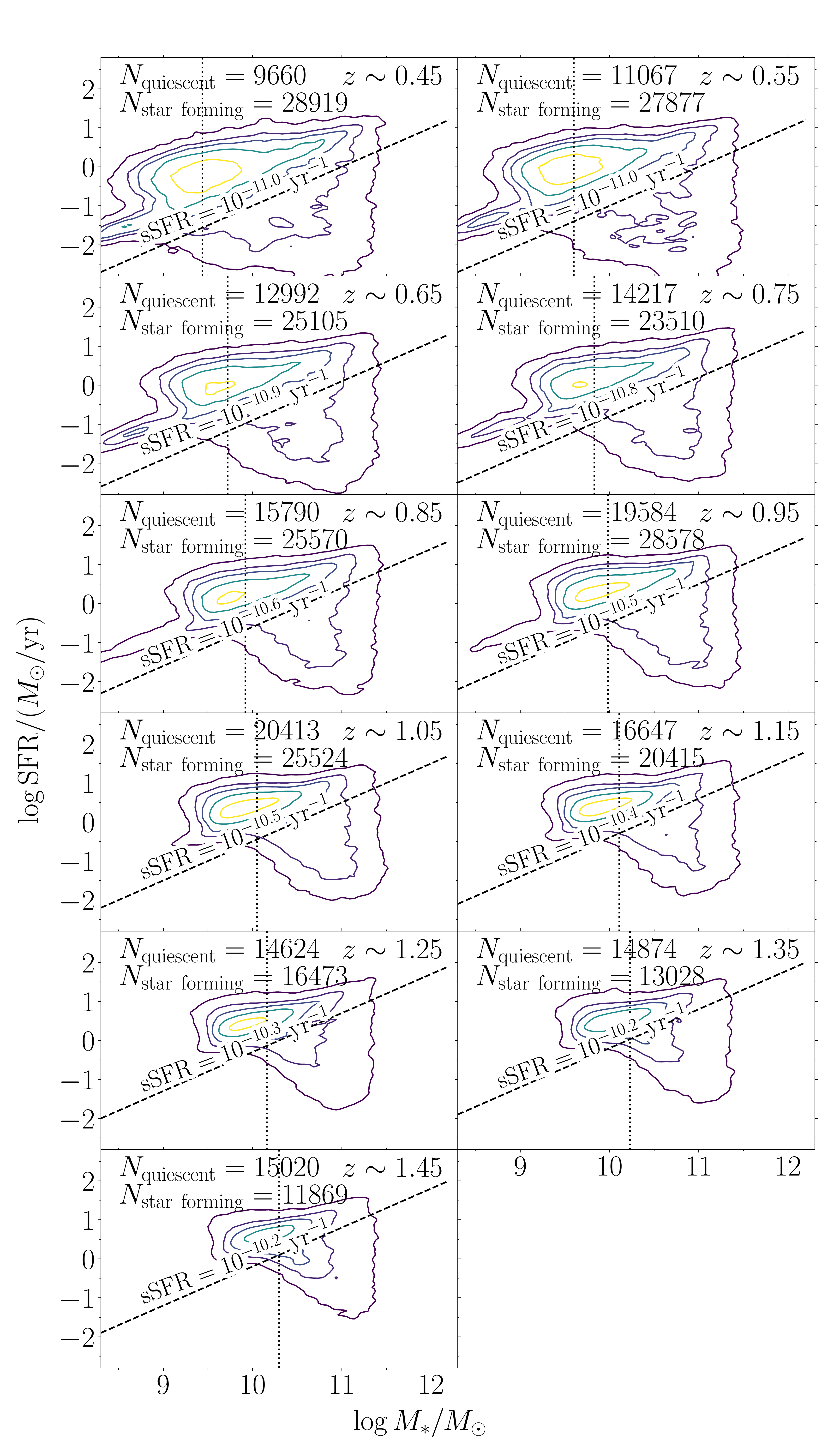}
	\caption{Star formation rate (SFR) vs. stellar mass in eleven bins of redshift from $z = 0.4-1.5$ based on our SHELA sample. We divide our sample into star-forming or quiescent according to whether a galaxy lies above or below the dashed line, respectively; this line is parallel to the star formation (SF) sequence and evolves with redshift. We also indicate the specific star formation rate (sSFR) corresponding to each quiescent/star-forming galaxies threshold.}
	\label{fig:sfrmass}
\end{figure}

\par We measure star formation rates (SFRs) for the galaxies in our sample using \emph{iSEDfit}. Similar to the stellar mass estimates, we adopt the median of the posterior probability distribution function (PDF) marginalized over the SFR as the best estimate.  We estimate SFRs for the four different sets of star formation history and SPS models with the same parameters as we did for our stellar mass estimates (Table~\ref{table:massestimates}).

\par We classify the galaxy population as either star-forming or quiescent based on whether they lie on or below the so-called \emph{star-forming main sequence} \citep{Noeske2007}. The star-forming (SF) main sequence is the correlation between SFR and stellar mass of star-forming galaxies that has been observed out to $z\sim2.5$ \citep[e.g.,][]{Rodighiero2011,Wuyts2011,Whitaker2012,Shivaei2015,Schreiber2016}. In Figure~\ref{fig:sfrmass} we plot SFR vs. stellar mass in eleven redshift bins from $z=0.4-1.5$ for our SHELA sample. The figure demonstrates the existence of a well-defined SF main sequence whose amplitude increases smoothly toward higher redshift, similar to other results \citep[see e.g.,][]{Speagle2014}. Additionally, we find a distinct  population of quiescent galaxies that lie below the SF main sequence at a given stellar mass. 
%
%

\par To classify the galaxy population, we use an evolving threshold of specific star formation rate (sSFR), computed as $\mathrm{sSFR} = \mathrm{SFR} / M_{\ast}$, to trace the lower envelope of the SF main sequence in each redshift bin. Specifically, we plot the distribution of SFR in each stellar mass and redshift bin, and model the bimodal distribution of each bin's SFR as the sum of two normal Gaussian functions \citep[e.g.,][]{Strateva2001,Baldry2004}. We then measure the mean ($\mu_{\mathrm{SF}}$) and dispersion ($\sigma_{\mathrm{SF}}$) of the distribution of the star-forming population. \edittwo{We define all galaxies whose SFRs lie below $3\sigma_{\mathrm{SF}}$ from $\mu_{\mathrm{SF}}$ to be quiescent. This results in the threshold for sSFRs evolving from  $10^{-11}~\mathrm{yr}^{-1}$ at $z=0.4$  to $10^{-10.2}~\mathrm{yr}^{-1}$ at $z=1.5$. We adopt the evolving threshold of sSFR to classify the galaxy population because the star forming main sequence evolves with redshift -- the main sequence as a whole moves to higher SFR as redshift increases \citep[e.g.,][]{Noeske2007}. However, in Section~\ref{sec:choiceofssfr}, we discuss the effect of using a non-evolving threshold of sSFR to classify galaxy population on the evolution of number densities of quiescent and star forming populations.} 

\editone{We also must emphasize an important caveat of our derived SFRs. At $z<0.5$, the $u$-band photometry samples the rest-frame wavelength of galaxies longer than near-ultraviolet (NUV), and we need NUV or Far-UV observations to probe the recent star formation of galaxies. As a result, in our lowest redshift bins, we may have less accurate SFRs and  also less reliable separation between star forming and quiescent populations. In Section~\ref{subsec:evonumberdensity}, we show that our interpretation of the number density and stellar mass density evolution may be impacted by this limitation.}

\par \edittwo{Finally, in Figure~\ref{fig:massredshift}, we show the distributions of redshift and stellar mass for our full SHELA sample and the subsamples of quiescent and star forming galaxies. Over $0.4 < z < 1.5$, the population of massive galaxies ($\log M_{\ast} / M_{\odot}>11$) are dominated by quiescent systems. In Section~\ref{sec:smfdepstarformation}, we will quantify how the evolution in the SMF of each subsample accounts for the evolution in the SMF for all massive galaxies.} 
%
%
%

\section{Methods: Forward Modeling the Galaxy Stellar Mass Function}
\label{sec:method}
\subsection{Accounting for Scatter in Stellar Mass Estimates}
\label{subsec:doubleschechter}

Errors in the stellar mass estimates ($M_{\ast}$) introduce a bias into the derived galaxy SMF, as random errors cause more objects to have ``upscattered'' stellar masses than ``downscattered'' values.  This is a form of \cite{Eddington1913} bias, and is especially problematic on the exponential (high-mass) end of the SMF due to the steep decline in the number of galaxies. 
%
%
Even a small fraction of these upscattered low-mass galaxies can dominate the number densities at high stellar masses \citep[]{Ilbert2013,Moster2013,Souza2015,Caputi2015,Grazian2015,Davidzon2017}.\footnote{We note that anything that causes stellar mass uncertainties or such ``upscattering'' will contribute to the Eddington bias. This includes galaxies with ``upscattered'' photometric redshifts, which produces an increase in their stellar mass that scales with the square of the distance.}
%
%
This Eddington-type bias in stellar mass may increase with increasing redshift because of the decreasing in signal-to-noise ratio with increasing redshift. Some previous studies have shown that Eddington bias can affect the interpretation of the evolution of the galaxy SMF \citep[see discussion in e.g.,][]{Fontanot2009, Moster2013,Bundy2017}.

\par Here, we account for a varying Eddington bias using a forward model method.  This method requires that we assume an intrinsic shape of a galaxy SMF, to which we then apply measurement (and systematic) uncertainties.   For our model, we first assume the galaxy SMF is well described by a double Schechter function \citep{Baldry2008} of the form,
\begin{equation}
\label{eq:dbschechter}
\begin{split}
\phi(M_{\ast}) = (\ln 10)\exp\left [ -\frac{M_{\ast}}{M^{\ast}} \right ] \times \left \{ \phi_{1}10^{(\alpha_{1}+1)(\log M_{\ast} - \log M^{\ast})} \right. \\  \left. +\phi_{2}10^{(\alpha_{2}+1)(\log M_{\ast} - \log M^{\ast})}  \right \}, 
\end{split}
\end{equation}
where $\alpha_{2} < \alpha_{1}$ are the faint end (power-law) slopes of the SMF (where the second term dominates at the low-mass end).    We denote the ``knee''  of the SMF with the characteristic stellar mass as, $M^{\ast}$, which marks the stellar mass above which the stellar mass function declines exponentially.  We require that both terms in the double Schechter function have the same $M^\ast$.

\begin{figure}
	\includegraphics[width=0.45\textwidth]{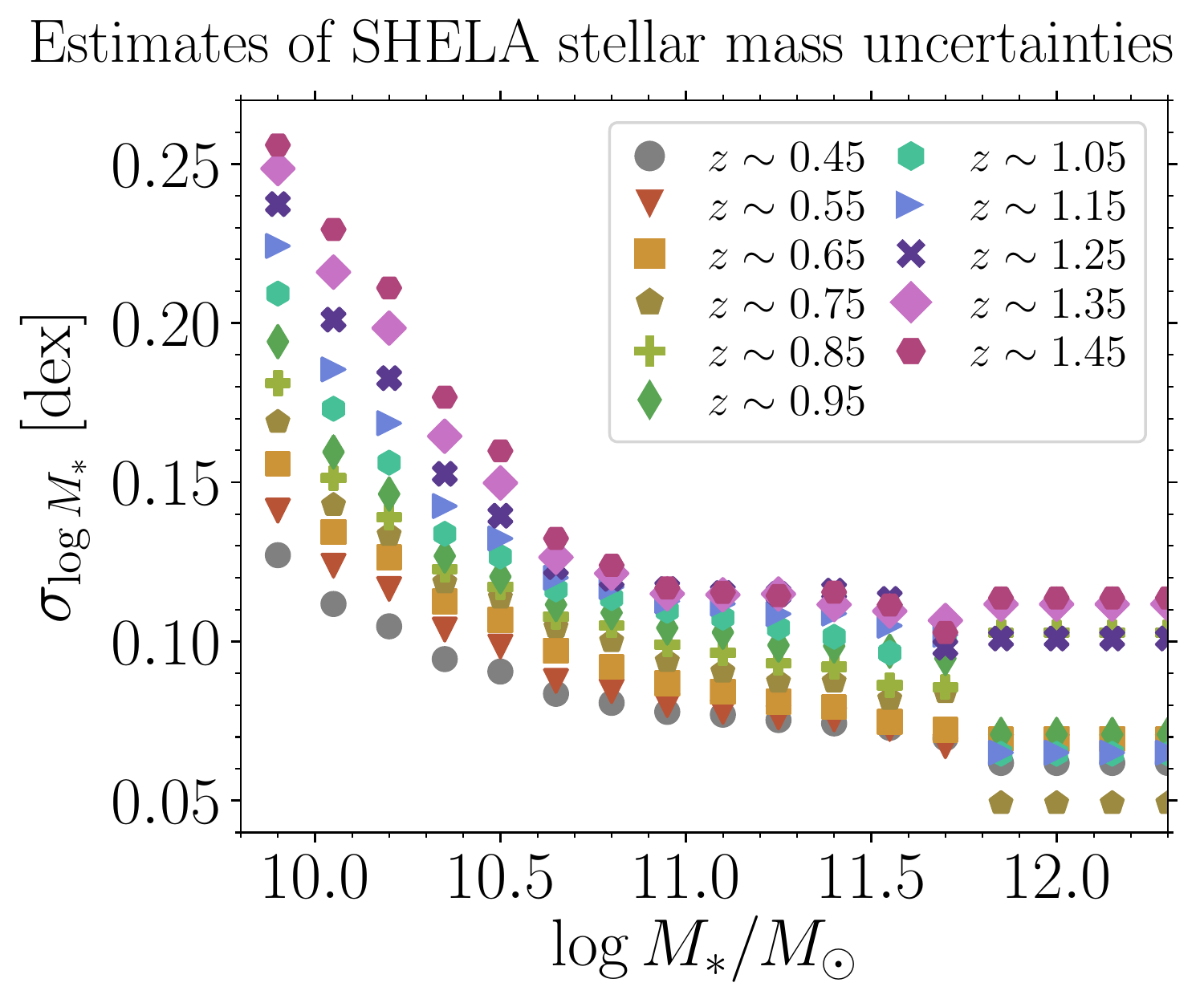}
	\caption{Stellar mass uncertainties ($\sigma_{M_{\ast}}(\log M_{\ast}, z)$) vs. stellar mass in eleven bins of redshift from $z=0.4-1.5$. These stellar mass uncertainties are derived from our SED fitting and are incorporated into the forward-modeling of the galaxy SMF (See Section~\ref{subsec:doubleschechter})}
	\label{fig:sigmalmass}
\end{figure}

We then construct a series of mock galaxy sample with the stellar mass
distribution following this double Schechter function.   We generate
500,000 mock datasets that
sample various parameters ranges of the double Schechter function, then we perturb the stellar mass
estimates using uncertainties drawn from a Gaussian distribution, with the width of the distribution equal to the $1\sigma$ uncertainty in stellar masses based on the SED fitting, $\sigma_{M_{\ast}}(\log M_{\ast}, z)$. This uncertainty is both 
stellar mass- and redshift-dependent. \editone{Figure~\ref{fig:sigmalmass} shows the stellar mass uncertainties that we used, including their dependence on redshift. } 

\begin{figure*}
	\includegraphics[width=1\textwidth]{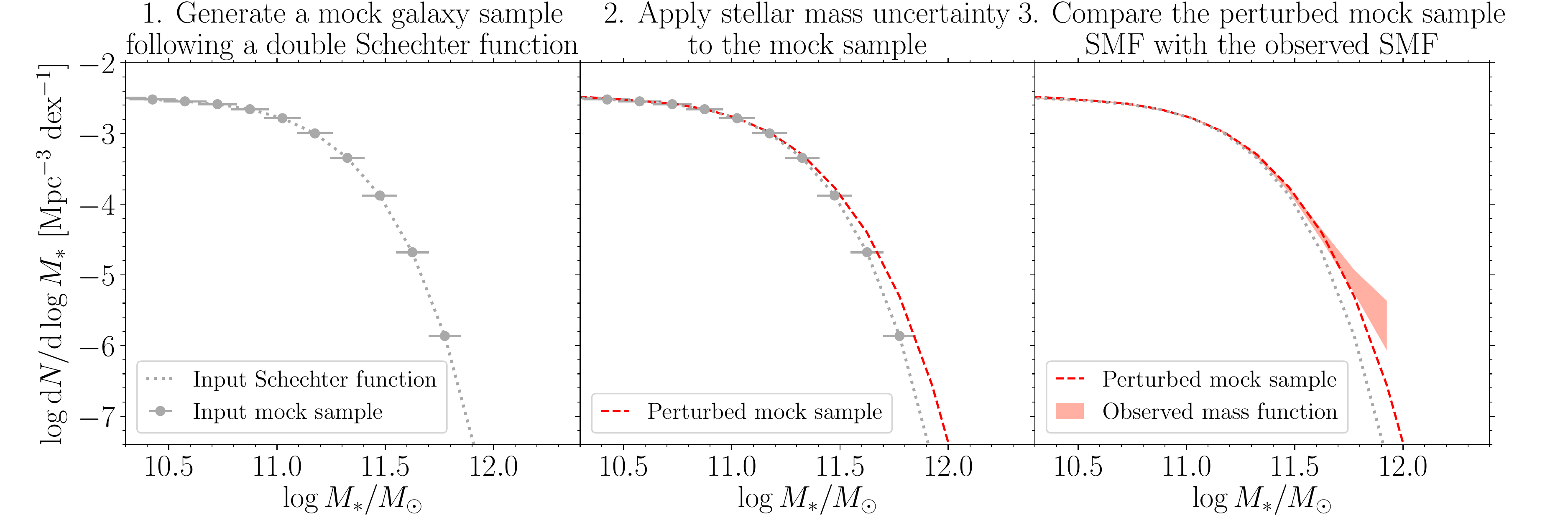}
	\caption{Demonstration of our method to forward model the galaxy SMF, taking into account the statistical and systematic uncertainties on stellar mass. The width of the error bar on the stellar mass for data points shows the associated stellar mass uncertainty for galaxies at that stellar mass.  In practice, we generate mock galaxy samples that sample various parameter ranges and compare them with the observed stellar mass function in an iterative approach.  In this figure we show one realization for illustration only.}
	\label{fig:forwardmodel}
\end{figure*}

\par We do not directly constrain the normalization $\phi_{1}$ and $\phi_{2}$ of the  double Schechter function. Instead, we define the parameter $\lambda_{\mathrm{mix}}$ that is varied between between 0 and 1 to indicate the relation between of the first term and the second term of the double Schechter form, such that $\lambda_{\mathrm{mix}}\propto \phi_1$  and  $(1-\lambda_{\mathrm{mix}}) \propto \phi_2$. We then evaluate the overall normalization factor $C$, of the stellar mass function of the mock sample $\phi(M_{\ast, \mathrm{mock}})$, such that the comoving number density of mock galaxies, $n_{\mathrm{mock}}$,
\begin{equation}
n_{\mathrm{mock}}  = C\int_{\log M_{\ast \mathrm{min}}}^{\log M_{\ast \mathrm{max}}}\phi(M_{\ast, \mathrm{mock}}) d \log M_{\ast},
\end{equation}
equals the observed comoving number density of SHELA galaxies, $n_{\mathrm{SHELA}}$, and $ M_{\ast \mathrm{min}}=M_{\mathrm{lim}}$ and $ M_{\ast \mathrm{max}}=10^{12.5} M_{\odot}$, where $M_{\mathrm{lim}}$ denotes the stellar mass completeness limit at a given redshift bin (see Table~\ref{table:fitparams}). We then bin the mock samples identically to the data. 
\par Finally, we constrain the double Schechter parameters by performing grid search over ranges of parameters ($M^{\ast}$, $\alpha_{1}$, $\alpha_{2}$, $\lambda_{\mathrm{mix}}$) and comparing the mock stellar mass functions with the observed one in each stellar mass bin. We demonstrate our method of forward modeling the galaxy stellar mass function in Figure~\ref{fig:forwardmodel}.

\subsection{Additional Sources of Uncertainty in the SMF} 
\label{sec:cvestimate}
\par In addition to the scatter in stellar mass estimates, we also consider the effects of Poissonian (counting) uncertainties, and ``cosmic variance''.  The latter are large-scale fluctuations in the spatial distribution of the number of galaxies in the universe.   While all of these effects contribute to the uncertainties in the SMF, cosmic variance is most significant for small fields and highly biased objects i.e., objects with strong spatial clustering, such as massive galaxies).   One advantage to our SHELA survey is that, since it spans such a large volume ($1.4\times 10^8$~Mpc$^3$ = 0.14 Gpc$^3$), the effects of cosmic variance on the number density of massive galaxies are mitigated, and we can estimate cosmic variance from the data itself. \edittwo{For example,  using the formalism set by \citet{Moster2011}, the relative cosmic variance, $\sigma_{v}$, of galaxies more massive than $10^{11}~M_{\odot}$ at $z=0.35$ is $36\%$ ($\sim0.1$~dex) for COSMOS ($\approx$2 deg$^2$), but only $\sim$17\% ($\sim0.07$~dex) for SHELA.} %


\par \cite{Bundy2017} estimated the cosmic variance in the S82MGC (140~$\mathrm{deg}^2$) using \edittwo{bootstrap resampling}. In the $0.3 < z < 0.65$ bin, their estimated $1\sigma$ error due to the cosmic variance is $\sim0.01$ dex (corresponding to $\sigma_{v}$ of $\sim2\%$) at $\log (M_{\ast}/M_{\odot})\sim11.0$, and 0.02-0.05 dex ($\sigma_{v}\sim 5\%-10\%$) at $\log (M_{\ast}/M_{\odot})\sim11.6$. 

\par We adopt the method of \cite{Bundy2017} to estimate the cosmic variance in our SHELA samples. We divide the SHELA survey into 150 roughly equal area regions and recompute stellar mass functions after resampling with replacement (see Appendix~\ref{sec:appendix}).  \edittwo{For low mass galaxies  ($\log (M_{\ast}/M_{\odot})<11.5$), the bootstrap resampling yields a cosmic variance, $\sigma_{v}$, of $5\%-12\%$ at $0.3 < z <1.5$. At higher masses ($\log (M_{\ast}/M_{\odot})>11.5$), the cosmic variance rises to 6\%-12\% (Figure~\ref{fig:sigmacv} in Appendix \ref{sec:appendix}).} 
%
%
In the redshift range where we overlap with \cite{Bundy2017}, our cosmic variance is larger.  This is expected as the area of S82MGC is 8 times larger than that of the SHELA.  


\par To test for other sources of systematic uncertainty, we compare our SMF from SHELA to that in S82-MGC \citep{Bundy2017} for all galaxies between $0.3 < z < 0.65$, where our samples overlap (see Section~\ref{sec:discussion} and Figure~\ref{fig:bundySMF}). 
The two estimates, which both use the forward modeling method, are in good agreement. The agreement is particularly good for the massive galaxies ($\log(M_{\ast}/M_{\odot})\gtrsim11.0$) between $0.4 < z < 0.6$. This suggests that we are not significantly affected by either cosmic variance or other systematics at this redshift range. However, between $0.3 < z < 0.4$, the normalization of SHELA SMF is lower than that of S82-MGC. \edittwo{In this redshift bin, the estimated relative cosmic variance \citep{Moster2011} in our SHELA survey is  20\% ($\sim0.1$~dex), compared to 7\% cosmic variance in S82-MGC survey.  Therefore, our results in this (smallest-volume) redshift bin may be effected by larger than typical cosmic variance.} In the following analysis, we omit this lowest redshift bin and only study the evolution of stellar mass function for galaxies at $0.4 < z < 1.5$. \edittwo{We also note that we do not include the effect of cosmic variance in our forward modeling method because the effects are correlated in stellar mass, and all mass bins should be equally affected by the same large-scale fluctuation. Consequently, the measurement of the galaxy SMF should be mainly affected by the random errors in the stellar mass estimates rather than cosmic variance. }


\subsection{The Impact of the Contamination from QSOs and AGNs on Galaxy Stellar Mass Function}
To we estimate the contamination from QSOs and AGNs on our SHELA galaxy SMFs, we first cross-match our sample with the SDSS spectral catalog (DR13; \citealt{Albareti2017}) to find 760 QSOs in our galaxy sample. 
%
%
Second, we cross-match the SHELA sample with the 31 $\mathrm{deg}^2$ Stripe 82X X-ray Catalog \citep{Lamassa2016viz}. \edittwo{The flux limits of this Stripe 82 X-ray survey are $8.7\times10^{-16}~\mathrm{erg}~\mathrm{s}^{-1}~\mathrm{cm}^{-2}$, $4.7\times10^{-15}~\mathrm{erg}~\mathrm{s}^{-1}~\mathrm{cm}^{-2}$, and  $2.1\times10^{-15}~\mathrm{erg}~\mathrm{s}^{-1}~\mathrm{cm}^{-2}$ in the soft ($0.5-2$ keV), hard ($2-10$ keV), and full X-ray bands ($0.5-10$ keV), respectively \citep{Lamassa2016}.} We found 1253 matched X-ray sources in our sample. in total, these sources account for less than 1\% of galaxies more massive than $10^{11} M_{\odot}$\footnote{\edittwo{The $\lesssim1\%$ X-ray source fraction represents only a lower limit to the true fraction of interlopers. X-ray surveys are not sensitive to all AGNs -- objects behind high column densities of neutral material may result in non-detections.}}
We then exclude these QSOs and X-ray sources from our galaxy sample and repeat our measurement of galaxy SMFs. Because these sources account for only a few percent of the total number of massive galaxies, the resulting SMFs and the inferred redshift evolution are not significantly affected by those contaminations.  Therefore, we conclude that our measurements are not adversely affected by the presence of AGN. \edittwo{For the rest of this paper, we exclude SDSS QSOs and X-ray sources from our galaxy sample.}

\section{Results} 
\label{sec:results}
\par
\begin{figure*}
\centering
\includegraphics[width=1.0\textwidth]{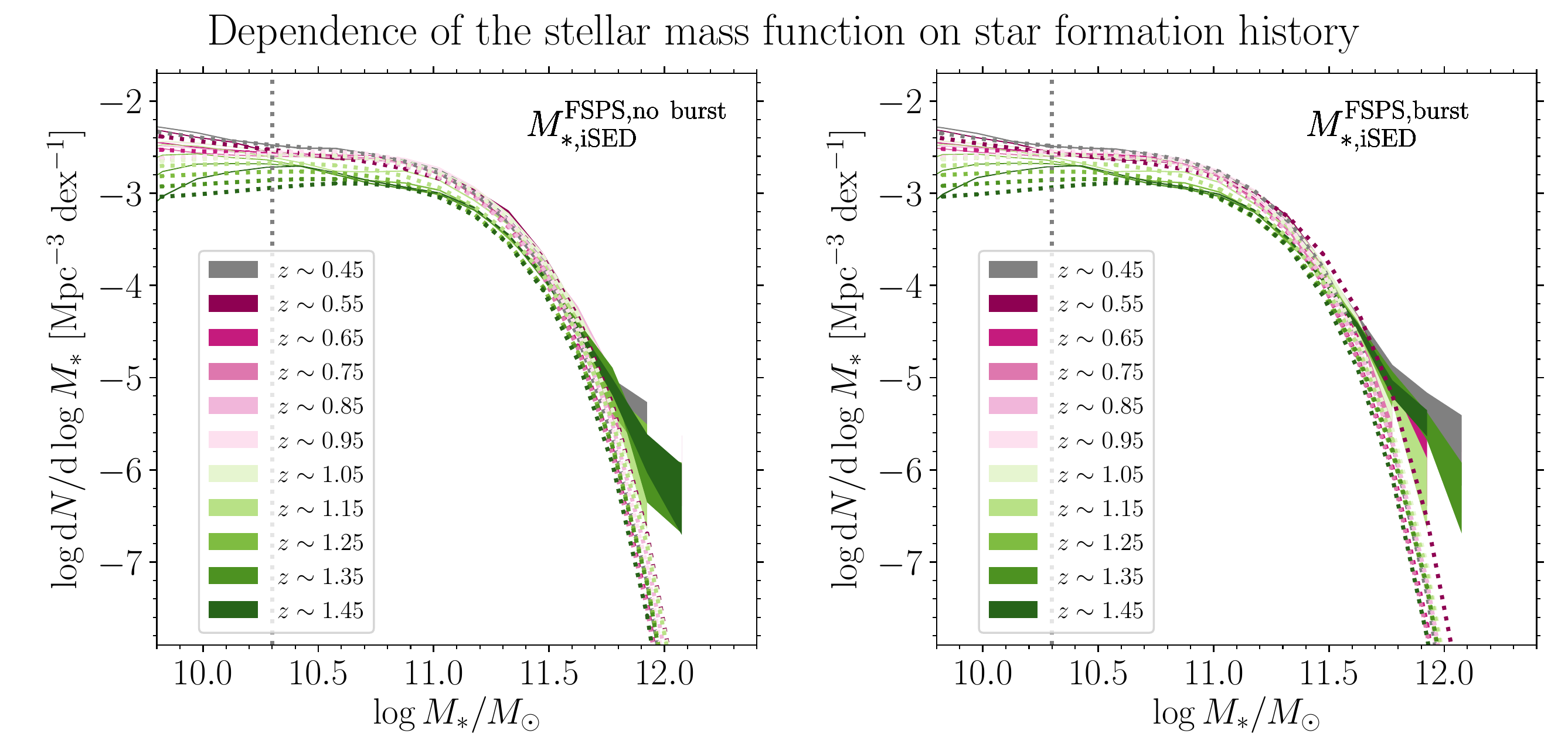}
\caption{SMFs derived using two $M_{\ast}$ estimators with different assumptions in star formation history (SFH). The shaded regions represent the observed SHELA SMFs and the corresponding Poissonian errors. The left panel corresponds to the stellar mass derived from no burst FSPS models ($M_{\ast,\mathrm{iSED}}^{\mathrm{FSPS,no\ burst}}$). The resulting mass function suggests no more than a $\lesssim0.1$~dex increase in the characteristic stellar mass ($M^{\ast}$) over the redshift range plotted.  The trend is similar for the stellar mass as derived from FSPS and including bursts ($M_{\ast,\mathrm{iSED}}^{\mathrm{FSPS,burst}}$; right panel). Forward-modeling results, which aim to account for (and thereby remove) biases caused by errors in the stellar mass, are shown as dotted curves in each panel. The vertical dotted line indicates our stellar mass completeness limit at $z=1.5$.}
\label{fig:massfuncsfh}
\end{figure*}

\begin{figure*}
\includegraphics[width=1.0\textwidth]{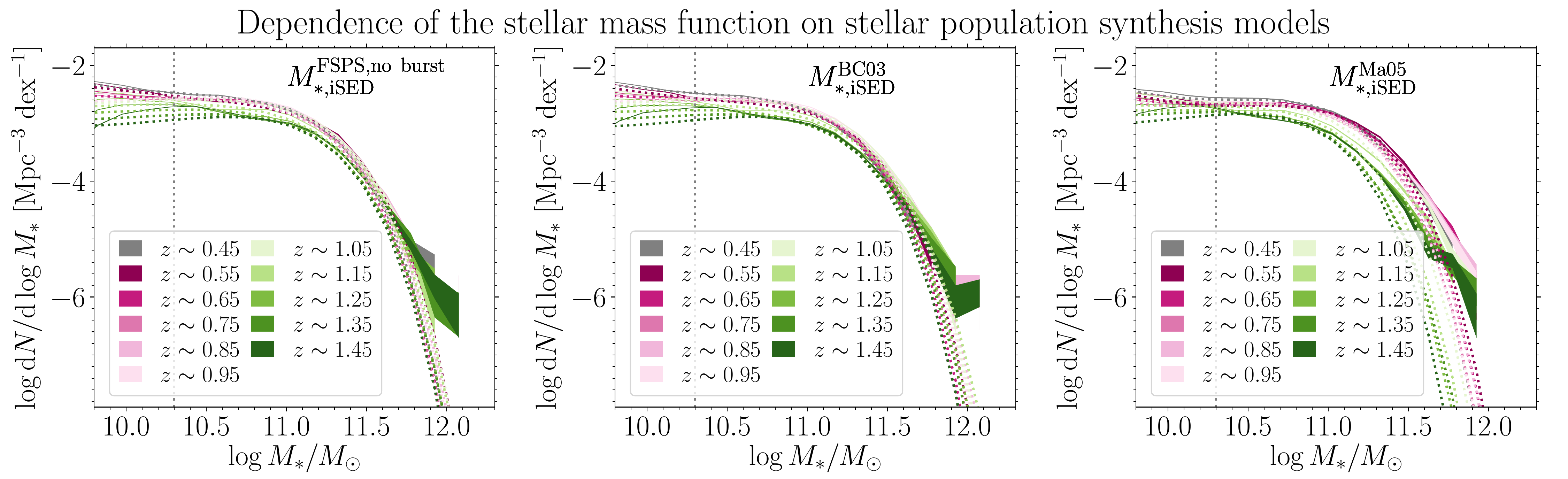}

\caption{SMFs derived using three $M_{\ast}$ estimators with different SPS models. All panels use the stellar masses based on the same SFH priors and without bursts. The shaded regions represent the observed SHELA stellar mass functions and the corresponding Poissonian errors.  We compare $M_{\ast,\mathrm{iSED}}^{\mathrm{FSPS,no\ burst}}$ (left panel), $M_{\ast,\mathrm{iSED}}^{\mathrm{BC03}}$ (middle panel), $M_{\ast,\mathrm{iSED}}^{\mathrm{Ma05}}$ (right panel). Forward-modeling results, which aim to remove an estimate of the biases caused by errors in the stellar mass, are shown as dotted curves in each panel. The vertical dotted line indicates our stellar mass completeness limit at $z=1.5$. The resulting mass function suggests no more than a $\lesssim0.1$~dex increase in the characteristic stellar mass ($M^{\ast}$) over the redshift range plotted, except for the SMF from $M_{\ast,\mathrm{iSED}}^{\mathrm{Ma05}}$, which exhibits $\sim0.2$ dex increase in $M^{\ast}$ from $z=1.5$ to $z=1.0$.}
\label{fig:massfuncpop}
\end{figure*}

 We begin with presenting the SMFs derived from individual stellar mass estimates with different assumptions on the star formation history (Figure~\ref{fig:massfuncsfh})  and SPS models (Figure~\ref{fig:massfuncpop}). In later sections, we average the results from the different SMFs to estimate the biases associated with assumptions from the different star formation history and SPS models.


\subsection{Assumption-averaged Estimate of the SMF}
\label{sec:assumpavgsmf}

We define the ``assumption-averaged'' SMF, i.e., the average of the SMFs derived using the different methods listed in Table~\ref{table:massestimates}, as our fiducial measurement.  In practice, we take the average number density in each bin of stellar mass using the different assumptions in star formation history and SPS model\footnote{We compute the average number density by binning the concatenated array of four different sets of $M_{\ast}$ estimates, and dividing by four times the corresponding volume of each redshift slice.}.  In this way, our results are a statistical mean, and we can study the variance in the SMF. These four $M_{\ast}$ estimates encompass the range of $M_{\ast}$ values obtained by adopting currently uncertain priors. Therefore, the assumption-averaged result represents a compromise among differing approaches. 
%
%
%

\begin{figure*}
\centering
\includegraphics[width=1.0\textwidth]{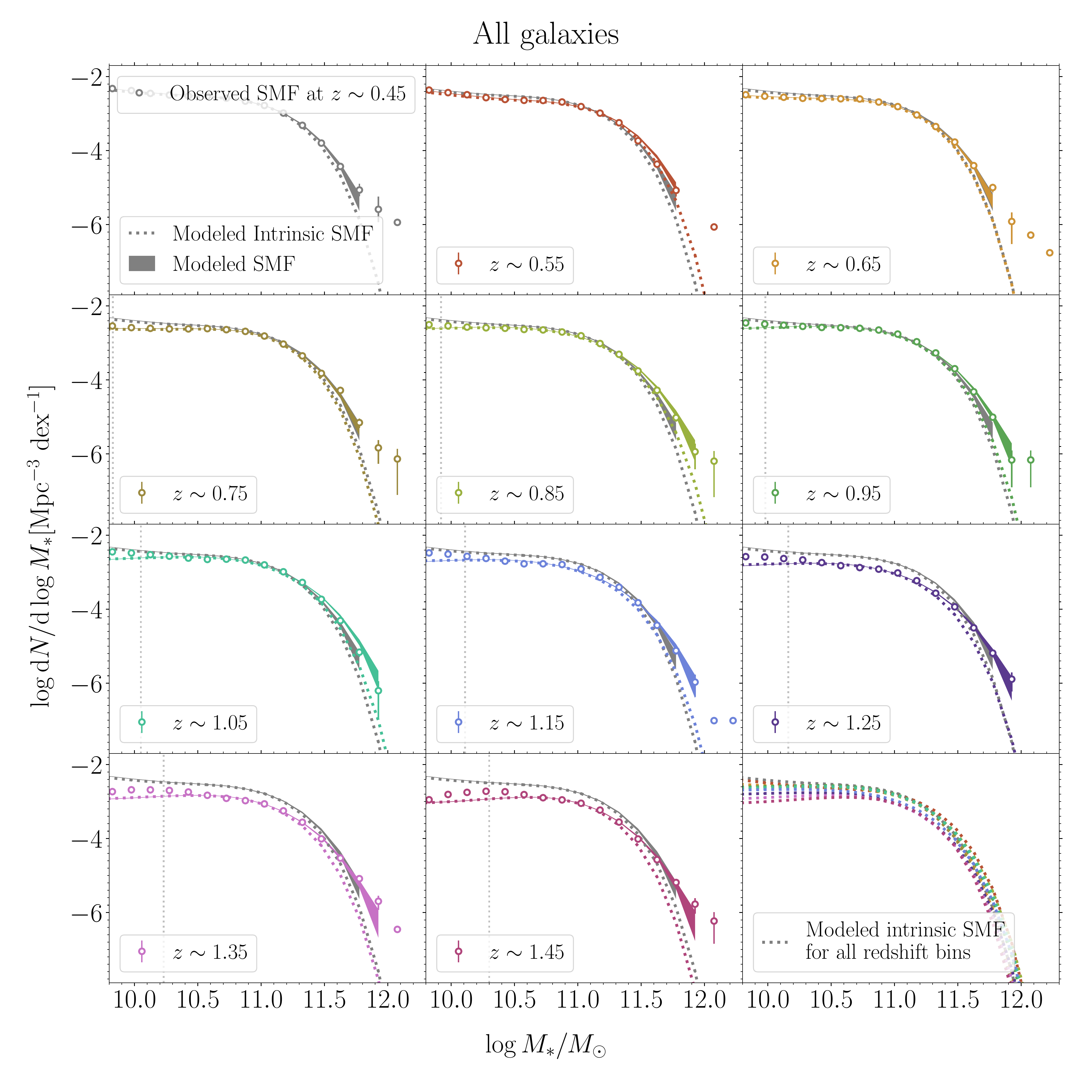}
\caption{Assumption-averaged estimate of the SHELA galaxy SMF between $0.4 < z < 1.5$ for all galaxies, defined by taking a mean of the SMFs found from the different methods (see text).  The circles represent the observed SHELA SMFs, and the error bars show the corresponding Poissonian uncertainties. The shaded regions represent the forward-modeled intrinsic SMFs. The dotted lines show the intrinsic models; these ``intrinsic SMFs'', aim to account for (and thereby remove) biases caused by scatter in stellar mass measurement.  The estimated stellar mass completeness at a given redshift bin is indicated by the vertical dotted line.  
  %
  %
  In each panel we show both the observed and the modeled SMF (from our forward modeling result) at the lowest redshift bin ($0.4 < z < 0.5$, grey shaded region and grey dotted curve) for comparison. \edittwo{The last panel shows the modeled intrinsic SMFs for all redshift bins.} We measure no ($\lesssim0.1$~dex) redshift evolution at the high-mass end ($\log M_{\ast} / M_{\odot}>11.0$) of the SMF between $0.4 < z < 1.5$.}
\label{fig:massfuncavg}
\end{figure*}

In Figure~\ref{fig:massfuncavg}, we show the observed galaxy SMF in each redshift bin with shaded regions corresponding to the Poisson errors \edittwo{computed by taking square root of the number of galaxies in stellar mass bin}. We indicate the stellar mass completeness limit at each redshift bin with a vertical dotted line. \edittwo{We present our measurements of the assumption-averaged stellar mass function and  the number of galaxies in each redshift bin in Table~\ref{table:smfavg} and Table~\ref{table:fitparams}, respectively.}

\begin{splitdeluxetable*}{lcccccBcccccc}
\tablecaption{Assumption-Averaged Stellar Mass Functions For All Galaxies\label{table:smfavg}}
\tablehead{
& \colhead{$0.4 < z < 0.5$} &  \colhead{$0.5 < z < 0.6$} & \colhead{$0.6 < z < 0.7$} & \colhead{$0.7 < z < 0.8$} & \colhead{$0.8 < z < 0.9$} &  \colhead{$0.9 < z < 1.0$} & \colhead{$1.0 < z < 1.1$} & \colhead{$1.1 < z < 1.2$} & \colhead{$1.2 < z < 1.3$} & \colhead{$1.3 < z < 1.4$} & \colhead{$1.4 < z < 1.5$}\\
\colhead{ $\log(M_{\ast}/M_{\odot})$} & \colhead{$\log(\phi/\mathrm{Mpc}^{-3}\mathrm{dex}^{-1}$)} &  \colhead{$\log(\phi/\mathrm{Mpc}^{-3}\mathrm{dex}^{-1}$)} & 
\colhead{$\log(\phi/\mathrm{Mpc}^{-3}\mathrm{dex}^{-1}$)} &
\colhead{$\log(\phi/\mathrm{Mpc}^{-3}\mathrm{dex}^{-1}$)} & 
\colhead{$\log(\phi/\mathrm{Mpc}^{-3}\mathrm{dex}^{-1}$)} &
\colhead{$\log(\phi/\mathrm{Mpc}^{-3}\mathrm{dex}^{-1}$)} &
\colhead{$\log(\phi/\mathrm{Mpc}^{-3}\mathrm{dex}^{-1}$)} &
\colhead{$\log(\phi/\mathrm{Mpc}^{-3}\mathrm{dex}^{-1}$)} &
\colhead{$\log(\phi/\mathrm{Mpc}^{-3}\mathrm{dex}^{-1}$)} &
\colhead{$\log(\phi/\mathrm{Mpc}^{-3}\mathrm{dex}^{-1}$)} &
\colhead{$\log(\phi/\mathrm{Mpc}^{-3}\mathrm{dex}^{-1}$)} \\
}  

\startdata
9.53 & $-2.22\pm0.01$ & ... & ... & ... & ... & ... & ... & ... & ... & ... & ... \\
9.68 & $-2.27\pm0.01$ & $-2.31\pm0.01$ & ... & ... & ... & ... & ... & ... & ... & ... & ... \\
9.83 & $-2.32\pm0.01$ & $-2.36\pm0.01$ & $-2.49\pm0.01$ & ... & ... & ... & ... & ... & ... & ... & ... \\
9.98 & $-2.38\pm0.01$ & $-2.43\pm0.01$ & $-2.53\pm0.01$ & $-2.59\pm0.01$ & $-2.54\pm0.01$ & ... & ... & ... & ... & ... & ... \\
10.13 & $-2.45\pm0.01$ & $-2.49\pm0.01$ & $-2.56\pm0.01$ & $-2.61\pm0.01$ & $-2.57\pm0.01$ & $-2.52\pm0.01$ & $-2.53\pm0.01$ & $-2.57\pm0.01$ & ... & ... & ... \\
10.28 & $-2.50\pm0.01$ & $-2.57\pm0.01$ & $-2.59\pm0.01$ & $-2.62\pm0.01$ & $-2.59\pm0.01$ & $-2.55\pm0.01$ & $-2.57\pm0.01$ & $-2.63\pm0.01$ & $-2.67\pm0.01$ & $-2.70\pm0.01$ & ... \\
10.43 & $-2.52\pm0.01$ & $-2.62\pm0.01$ & $-2.59\pm0.01$ & $-2.62\pm0.01$ & $-2.60\pm0.01$ & $-2.58\pm0.01$ & $-2.62\pm0.01$ & $-2.70\pm0.01$ & $-2.74\pm0.01$ & $-2.74\pm0.01$ & $-2.73\pm 0.01$ \\
10.58 & $-2.53\pm0.01$ & $-2.64\pm0.01$ & $-2.60\pm0.01$ & $-2.61\pm0.01$ & $-2.64\pm0.01$ & $-2.59\pm0.01$ & $-2.65\pm0.01$ & $-2.77\pm0.01$ & $-2.82\pm0.01$ & $-2.83\pm0.01$ & $-2.81\pm 0.01$ \\
10.73 & $-2.58\pm0.01$ & $-2.64\pm0.01$ & $-2.60\pm0.01$ & $-2.64\pm0.01$ & $-2.65\pm0.01$ & $-2.60\pm0.01$ & $-2.65\pm0.01$ & $-2.77\pm0.01$ & $-2.87\pm0.01$ & $-2.91\pm0.01$ & $-2.90\pm 0.01$ \\
10.88 & $-2.66\pm0.01$ & $-2.69\pm0.01$ & $-2.69\pm0.01$ & $-2.69\pm0.01$ & $-2.70\pm0.01$ & $-2.65\pm0.01$ & $-2.67\pm0.01$ & $-2.79\pm0.01$ & $-2.92\pm0.01$ & $-2.97\pm0.01$ & $-2.95\pm 0.01$ \\
11.03 & $-2.78\pm0.01$ & $-2.81\pm0.01$ & $-2.81\pm0.01$ & $-2.81\pm0.01$ & $-2.81\pm0.01$ & $-2.76\pm0.01$ & $-2.80\pm0.01$ & $-2.91\pm0.01$ & $-3.02\pm0.01$ & $-3.06\pm0.01$ & $-3.05\pm 0.01$ \\
11.18 & $-2.99\pm0.01$ & $-2.99\pm0.01$ & $-3.04\pm0.01$ & $-3.03\pm0.01$ & $-3.02\pm0.01$ & $-2.97\pm0.01$ & $-2.99\pm0.01$ & $-3.14\pm0.01$ & $-3.23\pm0.01$ & $-3.25\pm0.01$ & $-3.23\pm 0.01$ \\
11.33 & $-3.32\pm0.02$ & $-3.25\pm0.02$ & $-3.35\pm0.02$ & $-3.35\pm0.02$ & $-3.31\pm0.01$ & $-3.27\pm0.01$ & $-3.28\pm0.01$ & $-3.40\pm0.01$ & $-3.57\pm0.02$ & $-3.56\pm0.02$ & $-3.55\pm 0.01$ \\
11.48 & $-3.80\pm0.04$ & $-3.73\pm0.03$ & $-3.77\pm0.03$ & $-3.82\pm0.03$ & $-3.76\pm0.02$ & $-3.70\pm0.02$ & $-3.73\pm0.02$ & $-3.83\pm0.02$ & $-3.94\pm0.02$ & $-4.01\pm0.03$ & $-4.01\pm 0.02$ \\
11.63 & $-4.43\pm0.07$ & $-4.37\pm0.06$ & $-4.41\pm0.05$ & $-4.28\pm0.04$ & $-4.28\pm0.04$ & $-4.32\pm0.04$ & $-4.31\pm0.04$ & $-4.43\pm0.04$ & $-4.50\pm0.04$ & $-4.53\pm0.04$ & $-4.57\pm 0.05$ \\
11.78 & $-5.07\pm0.14$ & $-5.08\pm0.12$ & $-5.00\pm0.10$ & $-5.16\pm0.11$ & $-5.02\pm0.09$ & $-5.01\pm0.08$ & $-5.16\pm0.10$ & $-5.12\pm0.09$ & $-5.19\pm0.09$ & $-5.08\pm0.08$ & $-5.19\pm 0.09$ \\
11.93 & $-5.59\pm0.22$ & ... & $-5.92\pm0.24$ & $-5.84\pm0.21$ & $-5.94\pm0.22$ & $-6.16\pm0.26$ & $-6.20\pm0.26$ & $-5.97\pm0.20$ & $-5.89\pm0.19$ & $-5.70\pm0.15$ & $-5.77\pm 0.16$ \\
12.08 & $-5.94\pm0.30$ & $-6.07\pm0.30$ & $-6.28\pm0.33$ & $-6.14\pm0.28$ & $-6.20\pm0.28$ & $-6.16\pm0.26$ & ... & $-7.01\pm0.48$ & ... & $-6.46\pm0.30$ & $-6.23\pm 0.24$ \\
\enddata
\end{splitdeluxetable*}

\begin{deluxetable*}{cccccccc}
\tablecaption{Intrinsic Mass Function Shape Parameters from Forward Modeling for All Galaxies \label{table:fitparams}}
\tablehead{
\colhead{Redshifts} & \colhead{$\log(M_{\mathrm{lim}}/M_{\odot})$} & \colhead{$N_{gal}$} &
\colhead{$\log(\phi_{1}/ \mathrm{Mpc}^{-3}\mathrm{dex}^{-1})$ } & \colhead{$\log(\phi_{2}/ \mathrm{Mpc}^{-3}\mathrm{dex}^{-1})$} & \colhead{$\log(M^{\ast}/M_{\odot})$ } & \colhead{$\alpha_{1}$} & \colhead{$\alpha_{2}$} 
}
\colnumbers
\startdata
$(0.3,0.4)$ & 9.22 & 33848 & -2.03 & -2.98 & $10.70^{+ 0.01}_{- 0.03}$ & $-0.45\pm 0.17$ & $-1.70\pm 0.10$ \\
$(0.4,0.5)$ & 9.44 & 36283 & -2.15 & -2.75 & $10.76^{+ 0.02}_{- 0.03}$ & $-0.12\pm 0.25$ & $-1.50\pm 0.20$ \\
$(0.5,0.6)$ & 9.60 & 36389 & -2.25 & -2.72 & $10.82^{+ 0.02}_{- 0.00}$ & $ 0.00\pm 0.16$ & $-1.45\pm 0.05$ \\
$(0.6,0.7)$ & 9.72 & 36200 & -2.32 & -2.92 & $10.76^{+ 0.03}_{- 0.04}$ & $-0.10\pm 0.21$ & $-1.35\pm 0.11$ \\
$(0.7,0.8)$ & 9.83 & 36114 & -2.39 & -3.00 & $10.73^{+ 0.01}_{- 0.01}$ & $ 0.00\pm 0.03$ & $-1.25\pm 0.08$ \\
$(0.8,0.9)$ & 9.92 & 39318 & -2.36 & -3.31 & $10.87^{+ 0.02}_{- 0.02}$ & $-0.57\pm 0.05$ & $-1.27\pm 0.03$ \\
$(0.9,1.0)$ & 9.98 & 45584 & -2.33 & -3.29 & $10.82^{+ 0.02}_{- 0.03}$ & $-0.45\pm 0.09$ & $-1.30\pm 0.05$ \\
$(1.0,1.1)$ & 10.05 & 43303 & -2.37 & -3.32 & $10.84^{+ 0.02}_{- 0.02}$ & $-0.47\pm 0.07$ & $-1.27\pm 0.03$ \\
$(1.1,1.2)$ & 10.11 & 34655 & -2.45 & -3.40 & $10.86^{+ 0.01}_{- 0.01}$ & $-0.65\pm 0.03$ & $-1.25\pm 0.08$ \\
$(1.2,1.3)$ & 10.16 & 28914 & -2.55 & -3.50 & $10.82^{+ 0.02}_{- 0.03}$ & $-0.55\pm 0.05$ & $-1.25\pm 0.08$ \\
$(1.3,1.4)$ & 10.23 & 25776 & -2.63 & -3.58 & $10.78^{+ 0.02}_{- 0.00}$ & $-0.35\pm 0.10$ & $-1.25\pm 0.08$ \\
$(1.4,1.5)$ & 10.30 & 24697 & -2.70 & -3.65 & $10.74^{+ 0.03}_{- 0.02}$ & $-0.10\pm 0.13$ & $-1.25\pm 0.08$ \\
\enddata
\tablecomments{(1) Redshift range used for the SMF, (2) $\log(M_{\mathrm{lim}}/M_{\odot})$ denotes the stellar mass completeness limit of each redshift bin, (3) $N_{gal}$ denotes the number of galaxies with stellar mass above the stellar mass completeness limit at each redshift bin, (4) the normalization, $\phi_{1}$, of a double Schechter function, (5) the normalization, $\phi_{2}$, (6) the characteristic stellar mass, (7) the power-law slope of the high-mass end, and (8)  the power-law slope of the low-mass end. The uncertainties of the double Schechter parameters are from the forward modeling fits to the assumption-averaged  SMF. The normalization of a double Schechter function ($\phi_{1}$ and $\phi_{2}$) do not have error bars because we do not directly constrain the normalization. Instead, we define the parameter $\lambda_{\mathrm{mix}}$ that is varied between 0 and 1 to indicate the relation between of the first term and the second term of the double Schechter form (Equation~\ref{eq:dbschechter}). We then scale the stellar mass function of the mock sample so that its comoving number density is equal to the observed comoving number density of SHELA galaxies in each redshift bin (see Section~\ref{subsec:doubleschechter}).}

\end{deluxetable*}

\par We further perform forward-modeling on the observed galaxy SMF in each redshift bin as described in Section~\ref{sec:method}.
Figure~\ref{fig:massfuncavg} compares the measured SMF, which includes the measurement uncertainties, and the fitted intrinsic SMF. The forward-modeling involves random draws from estimated error distributions of stellar masses ($\sigma_{M_{\ast}}$); as a result, the intrinsic models can vary from run to run with a scatter consistent with the error bars indicated on the observed stellar mass functions.
%

\par
\begin{figure*}
\centering
\includegraphics[width=1.0\textwidth]{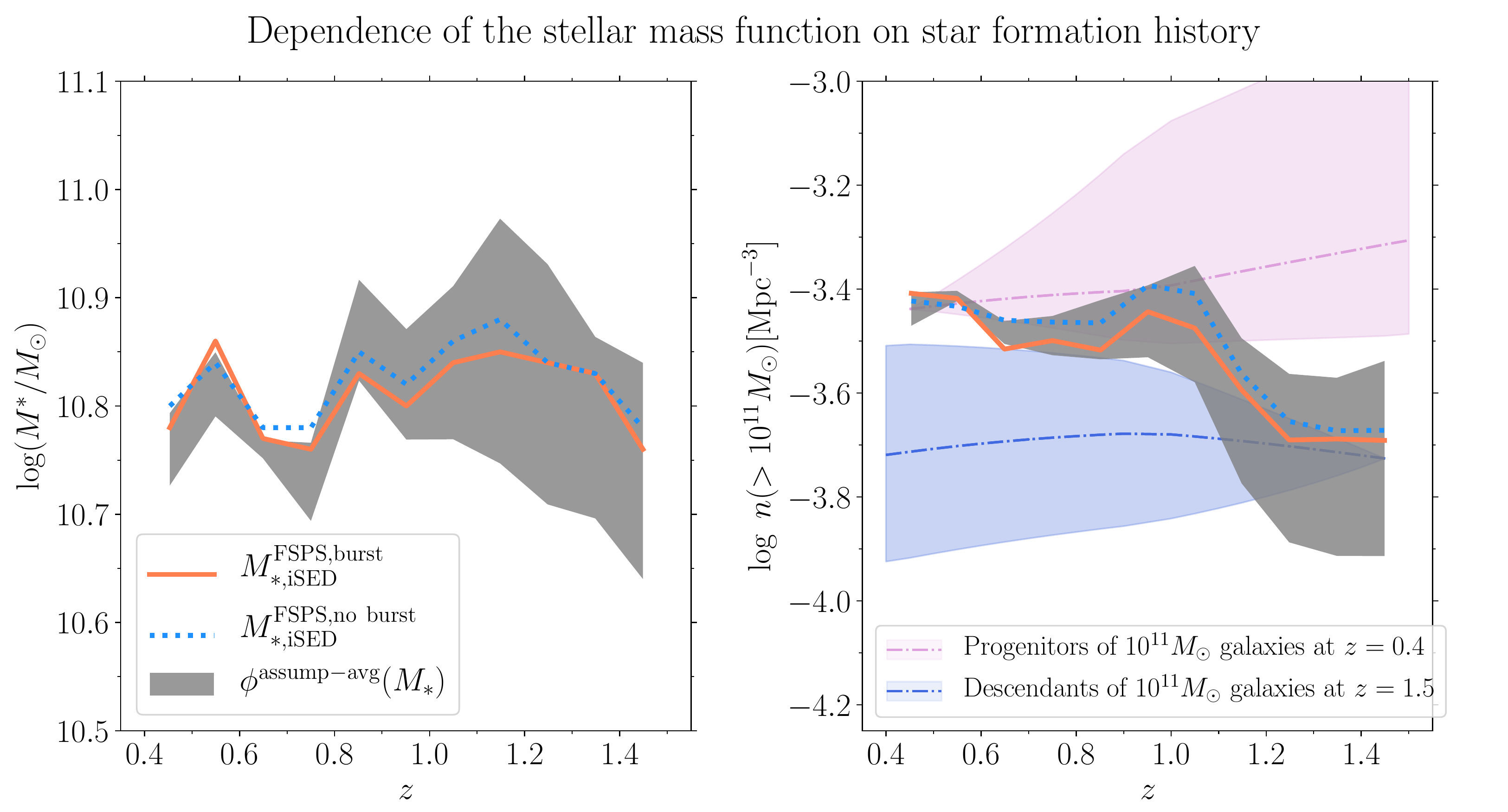}
\caption{\emph{Left:} The redshift evolution of the characteristic mass ($M^{\ast}$) of the galaxy stellar mass function (SMF) resulting from the forward-modeled SMF for two $M_{\ast}$ estimates as indicated. \emph{Right:} The redshift evolution of the cumulative comoving space density of galaxies more massive than $10^{11}~M_{\odot}$ resulting from integrating each of the forward-modeled stellar mass functions. In each panel, the grey shaded region shows the result from the assumption-averaged stellar mass function (Figure~\ref{fig:massfuncavg}) and the 68\%-tile range for all four $M_{\ast}$ estimators used to compute the assumption-averaged SMF. The individual evolutionary trends are generally consistent with the 68\%-tile range. Even with the systematic uncertainty due to the different stellar mass estimators, we find an increase in the number density of massive galaxies ($>10^{11}~M_{\odot}$) from $z=1.5$ to $z=1.0$. However, we measure \emph{no redshift evolution} in either the characteristic stellar mass of the stellar mass function ($M^{\ast}$) or the cumulative number density of massive galaxies ($>10^{11}~M_{\odot}$) from $z=1.0$ to $z=0.4$, even after accounting for the bias caused by random errors in stellar mass measurement. The pink dotted dashed line and shaded region indicate the evolution of the median and 68\%-tile range of the cumulative number density of the progenitors of $10^{11}~M_{\odot}$ galaxies at $z=0.4$ and the descendants of $10^{11}~M_{\odot}$ galaxies at $z=1.5$ (blue shaded region)  estimated using the abundance matching technique  (\citealt{Behroozi2013,Behroozi2013nov}; see text for details).}
\label{fig:evolm0sfh}
\end{figure*}

  To quantify evolution in the SMF, we first present the evolution of the characteristic stellar mass ($M^{\ast}$) resulting from the forward-modeling of the assumption-averaged mass function (Figure~\ref{fig:evolm0sfh}). Within the systematic uncertainty due to the different stellar mass estimators, we detect no redshift evolution in, $M^{\ast}$ ($\lesssim0.1$~dex) from $z=1.5$ to $z=0.4$ even after accounting for the Eddington bias caused by random errors in stellar mass measurement.

 \par Second, we derive the cumulative number density of galaxies with stellar mass greater than $10^{11}~M_{\odot}$ by integrating the intrinsic stellar mass function inferred from the forward modeling,
 \begin{equation}
     n(>M_{\ast 11}) = \int_{M_{\ast 11}}^{\infty}\phi(M_{\ast}) dM_{\ast},
 \end{equation}
 
 \noindent In practice we use an upper limit of the integral of $M_{\ast \mathrm{max}} = 10^{12.5}~M_{\odot}$, because our catalog contains no objects at higher stellar mass.
 \par We note that the cumulative number density is less sensitive to the degeneracy between the characteristic stellar mass, $M^{\ast}$, and other derived Schechter parameters. In Figure~\ref{fig:evolm0sfh} we plot the cumulative number density of galaxies with stellar mass greater than $10^{11}~M_{\odot}$ and the corresponding 68\% range. We find no significant evolution ($\lesssim0.1$~dex) in these densities out to $z<1$. In contrast, at higher redshift, the cumulative number density of massive galaxies increases by \edittwo{$\lesssim0.3$~dex from $z=1.5$ to $z=1.0$.}
 
 \par We further use the abundance matching technique to \edittwo{identify galaxy cumulative number densities with dark matter halo cumulative number densities} and estimate the evolution in the median cumulative number density of the progenitors of $10^{11}~M_{\odot}$ galaxies at $z=0.4$. We specifically implement the Number Density Redshift Evolution Code (NDE) \footnote{https://code.google.com/archive/p/nd-redshift/} \citep{Behroozi2013nov,Behroozi2013} to convert the cumulative number density of $10^{11}~M_{\odot}$ galaxies at $z=0.4$ (resulting from integrating the modeled intrinsic assumption-averaged SMF) to number densities at higher redshifts. Similarly, we estimate the evolution in median cumulative number density of the descendants of $10^{11}~M_{\odot}$ galaxies at $z=1.5$. The predicted evolution in cumulative number density of the progenitors of $10^{11}~M_{\odot}$ galaxies at $z=0.4$ is consistent with that found from the forward-modeling the SHELA SMF (Figure~\ref{fig:evolm0sfh}) at $z<1$. However, the predicted evolution in the cumulative number density of the progenitors of $10^{11}~M_{\odot}$ galaxies stays constant out to $z=1.5$, which we do not observe.
 
 \par \editthree{The discrepancy between the predicted evolution in the cumulative number density of the progenitors of $10^{11}~M_{\odot}$ galaxies at $z=0.4$ and the observed evolution, particularly at $z>1$, could arise from assumptions of the NDE.
This code ignores scatter in mass accretion and galaxy-galaxy mergers  histories. This can lead to errors when comparing the evolution of galaxies over large redshift ranges ($\Delta z > 1$). This is evidenced by comparing the evolution over the redshift range of $0.4 < z < 1.5$ (but the predicted evolution in the number density of the progenitors and the observed evolution are more consistent at $0.4 < z <1$). The scatter in stellar mass at fixed halo mass will influence the inferred $1\sigma$ range of cumulative number densities for galaxy progenitors and descendants. The NDE assumes that the growth in the differences in the ranked order of galaxy stellar mass are the same as the growth in the differences in the ranked order of halo mass as function of time. In reality, this would be the case if the star formation efficiency in individual galaxies depends much more on halo mass than on cosmic time or environment \citep{Behroozi2013jan}. }

 \par
\begin{figure*}
\epsscale{1.1}
\plottwo{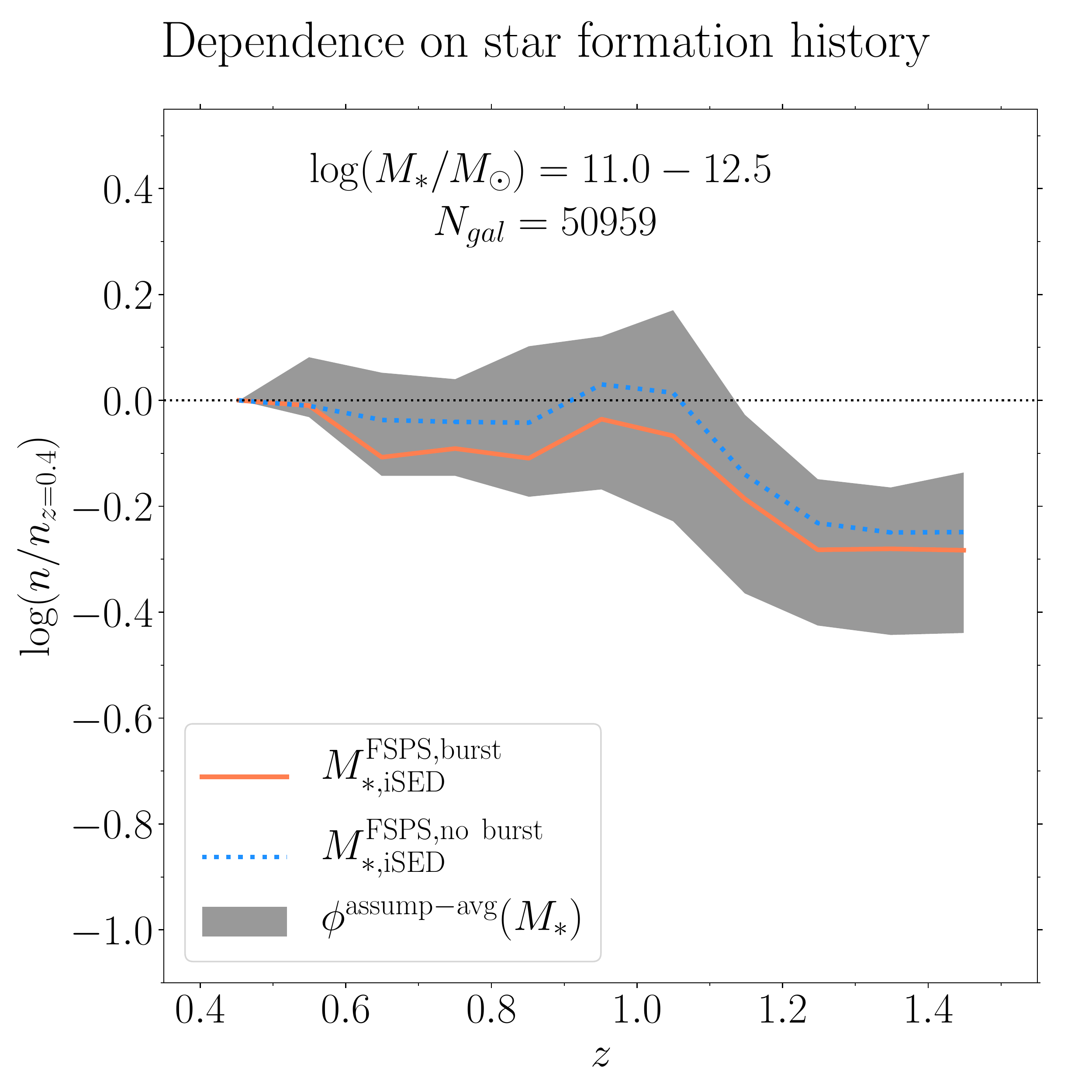}{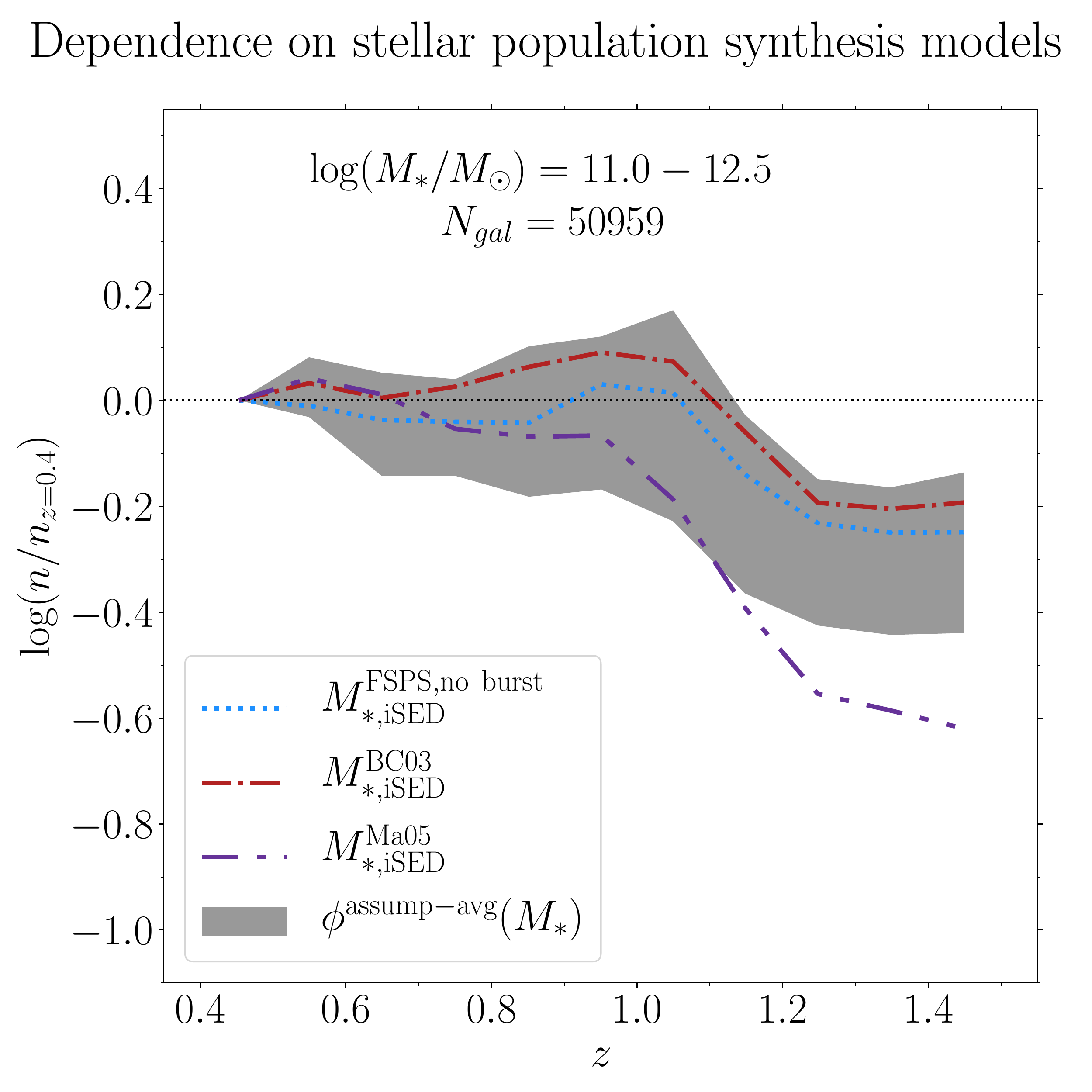}
\caption{\emph{Left:} The redshift evolution of the number density for $M_{\ast} > 10^{11}~M_{\odot}$ galaxies resulting from the forward-model fits of the SMFs with different assumptions in star formation history (SFH). Each relation has been normalized by the number density at  $z=0.4$ to compare the relative evolutionary trend. The grey shaded region shows the result from the assumption-averaged SMF (Figure~\ref{fig:massfuncavg}) and the 68\%-tile range over all four $M_{\ast}$ estimators used to compute the assumption-averaged SMF. 
\emph{Right:} Similar to the left panel but for a set of different SPS models.  The individual evolutionary trends are generally consistent with the 68\%-tile range, except for the stellar mass derived using \cite{Maraston2005} model without bursts ($M_{\ast,\mathrm{iSED}}^{\mathrm{Ma05}}$), which exhibits a stronger increase in the number density of massive galaxies with decreasing redshift ($\lesssim0.5$~dex). Overall, given the systematic uncertainty associated with the different assumptions in SFH and SPS models, the cumulative number density galaxies more massive than $10^{11}~M_{\odot}$ increases by $\sim0.4$~dex from $z\sim1.5$ to $z\sim1.0$. In contrast, at $z<1$ we detect no evolution in the cumulative number density of massive galaxies.}
\label{fig:rationdensfh}
\end{figure*}

 To better compare the relative evolutionary trend, we normalized the cumulative number density of galaxies with mass $\log M_{\ast} / M_{\odot}>11.0$ at each redshift bin to that at $z=0.4$ (Figure~\ref{fig:rationdensfh}). This is the lowest bin where the comparison between SHELA SMFs and S82-MGC \citep[$140~\mathrm{deg}^2$]{Bundy2017} shows that the SHELA sample is not significantly affected by cosmic variance (see Figure~\ref{fig:bundySMF}).  The assumption-average mass function suggests no more than a $\lesssim0.1$~dex increase in the cumulative number density of these massive galaxies since $z=1.0$ relative to those at $z=0.4$. On the other hand, the number density of galaxies increases by \edittwo{$\lesssim0.3$~dex} from $z=1.5$ to $z=1$ relative to those at $z=0.4$. In the following section, we further explore the impact of the different star formation history priors and stellar population synthesis models on the stellar mass functions and their redshift evolution.

\subsection{Dependence on Star Formation History}
\label{sec:dependsfh} In the previous section, we derive galaxy SMFs
by averaging all different sets of $M_{\ast}$ measurements that
include our various assumptions for the star formation history (SFH) and SPS models.  Within the systematic
uncertainty due to the different stellar mass estimators, we detect no
redshift evolution in either the characteristic stellar mass
($M^{\ast}$) or the cumulative number density of massive galaxies
($\log (M_{\ast}/M_{\odot})>11$) over $0.4 < z < 1.0$ (even after
accounting for the Eddington bias using the forward-modeling
method). In this section, we further investigate the evolution of the SMFs
derived using specific sets of $M_{\ast}$ measurements. 

\par We first consider how the SMF changes based on different
assumptions for the SFH.  In Figure~\ref{fig:massfuncsfh}, we show galaxy
SMFs derived using stellar masses from different SFHs, including the effects of bursts.
%
%

It is clear that the effects \edittwo{of star formation history, at least among the set of stellar mass estimates used here,} on the SMF are minor.  In Figure~\ref{fig:evolm0sfh}, we illustrate this by comparing the SMFs based on FSPS models with bursts
($M_{\ast,\mathrm{iSED}}^{\mathrm{FSPS,burst}}$) with those with no bursts
($M_{\ast,\mathrm{iSED}}^{\mathrm{FSPS,no\ burst}}$).  While there is a slight ($\lesssim 0.1$ dex) increase in characteristic mass derived from SFHs that allow bursts, this falls within the range of uncertainties, and we do not consider it significant.  In both cases where we measure the SMF from stellar masses with SFHs that allow and disallow bursts, we find the characteristic mass shows little evolution from $z=0.4$ to 1.5.   
%

\par In Figure~\ref{fig:rationdensfh} we show the cumulative comoving number density of massive galaxies ($M_{\ast} > 10^{11}M_{\odot}$) derived using stellar masses with different SFHs. In both cases we normalize the results to the measurement at $z=0.4$.  Over $0.4 < z < 1.0 $, the normalized comoving number density of massive galaxies is approximately constant regardless of SFH.
%
%
%
Similarly, the number densities drop by $\lesssim0.3$~dex from $z=1.0$ to $z=1.5$ relative to that at $z=0.4$ if we use stellar masses from SFHs that allow bursts.   We conclude that the systematic uncertainties arising from the choice of star formation history contribute $<0.1$~dex to the error budget in the growth of the characteristic stellar mass of massive galaxies, which we determined from the combined assumption-average mass function. 

\subsection{Dependence on Stellar Population Synthesis Models}
\label{sec:dependpop}
\par In Figure~\ref{fig:massfuncpop} we evaluate how three choices for the stellar population models underlying \emph{iSEDfit} $M_{\ast}$ estimates impact the derived SMFs and constraints on the growth of massive galaxies. In all cases, we compare only models with smoothly varying SFHs (e.g., no bursts).  We show again the FSPS $M_{\ast,\mathrm{iSED}}^{\mathrm{FSPS,no\ burst}}$ SMF in the left panel. We compare these to the SMFs based on \cite{Bruzual2003} masses ($M_{\ast,\mathrm{iSED}}^{\mathrm{BC03}}$, middle panel) and \cite{Maraston2005} masses ($M_{\ast,\mathrm{iSED}}^{\mathrm{Ma05}}$, right panel). The different SPS models lead to different trends in terms of the redshift of evolution of both the characteristic mass ($M^{\ast}$) of the SMF and comoving number density of galaxies with stellar mass $> 10^{11}M_{\odot}$. 

\par \editthree{There are competing claims as to the ability of the treatment of TP-AGB stars in the \citet{Maraston2005} models to reproduce the colors of galaxies and star clusters.   
 \cite{Kriek2010} found the \cite{Maraston2005} models could not simultaneously reproduce the rest frame optical and near-IR portions of galaxy SEDs. Similarly, \cite{Conroy2010} showed that the \cite{Maraston2005} models produce redder colors at intermediate ages inconsistent with the colors of star clusters in the Magellanic Clouds. However, \cite{Capozzi2016} argued that the \citet{Maraston2005} models fit better the SEDs of a sample of high redshift galaxies in COSMOS with spectroscopic redshifts, and that therefore the contribution from TP-AGB stars remains an important component in galaxy models. These points illustrate that  uncertainties in stellar population models (in particulate the treament of TP-AGB stars) is an important component to the total error budget in the evolution of the SMF.  We therefore include the results from \cite{Maraston2005} fits with those from the FSPS and \cite{Bruzual2003} models in our assumption-averaged stellar mass function (see below) to estimate the effect of uncertaintites in the stellar population models to our results.}

\par
\begin{figure*}
\centering
\includegraphics[width=1.0\textwidth]{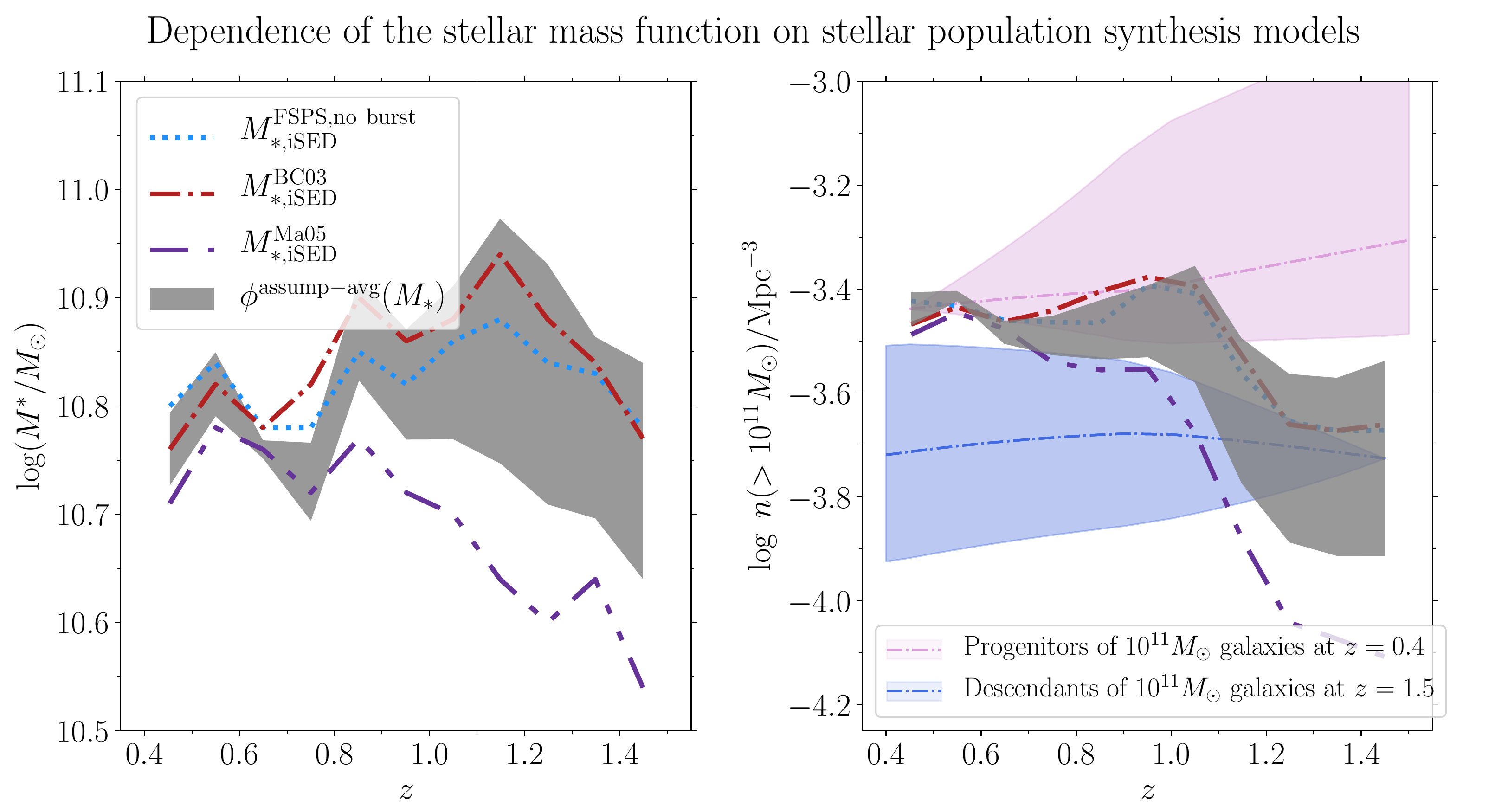}
\caption{\edittwo{Similar to Figure~\ref{fig:evolm0sfh} but for a set of different SPS models. The SMF based on \cite{Maraston2005} masses ($M_{\ast,\mathrm{iSED}}^{\mathrm{Ma05}}$) exhibits a $\sim 0.2$ dex increase in characteristic mass and a $\sim 0.4$ dex  increase in cumulative number density from $z=1.5$ to $z=1.0$.  The evolution is  milder for the stellar masses based on \cite{Bruzual2003} masses ($M_{\ast,\mathrm{iSED}}^{\mathrm{BC03}}$) and those based on FSPS ($M_{\ast,\mathrm{iSED}}^{\mathrm{FSPS,no\ burst}}$).}}
\label{fig:evolm0pop}
\end{figure*}

  The SMFs based on all SPS models we are using in this study exhibit a $\lesssim 0.1$ dex change in characteristic stellar mass from $z=1.0$ to $z=0.4$. On the other hand, at $z>1$, the SMF based on \cite{Maraston2005} masses ($M_{\ast,\mathrm{iSED}}^{\mathrm{Ma05}}$) exhibits a $\sim 0.2$ dex increase in characteristic mass (Figure~\ref{fig:evolm0pop}) from $z=1.5$ to $z=1.0$.  This evolution is milder for the stellar masses based on the \cite{Bruzual2003} models ($M_{\ast,\mathrm{iSED}}^{\mathrm{BC03}}$) and those based on FSPS  ($M_{\ast,\mathrm{iSED}}^{\mathrm{FSPS,no\ burst}}$). 

\par  We derive the cumulative number density of massive galaxies ( $> 10^{11}M_{\odot}$) normalized to that at $z=0.4$ for each SPS model (Figure~\ref{fig:rationdensfh}). Similar to the observed evolution in the characteristic stellar mass, at $z<1.0$ we find almost no evolution in the cumulative number density of massive galaxies with stellar mass $> 10^{11}M_{\odot}$ based on the stellar masses from any of the stellar population models.  On the other hand, at $z > 1.0$, \citet{Maraston2005} models exhibit a $0.4$~dex increase in the number density of massive galaxies since $z=1.5$.

%

\par \editthree{The larger evolution of the SMF based on \citet{Maraston2005} models likely arises from the different prescriptions for the TP-AGB stars in these models (compared to the assumptions used by \citealt{Bruzual2003} and \citealt{Conroy2010ascl} FSPS models).    For galaxies at $z > 0.8$, the fits using the \citet{Maraston2005} models have ages near 0.5--2~Gyr where the effects of the TP-AGB stars are most pronounced.  This lowers the stellar $M/L$ ratios of the models \citep{Maraston2006}. As a result, SED-fitting with these smodels produce fits with lower stellar masses ($M_{\ast,\mathrm{iSED}}^{\mathrm{Ma05}}$) than those of the other stellar population models (see Figure~\ref{fig:comparefastised}).} 
This reduces the number density of massive galaxies at the high mass end, yielding increased evolution in both characteristics stellar mass $(M^{\ast})$ and the number density.   Additionally, $M_{\ast,\mathrm{iSED}}^{\mathrm{Ma05}}$ has a larger spread relative to the other mass estimates, and this may contribute to the evolution in the SMF.  


\par \editthree{  In our analysis of the evolution of the SMF, we average the results from the different stellar population codes (\citealt{Conroy2010ascl}, FSPS; \citealt{Bruzual2003}, and \citealt{Maraston2005}).   Given the different treatments in the prescription of the TP-AGB phases (which leads to stronger redshift evolution in the number density of massive galaxies), this highlights how our uncertainties in the details of the later stages of stellar evolution propagate into uncertainties on measures of galaxy evolution such as the galaxy SMF.}

%
%

%

\par We conclude from the combined assumption-average SMF in Figure~\ref{fig:evolm0pop} (left panel) that systematic uncertainties arising from the choice of SPS model contribute $\lesssim0.1$~dex to the error budget in the growth of the characteristic stellar mass of massive galaxies at $z<1$, and $0.2$~dex at $1.0 < z < 1.5$.    In addition,  at least among the set of stellar mass estimates used here, the difference in SPS models are more important and lead to larger variance in the implied number density evolution than the assumptions in the star formation history.

\subsection{Dependence on Galaxy Stellar Mass}
\label{subsec:smfdepmass}
\par
\begin{figure*}
\centering
\includegraphics[width=1.0\textwidth]{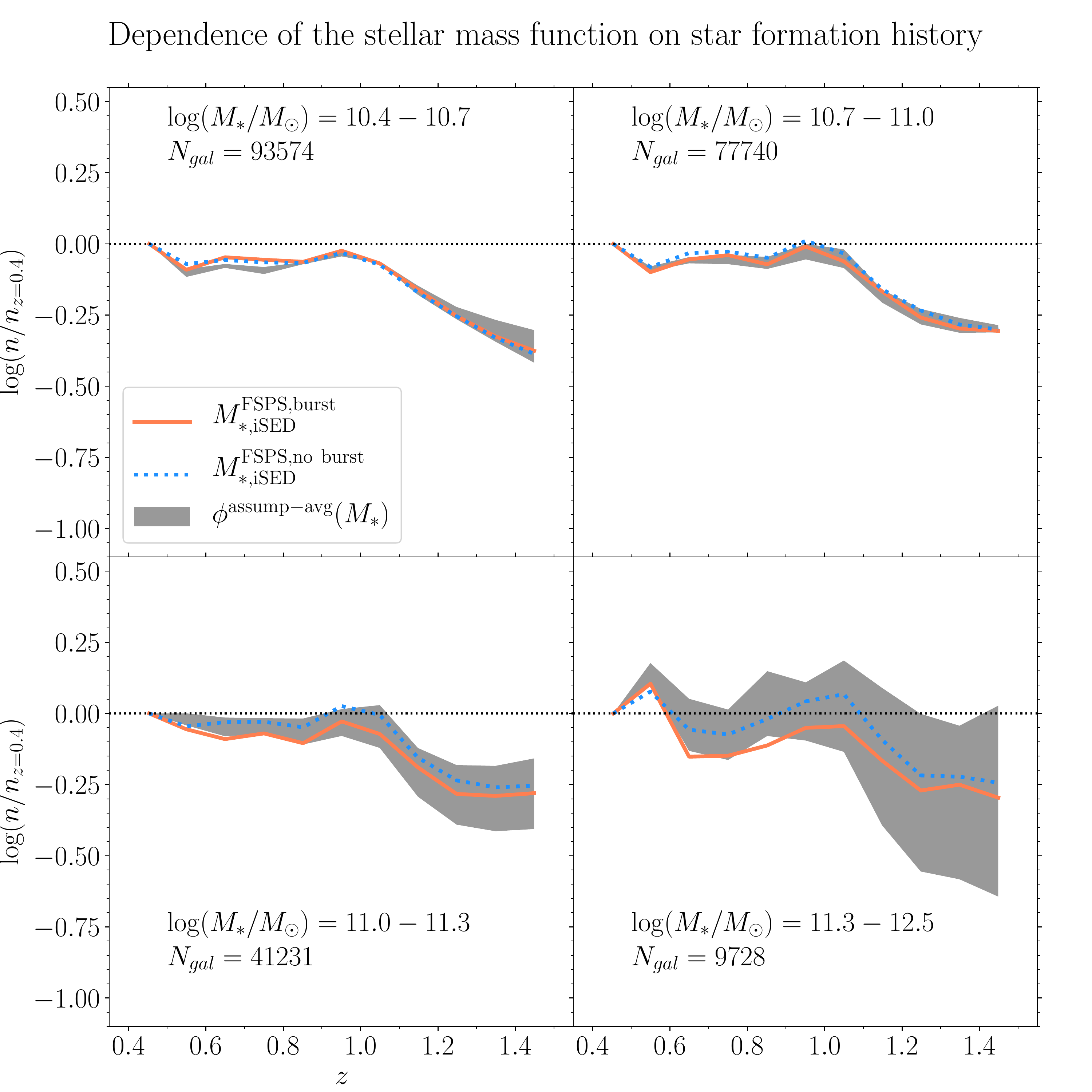}
\caption{
The relative number density of galaxies in four bins of stellar mass between $10^{10.4}-10^{12.5}~M_{\odot}$ based on different assumptions in star formation history as a function of redshift. Each relation has been normalized to the number density at  $z\sim0.4$.  In each panel, the grey shaded region shows the result from the assumption-averaged SMF (Figure~\ref{fig:massfuncavg}) and the 68\%-tile error range over all four $M_{\ast}$ estimators used to compute the assumption-averaged SMF. 
At all stellar masses,  we find $\lesssim0.5$~dex increase in the number density of galaxies more massive than $10^{10.4}M_{\odot}$ from $z=1.5$ to $z=1.0$.   At lower redshifts, $z=1$ to 0.4, we find evolution in the cumulative number density only for galaxies less massive that $10^{11}$~$M_\odot$.  Galaxies at higher stellar mass show no significant evolution in number density from $z=1.0$  to $z=0.4$.}
\label{fig:rationdensfh2}
\end{figure*}

\begin{figure*}
\centering
\includegraphics[width=1.0\textwidth]{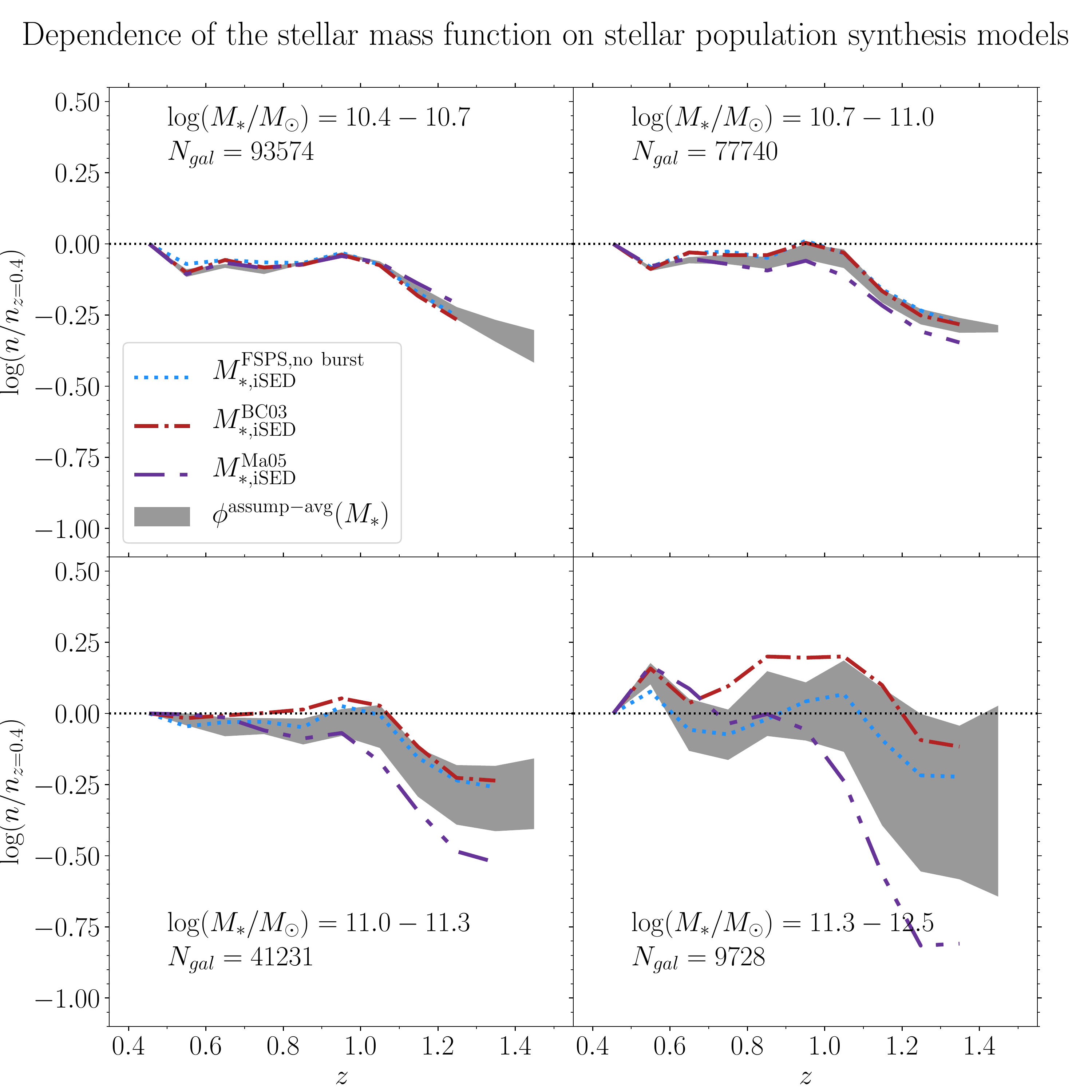}
\caption{Similar to Figure~\ref{fig:rationdensfh2} but for a set of different SPS models. In each stellar mass bin, the individual evolutionary trends are generally consistent with the 68\%-tile error range. However, for massive galaxies with $\log(M_{\ast}/M_{\odot}) > 11.0$, the stellar mass derived using the \cite{Maraston2005} models without bursts ($M_{\ast,\mathrm{iSED}}^{\mathrm{Ma05}}$), exhibits a steeper redshift dependence in the number density of massive galaxies.}
\label{fig:rationdenpop2}
\end{figure*}

 We further quantify how the evolution in SMF depends on different SFH priors in various stellar mass bins. In Figure~\ref{fig:rationdensfh2} and Figure~\ref{fig:rationdenpop2}, we plot the number density of galaxies versus redshift in four bins with stellar mass between $10^{10.4}-10^{12.5}~M_{\odot}$. Each redshift bin is normalized to that at $z=0.4$. These figures show the evolutionary trend for a set of different priors in star formation history (SFH)  and a set of different SPS models, respectively. The shaded region shows the 68\%-tile range allowed by the different model assumptions in deriving the stellar masses.
The number density of galaxies increases at different rates, with a dependence on stellar mass.

\par This leads to one of the main conclusions in this work:  \textit{The number density of galaxies  more massive than $10^{11}~M_{\odot}$ does not change significantly over $0.4 < z < 1.0$ where our sample is complete.}   Galaxies with masses between $10^{10.4}$ and $10^{11}$~$M_\odot$ show weak evidence for number density evolution.    
%
%
In contrast, there is a decline in the number density of these massive galaxies of $\simeq$0.25-0.50~dex from $z=1$ to 1.5.  
In each stellar mass bin, the individual evolutionary trends are consistent with one another at the $\pm1\sigma$ level for the different choices of SFH priors and SPS models used in this study (with the exception of the results using the \citet{Maraston2005} models). We will discuss the implications of this in the next section. 


\subsection{Dependence on Galaxy Star Formation Activity}
\label{sec:smfdepstarformation}
\par In the previous subsections we measured the evolution of the SMF for the global population of galaxies from $z=0.4-1.5$. We found no significant change in both the characteristic stellar mass and the cumulative number density of galaxies more massive than $10^{11}~M_{\odot}$ at $z<1.0$.  At these redshift, the evolution at lower stellar masses $10^{10.4}-10^{10.7}~M_{\odot}$ is largest ($\lesssim0.1$~dex) relative to the higher mass bins.  One explanation for this difference is that a higher fraction of the lower mass galaxies are still star-forming (and therefore the number density of galaxies at fixed stellar mass grows with time). We therefore compare the evolution in the SMF for galaxies that are star-forming and those that are quiescent. To make this classification, we use median of the star formation rate posteriors reported by \emph{iSEDfit} and compute the specific SFR as, $\mathrm{sSFR} = \mathrm{SFR} / M_{\ast}$. We then divide the sample into quiescent and star-forming galaxies using the evolving sSFR threshold described in  Section~\ref{sec:selectquisf}.


\begin{figure*}
\centering
\includegraphics[width=1.0\textwidth]{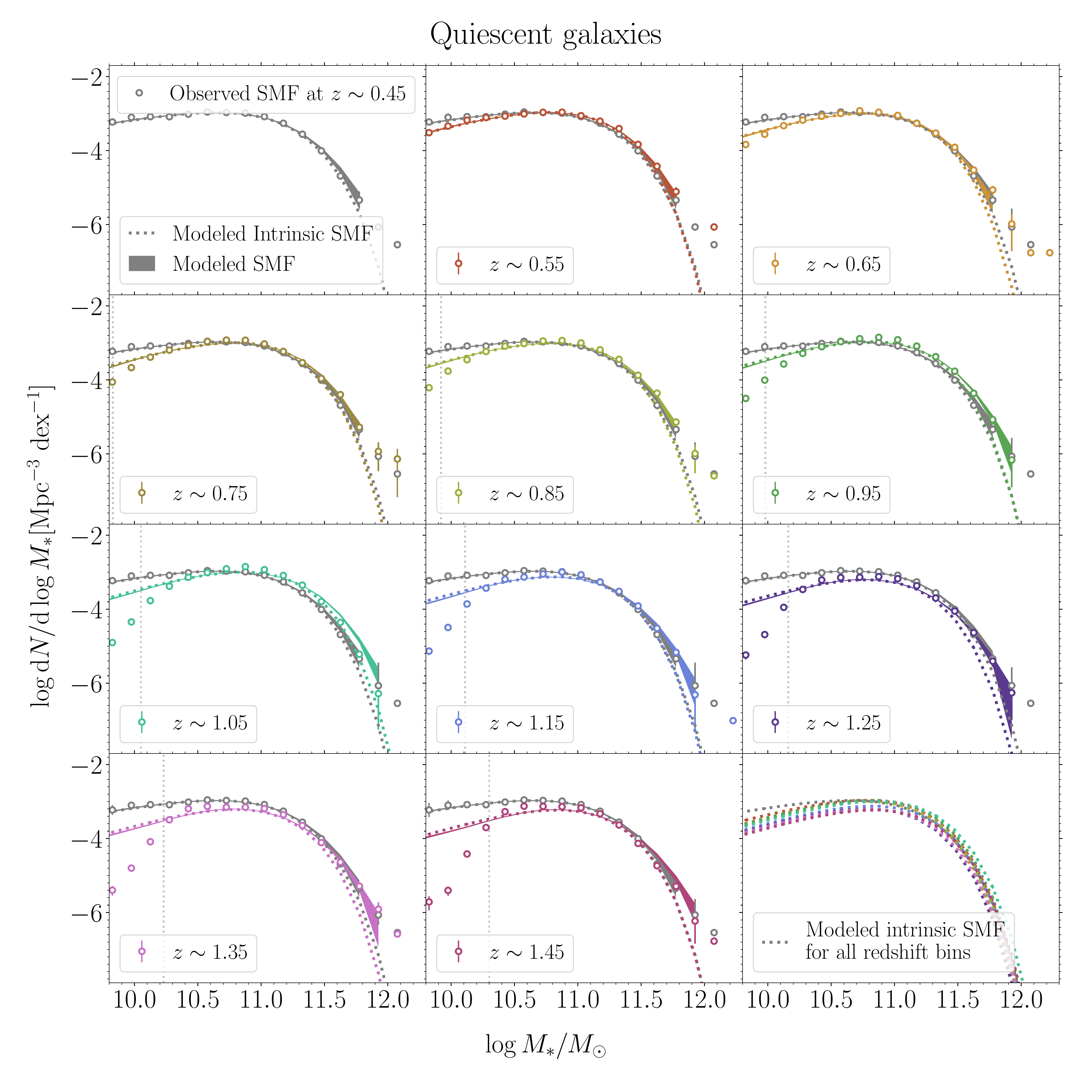}
\caption{Assumption-averaged estimated SMFs for quiescent SHELA galaxies resulting from taking a mean of the SMFs from the separate stellar mass estimators. The circles with error bars represent the observed SHELA SMFs and the corresponding Poissonian uncertainties in each redshift bin. The shaded regions represent the modeled SMFs. The estimated stellar mass completeness corresponding to each redshift bin is indicated by a vertical dotted line. Forward-modeling results, which aim to account for (and thereby remove) biases caused by scatter of stellar mass measurement, are shown as dotted curves. In each panel we show both modeled SMF and  modeled intrinsic SMF at the lowest redshift bin ($0.4 < z < 0.5$, grey shaded region and grey dotted curve) for comparison. \edittwo{The last panel shows the modeled intrinsic SMFs for all redshift bins.} The population of massive galaxies ($\gtrsim10^{11}~M_{\odot}$) are dominated by quiescent objects, which exhibits no growth ($\lesssim0.1$~dex ) in the characteristic stellar mass at fixed number density between $z=1.5$ to $z=0.4$.}
\label{fig:massfuncsqui}
\end{figure*}

\begin{figure*}
\centering
\includegraphics[width=1.0\textwidth]{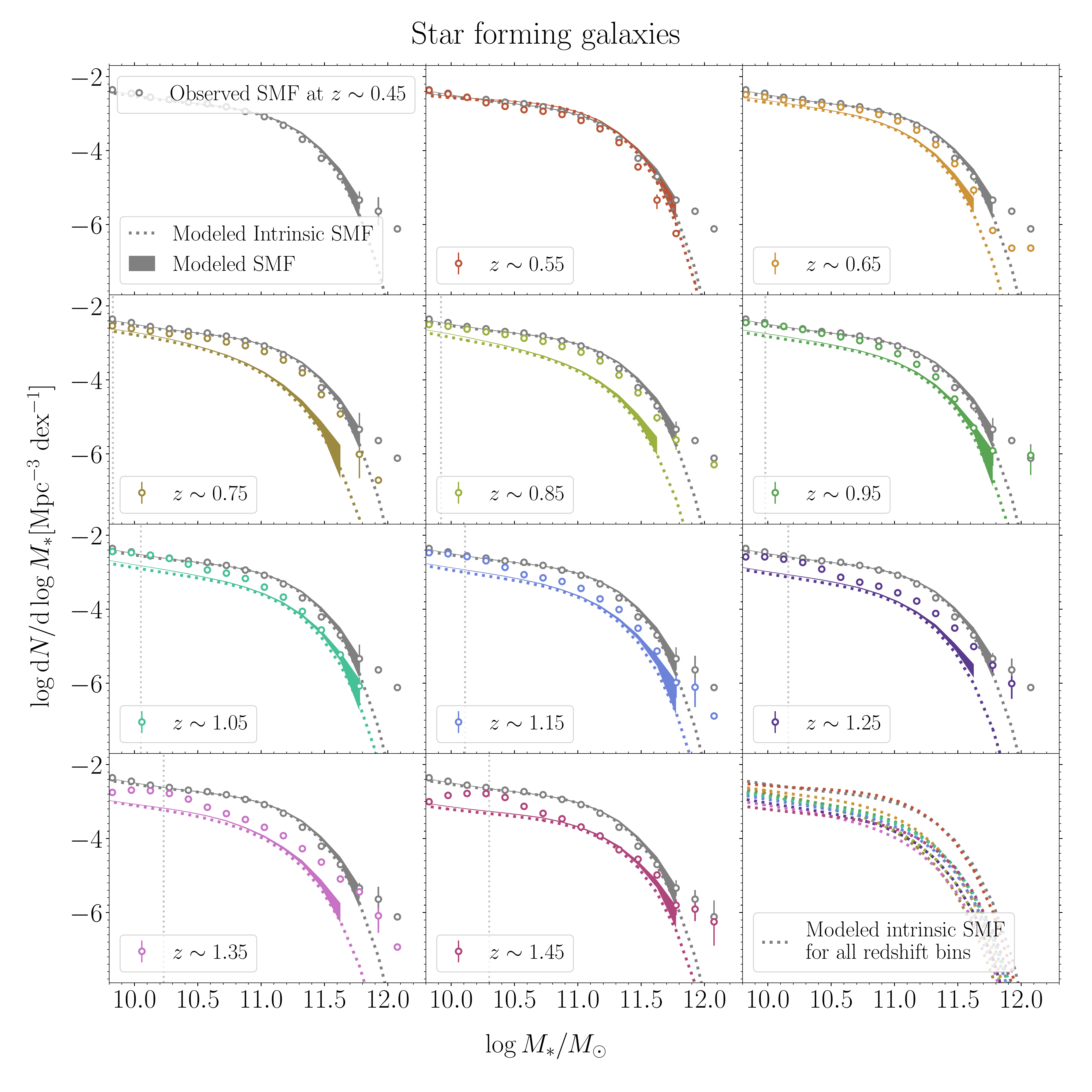}
\caption{Similar to Figure~\ref{fig:massfuncsqui} but for star-forming galaxies. At any redshift bin, the star-forming population shows moderate ($\lesssim0.2$~dex) growth in the characteristic stellar mass relative to that at $z=0.4$.}
\label{fig:massfuncssf}
\end{figure*}

\begin{figure}
\centering

\includegraphics[width=0.47\textwidth]{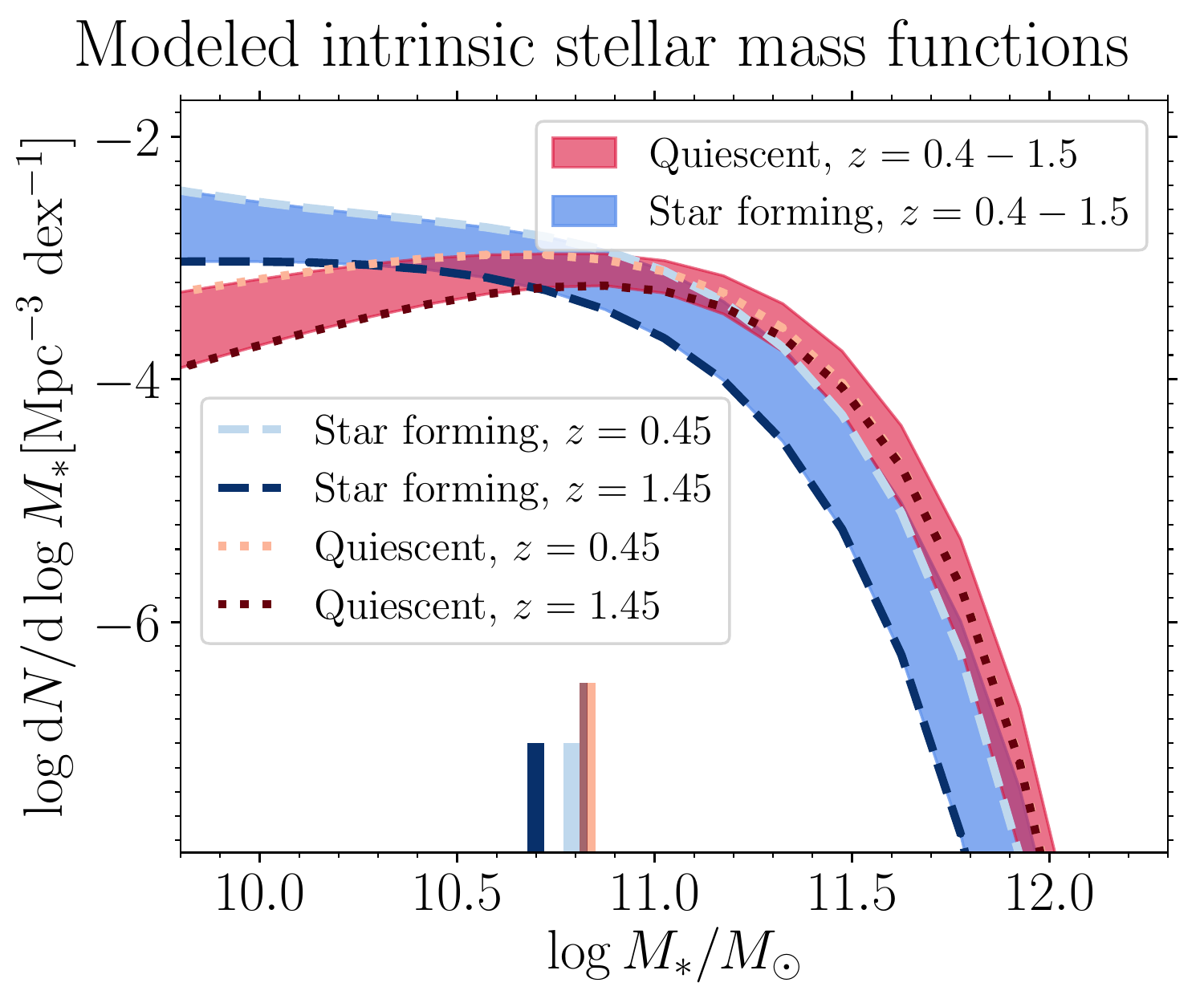}
\caption{The forward-modeled intrinsic SMFs for quiescent and star forming SHELA galaxies. For clarity, we only show the SMFs for galaxies in the lowest ($0.4 < z < 0.5$) and the highest redshift bins ($1.4 < z < 1.5$) for quiescent galaxies (red dotted curves) and star forming galaxies (blue dashed curves). The intrinsic SMFs for quiescent and star forming galaxies in all redshift bins are indicated by red and blue shaded regions, respectively.  The characteristic stellar mass ($M^{\ast}$) for each population at $z=0.4$ and at $z=1.5$ are shown as thick vertical lines on the abscissa. Quiescent galaxies exhibit no growth ($\lesssim0.1$~dex) in $M^{\ast}$ between $z=1.5$ to $z=0.4$. In contrast, star forming galaxies exhibit moderate growth ($\sim0.1-0.2$~dex) over the same redshift range.}
\label{fig:intrinmassfuncsquisf}
\end{figure}

\par We recompute the SMFs for the quiescent galaxies and star-forming galaxies using our forward modeling method. \edittwo{We plot the results in Figure~\ref{fig:massfuncsqui} and Figure~\ref{fig:massfuncssf} and present our measurements in Appendix~\ref{sec:appendix2}. Additionally, we compare the evolution of forwarded-modeled intrinsic SMFs for both populations in Figure~\ref{fig:intrinmassfuncsquisf}.} For massive quiescent galaxies with stellar mass $\gtrsim10^{11}~M_{\odot}$, we do not detect growth ($\lesssim0.1$~dex ) in the characteristic stellar mass from $z=1.5$ to $z=0.4$. However, there is strong evolution in the number density of \textit{lower mass} quiescent galaxies, similar to that seen in other studies \citep[e.g.,][]{Tomczak2014,Moutard2016}.  This build up in low-mass quiescent galaxies is expected to occur from the quenching of satellites \citep[see discussion in][]{Kawinwanichakij2017,Papovich2018}. 

For the star-forming population, we find moderate growth of $\sim0.1-0.2$ dex in the characteristic stellar mass from $z=1.5$ to $z=0.4$, and this growth in $M^{\ast}$ is larger at higher redshift. This is also consistent with previous studies \citep[e.g.,][]{Tomczak2014,Moutard2016}.  However, in this paper, we focus on the evolution of the total SMF, and we save a detailed comparison of the SMF as a function of star-formation activity for a future paper. 

\par Figures~\ref{fig:massfuncsqui},  ~\ref{fig:massfuncssf}, and \edittwo{\ref{fig:intrinmassfuncsquisf}} also show that, over $0.4 < z < 1.5$, the population of massive galaxies ($\log M_{\ast} / M_{\odot}>11$) are dominated by quiescent systems. It is therefore the evolution of this population that must account for the (lack of) evolution in the SMF for all massive galaxies.    The evolution of quiescent galaxies will involve mass losses from stellar evolution processes \citep{Girardi2000}, and we expect additional mass growth either by mergers, and/or the quenching of massive star-forming galaxies, \editone{ as there appear to be too few of the latter at $z < 1$.} 
In the next section, we discuss the implication of our finding on the rate of mass growth for these massive galaxies through merging.

\section{Discussion}
\label{sec:discussion}

\par A main conclusion from this work is that there is little observed evolution in the number density of massive galaxies, $\log M_\ast / M_\odot > 11$, from $z=1$ to $0.4$.   This result appears robust even when considering our uncertainties. Given the size of our samples, systematics dominate our uncertainties.  Of these, the most significant uncertainty comes from redshift-dependent biases in stellar mass under different assumptions for both the star formation history and SPS models (see Section~\ref{sec:dependpop}). However, these too are small to explain the results, as the models of \cite{Bruzual2003} and FSPS  are internally consistent.  
While the stellar population models of \cite{Maraston2005} would lead to stronger evolution, we consider these less favored for the reasons discussed above (Section~\ref{sec:dependpop}).  Therefore, we estimate that systematic uncertainties contribute  $<0.1$~dex to the error budget in the growth of $M^\ast$ from $z=1.0$ to $z=0.4$.   We then reach the inescapable conclusion that  there is very little (possibly no) evolution in both the characteristic mass and the cumulative number density of massive galaxies ($>10^{11}$~$M_\odot$).   

In the following subsections, we consider the implications that this conclusion has for galaxy evolution, including on the galaxy merger rate. 

%


\subsection{The Lack of Number Density Evolution: Implications for Galaxy Evolution and Galaxy Mergers}


\par The lack of evolution in the SMF of massive galaxies places constraints on models of galaxy growth and evolution.   According to the two-phase formation scenario for the formation of massive galaxies \citep[e.g.,][]{Oser2010,Oser2012}, mass assembly at late times is dominated by minor mergers \citep[e.g.,][]{Hilz2013,Oogi2013,Bedorf2013,Laporte2013}.  We also expect some mass loss of quiescent galaxies from stellar evolutionary processes and dynamical processes in clusters. The lack of observed evolution in the SMF of massive galaxies could be a balance between the build-up of stellar mass through mergers and mass loss due to stellar evolution process.

%
\par We can estimate the rate at which these massive galaxies grow by mergers at this late epoch using the results of \citet{Moster2013}, who provided a parametrization for the star formation history and mass accretion for galaxies of arbitrary present-day stellar mass. We integrate the \citet{Moster2013} fitting functions with respect to time and account for mass losses from passive stellar evolution (see \citealt{Moster2013}, their Equation 16) to derive the expected stellar mass evolution of galaxies. For systems with a present day stellar mass of $\log M_{\ast} / M_{\odot}=11$, the fraction of stellar mass loss relative to the stellar mass growth (both due to star formation and mass accretion) from $z=1.0$ to $z=0.4$ is 39\% - 45\%.  To be consistent with our measurements and to account for the lack of evolution of the SMF of massive galaxies, the upper limit on the amount of mass growth from mergers from $z=1.0$ to $z=0.4$ must be $\sim45\%$ ($\simeq0.16$~dex). 
%


%
\par Our estimate of mass growth by mergers is in good agreement with a study by \citet{vanDokkum2010}, who used a stacking analysis to study the growth of massive galaxies with a constant number density of $2\times10^{-4}~\mathrm{Mpc}^{-3}$, corresponding to a galaxy with a stellar mass of $3\times10^{10}~M_{\odot}$.  At $0.6 < z < 0.1$, \citeauthor{vanDokkum2010} found $\sim0.1$~dex mass growth for these massive systems. In addition, \citet{Marchesini2014} used the UltraVISTA catalogs to investigate the evolution of the progenitors of local ultra-massive galaxies ($\log(M_{\ast}/M_{\odot}) \approx 11.8$; UMGs). They selected progenitors with the semi-empirical approach of abundance matching, and found a growth in stellar mass of $0.27^{+0.08}_{-0.12}$ dex from $z=1$ to $z=0$ after including the scatter in the progenitor's number density in the error budget. \citeauthor{Marchesini2014} also found that half of the assembled stellar mass of local UMGs is formed primarily by merging over this redshift range. Our infer stellar mass growth and that of \citet{Marchesini2014} is consistent within the range of the uncertainties. 

\par On the other hand, \citet{Ownsworth2014} presented a study on the stellar mass growth for the progenitors of galaxies with $M_{\ast} = 10^{11.24} M_{\odot}$ at $z=0.3$, and showed that these massive galaxies have grown by a factor of \edittwo{$\sim1.8$ ($\sim0.25$~dex)} in total stellar mass since $z=1.0$. They also found that, on average, major and minor mergers account for $\sim17\%$ and $\sim34\%$ of the mass assembled to galaxies at $z=0.3$, respectively. \edittwo{In contrast, the process of star formation accounts for $\sim24\%$ of the total stellar mass.}  We observe a lower rate of mass growth from mergers compared to that from the \citet{Ownsworth2014}, and this discrepancy may result from the different SED-modeling assumptions, and the manner with which we have estimated the effects of the  Eddington bias.


\begin{figure*}
\centering
\includegraphics[width=1.0\textwidth]{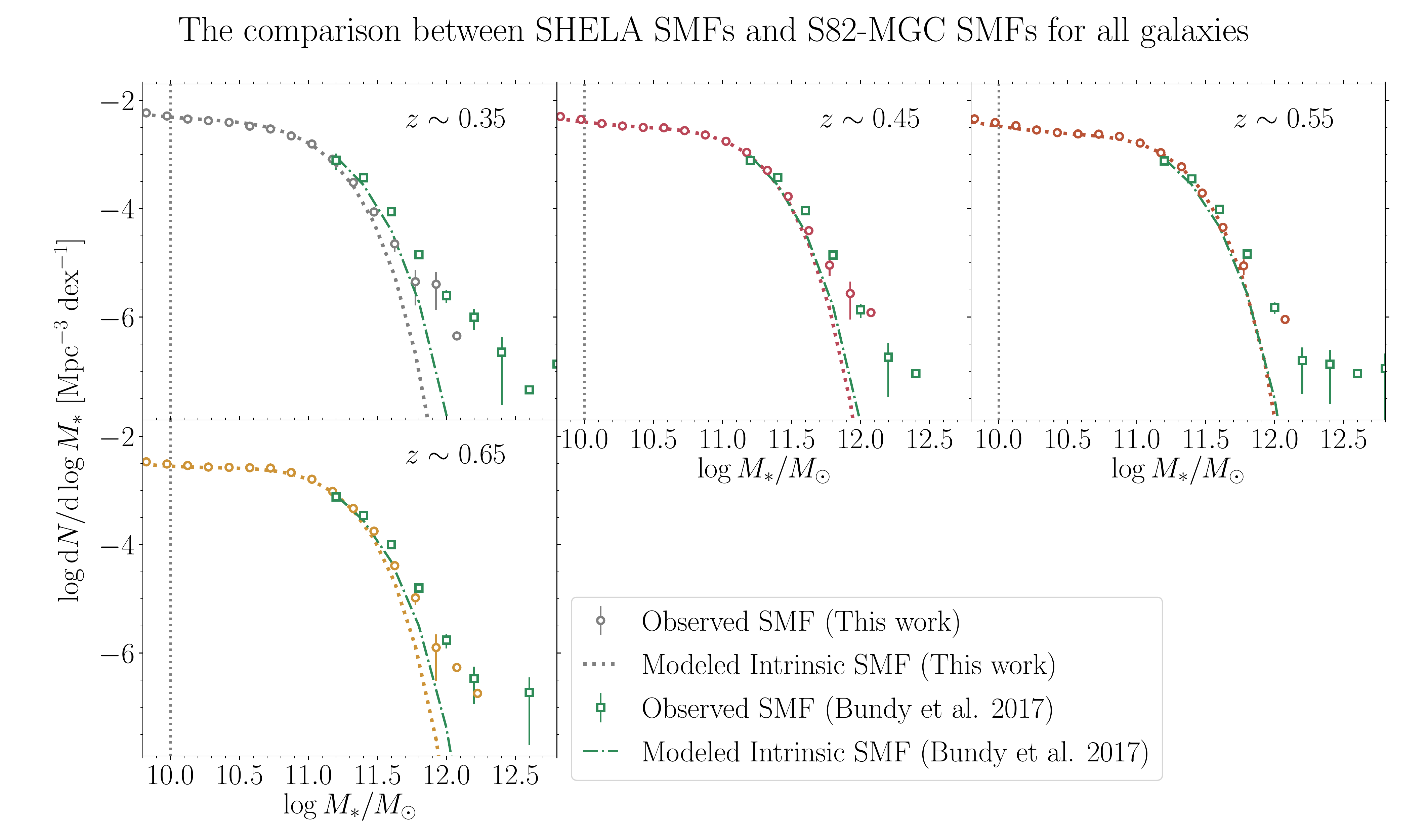}
\caption{The comparison of SMFs from SHELA and S82-MGC \citep{Bundy2017} for all galaxies between $0.3 < z < 0.65$. We reproduce our assumption-averaged SMF results from Figure~\ref{fig:massfuncavg}, with circle indicating observed SMF (and the associated Poissonian errors) and the dotted curves indicating the modeled intrinsic SMF (from the forward-model fitting results after accounting for measurement scatter). In each panel we show the observed SMF (green squares) and modeled intrinsic SMF (green dashed dotted curve) from \citeauthor{Bundy2017} at redshift $0.3 < z < 0.65$ and for the completeness limit of $\log M_{\ast} / M_{\odot}=11.2$. We are able to recover the observed S82-MGC SMF, particularly  in the $z\sim0.46$ and $z\sim0.55$ bins, suggesting that we are not significantly affected by cosmic variance at these redshifts.}
\label{fig:bundySMF}
\end{figure*}
\par Our finding can be directly compared to the recent study by \citet{Bundy2017}, who followed a similar analysis as we have here. \citet{Bundy2017} detected no growth (with an uncertainty of 9\%)  in the characteristic stellar mass of massive galaxies ($\log(M^{\ast}/M_{\odot}) > 11.2$) from $z=0.65$ to $z=0.3$ in S82-MGC. We reproduce their observed SMFs and find a consistent intrinsic SMF after accounting for the Eddington bias. 
Our number density of massive galaxies ($\log(M_{\ast}/M_{\odot}) > 11.5$) is lower than that of \citeauthor{Bundy2017}, but because of the larger volume probed by the S82-MGC, particularly at $0.3 < z < 0.4$, and the possible effects of cosmic variance, our results are still consistent. %

\editone{\cite{Capozzi2017} studied the evolution of the galaxy stellar mass function since $z=1$, using $\sim155~\mathrm{deg}^2$ of the Dark Energy Survey. In good agreement with our finding, \citeauthor{Capozzi2017} find that the number densities of galaxies with $\log(M^{\ast}/M_{\odot}) > 11$ are constant from $z\sim1$ to $z\sim0.2$. In addition, these authors also find the mass-dependence of the galaxy number density -- less massive galaxies exhibit larger evolution in the number density compared to more massive galaxies. Again, this is qualitatively consistent with our finding (see Section~\ref{subsec:smfdepmass}), and we are able to verify the robustness of these results by fully accounting for the statistical and systematic uncertainties on stellar mass estimates. Also, the deep mid-infrared photometry from Spitzer/IRAC allows us to better constrain stellar masses and improve the uncertainties.}

\par \citet{Moutard2016} presented an analysis on the evolution of the SMF from redshift $z=0.2$ to $z=1.5$ of a $K_{s} < 22$-mag selected sample, over an effective area of $\sim22.4~\mathrm{deg}^2$ \editone{of the VIPERS Multi-Lambda Survey}. \editone{To account for scatter in the stellar mass measurements, \citet{Moutard2016} corrected the SMF during their fitting procedure by convolving the parametric form of the SMF with the stellar mass uncertainty \citep{Ilbert2013}.  \citeauthor{Moutard2016} showed that the number density of the most massive galaxies ($\log(M_{\ast}/M_{\odot}) > 11.5$) increases by a factor of $\sim2$ from $z\sim1$ to $z\sim0.3$. The higher number density of massive galaxies inferred by \citeauthor{Moutard2016} compared to our finding could arise from the different methods used to account for the scatter in stellar mass measurement and our inclusion of Spitzer/IRAC mid-IR measurements in our stellar mass determinations.
\citeauthor{Moutard2016} also demonstrated that the quiescent population largely dominates the massive galaxies population since $z\sim1$; this agrees with our result that the massive galaxies assemble their stellar masses through mergers.}

%



\subsection{The Evolution of the Cumulative Number Density of Massive Galaxies}
\label{subsec:evonumberdensity}
\begin{figure*}
\centering
\includegraphics[width=1.0\textwidth]{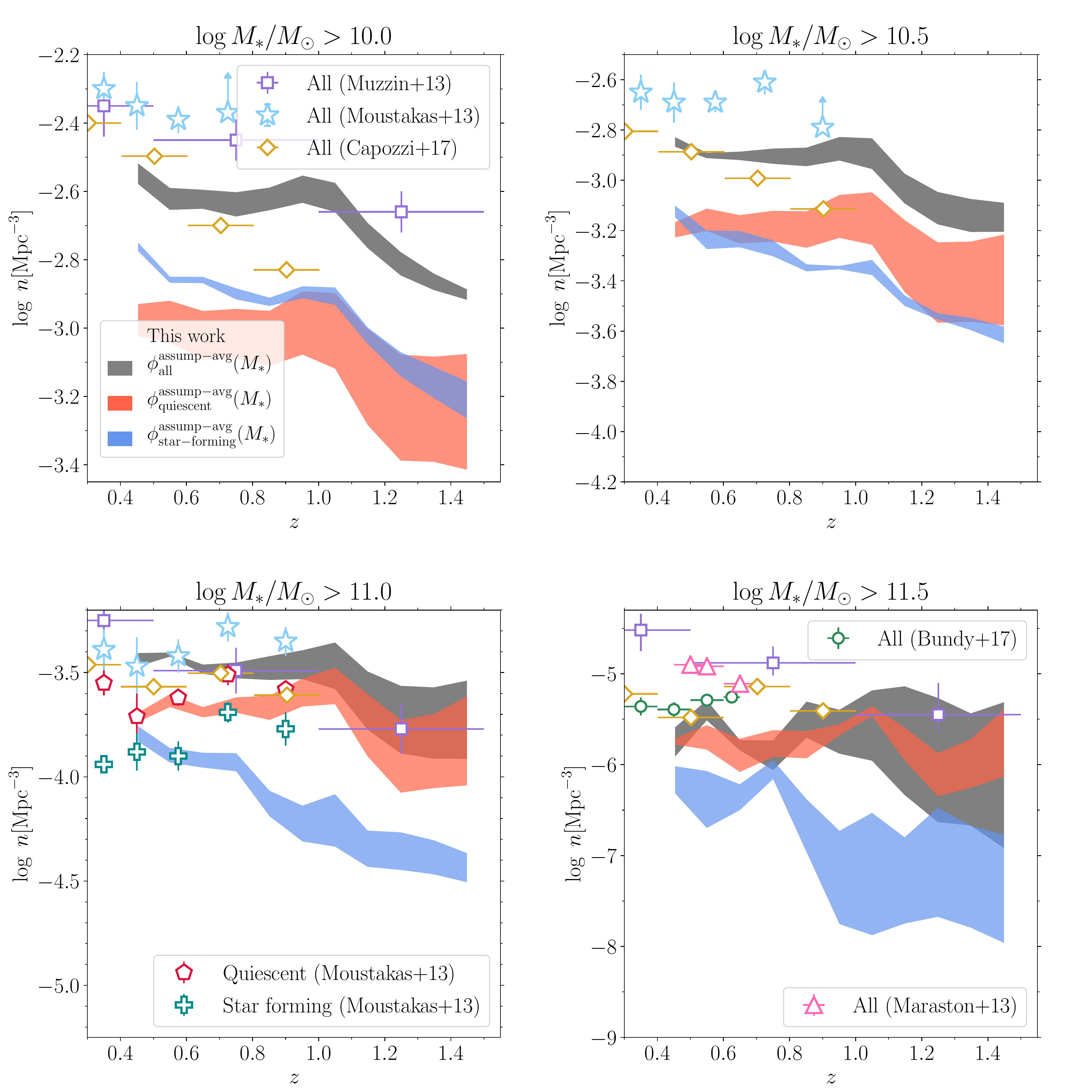}
\caption{The evolution of the cumulative number density of galaxies above a fixed mass limit and the 68\%-tile range over all four stellar mass estimators. The results from the assumption-averaged SHELA SMF for all, quiescent, and star-forming populations are shown as grey, red, and blue shaded regions, respectively. The purple squares, light blue stars, yellow diamonds, green circles, and pink triangle are from UltraVISTA \citep{Muzzin2013}, PRIMUS \citep{Moustakas2013}, \edittwo{DES \citep{Capozzi2017}, S82-MGC \citep{Bundy2017}, and BOSS \citep{Maraston2013}}, respectively, for all galaxy population. The red pentagons and blue crosses are from PRIMUS for quiescent and star-forming populations with stellar masses of $10^{11-11.5}M_{\odot}$. The error bars of PRIMUS represent the quadrature sum of the Poisson and cosmic variance uncertainties. Similarly, the error bars of UltraVISTA represent the quadrature sum of the Poisson, cosmic variance and the errors from photometric uncertainties. The error bars from S82-MGC are Poisson uncertainties.
}
\label{fig:muzzincumudensity}
\end{figure*}

\par In this section, we compute the evolution of the cumulative number density of galaxies as a function of  stellar mass threshold and compare it with other studies. Here, we are particularly interested in the  redshift evolution measured within our SHELA field, as it mitigates against systematic uncertainties in the analysis between other surveys (and our survey is one of the few that attempts to forward-model the SMF).

\par We integrate our best-fitting intrinsic SHELA SMF (i.e, the best
fit SMF derived from the forward modeling of assumption-averaged SMF)
for stellar masses greater than $\log M_\ast / M_\odot$ =  10, 10.5, 11, and 11.5. In Figure~\ref{fig:muzzincumudensity} we show
the cumulative number density of galaxies in the four mass bins.
For galaxies at all masses ($\log M_\ast/M_\odot > 10$) the cumulative number density in our sample is consistent with little evolution (as we have showed above).   We see here that for galaxies more massive than $\log M_\ast/M_\odot > 10.5$, this (lack of) evolution in number density is primarily due to quiescent galaxies, which show a constant number density (out to $z < 1$).  
In contrast, star-forming galaxies show an increase in cumulative number density at all redshifts and stellar masses.  This is consistent with the scenario that these galaxies continue to form stars (and stellar mass) and build up their number densities at later cosmic times.

For all $\log M_\ast/M_\odot > 10.5$ galaxies, quiescent systems dominate by number density for all redshifts considered here ($z \lesssim 1.5$). 
For galaxies more massive than $\log M_\ast / M_\odot > 11$, quiescent galaxies at fixed mass outnumber star-forming galaxies by a factor of $\sim$3:1.  Quenching of star-forming galaxies can therefore at most contribute to roughly a 33\% increase in the number density of quiescent galaxies at $z < 1.5$.


We can additional gain insight by comparing our results to other studies. These cumulative number densities are plotted in Figure~\ref{fig:muzzincumudensity} at the redshifts and stellar masses where they overlap with our study.  We begin by comparing our data to that of S82-MGC \citep[$\sim140~\mathrm{deg}^2$]{Bundy2017}. Because both their study and our study use the  forward modeling method to account for the Eddington bias, we can directly take their best-fit intrinsic SMF and integrate it to compute the cumulative number density, at least to $>10^{11.2}~M_{\odot}$, where \citeauthor{Bundy2017} are complete.   %
%
%

Given that our result is in excellent agreement with that of the larger survey area of S82-MGC, this strongly suggests that our 17.5~$\mathrm{deg}^2$ SHELA survey comoving volume of $\sim0.15~\mathrm{Gpc}^3$ in the redshift range of $0.4 < z < 1.5$, is sufficient to mitigate the effects of cosmic variance, even for very rare galaxy populations.  We can thus put strong constraints on the evolution of the galaxy SMF down to a stellar mass of $10^{10.3}~M_{\odot}$. 
%

\par The lack of evolution ($\lesssim0.1$~dex) seen in our cumulative number density of all galaxies more massive than $M_{\ast} > 10^{10}~M_{\odot}$ is also in agreement with the result of \citet{Moustakas2013} (Figure~\ref{fig:muzzincumudensity}). By integrating the observed SMF of the $5.5~\mathrm{deg}^2$ PRIsm MUlti-object Survey \citep[PRIMUS;][]{Coil2011}, these authors find  a $\lesssim10\%$ change in the number density of $M_{\ast} > 10^{11}~M_{\odot}$ galaxies since $z\approx1$.  It is interesting that when \citet{Moustakas2013} divided their sample into quiescent and star-forming galaixes, they found that the number density of quiescent galaxies with $10^{11-11.5}M_{\odot}$ has changed relatively little since $z=1$ and the decline in the number density of massive star-forming galaxies is $\lesssim0.2$ dex. This is in agreement with our finding.  We do observe slight offsets in the normalization of the SMF compared to our result, which may be a result of systematics in the sample selection or SED analysis. 

\par In contrast, the cumulative number density evolution observed in the UltraVISTA \citep[1.62 $\mathrm{deg}^2$]{Muzzin2013} survey finds $\sim0.2-0.4$~dex growth in the number density of galaxies with stellar mass above $10^{10},10^{11}$, and $10^{11.5}~M_{\odot}$, respectively, from $z=1$ to $z=0.2$.  While on the surface this runs counter to our results, an inspection of Figure~\ref{fig:muzzincumudensity} shows that we are in agreement (within the uncertainties) for regions where both surveys are complete:  higher mass galaxies with $\log M_\ast / M_\odot > 11$ and redshifts, $z > 0.4$.   At higher masses, $\log M_\ast / M_\odot > 11.5$, it is likely that UltraVISTA is \edittwo{limited by the cosmic variance}. While at $z < 0.4$, we have already argued that SHELA may be incomplete (in comparison to \citeauthor{Bundy2017}) and this may be true as well for the much smaller-area UltraVISTA survey. \editone{Additionally, as we already noted in Section~\ref{sec:selectquisf}, our interpretation of the number density and stellar mass density evolution may be impacted by less accurate SFRs at $z<0.5$ because the $u-$band does not sample the rest-frame NUV \citep[see][who perform similar analysis including UV photometry with the GALEX satellite]{Moutard2016}.} 

\par \editthree{Figure~\ref{fig:muzzincumudensity} compares the results from our study to those from \cite{Capozzi2017}, using galaxies from at $0.1 < z < 1$ observed in DES. \citealt{Capozzi2017} find that the cumulative number density evolves grows by $\sim0.4$ (0.3) dex over this redshit range for galaxies with stellar mass above $10^{10}$ ($10^{10.5} M_{\odot}$).  The larger evolution in number density could arise from a difference in survey selection: the \citeauthor{Capozzi2017} sample is optical $i-$band-selected ($i<23$ mag), and this could impact their stellar mass completeness limit particularly at lower masses and higher redshifts. This is where we observed the greatest discrepancy in the evolution of the cumulative number density. Additionally, the discrepancy could arise from the different methods used to account for the statistical and systematic uncertainties in stellar mass measurements. Interestingly, the number density of galaxies with $M_{\ast} > 10^{11.5}$ of \cite{Capozzi2017} show no evolution since $z\sim1$, in good agreement with our finding and the result at $z<0.06$ from GAMA \citep[covering 143 deg$^2$, ][see also Figure 15 of \citealt{Capozzi2017}]{Baldry2012}. } 

\par \editthree{Figure~\ref{fig:muzzincumudensity} also shows number densities from the integrated SMF from BOSS \cite{Maraston2013} for galaxies with $M_{\ast} > 10^{11.5}$, out to $z\sim 0.7$ where their survey is complete in stellar-mass.   The lack of evolution between $0.45 < z < 0.7$ in the cumulative number density of BOSS galaxies in the highest mass bin ($M_{\ast} > 10^{11.5}$) is in good agreement with our finding here. There is a slight offset in the normalization of the cumulative number density of \cite{Maraston2013} compared to our results, but that may arise from the systematics in the sample selection or SED analysis.}

\par \editthree{Finally, our results are broadly in agreement with (smaller area) surveys selected using deep near-IR data.  \cite{Mortlock2015} find little evolution in the characteristic stellar mass  ($M^{\ast}$) of the SMF from $z=3$ to $z=0.3$ in their analysis of the combination of deep near-IR data from the Ultra Deep Survey (UDS), and the Cosmic Assembly Near-infrared Deep Extragalactic Legacy Survey (CANDELS) UDS and GOODS-S fields (see Figure~\ref{fig:evolm0sfh}). Additionally, \citet{Conselice2007} presented the evolution of massive galaxies at $z\sim0.4-2$ by combining wide and deep NIR imaging from the Palomar telescope with DEEP2 spectroscopy. They found that galaxies with $\log(M_{\ast}/M_{\odot}) > 10.5$ exhibit no significant evolution in the number density since $z<1$, but show an increase in the number density of galaxies with $11 < \log M_\ast / M_\odot < 11.5$ from $z\sim 1$ to 1.5 (see their Figure 4). At higher masses, $\log M_\ast / M_\odot > 11.5$, it is likely that their sample is limited by the cosmic variance.  These results are consistent with our findings.}

Taken together, our analysis, combined with results in the literature, paint a picture where the number density of massive galaxies, $\log M_\ast / M_\odot \gtrsim 11$, is roughly constant out to $z\sim 1$.  Additionally, this evolution is dominated by the number density of quiescent galaxies.  

\subsection{The Evolution of the Total Stellar Mass Density.}
\label{subsec:evomassdensity}
\par We compute the total stellar mass density by integrating our best-fitting intrinsic (assumption-averaged) SHELA SMF for stellar masses greater than $10^{9}~M_{\odot}$, and compare our results with other studies. Again, we are particularly interested in the internal redshift evolution, as there are significant discrepancies in the normalization of the stellar mass density between the different studies.

\begin{figure}
\centering
\includegraphics[width=0.47\textwidth]{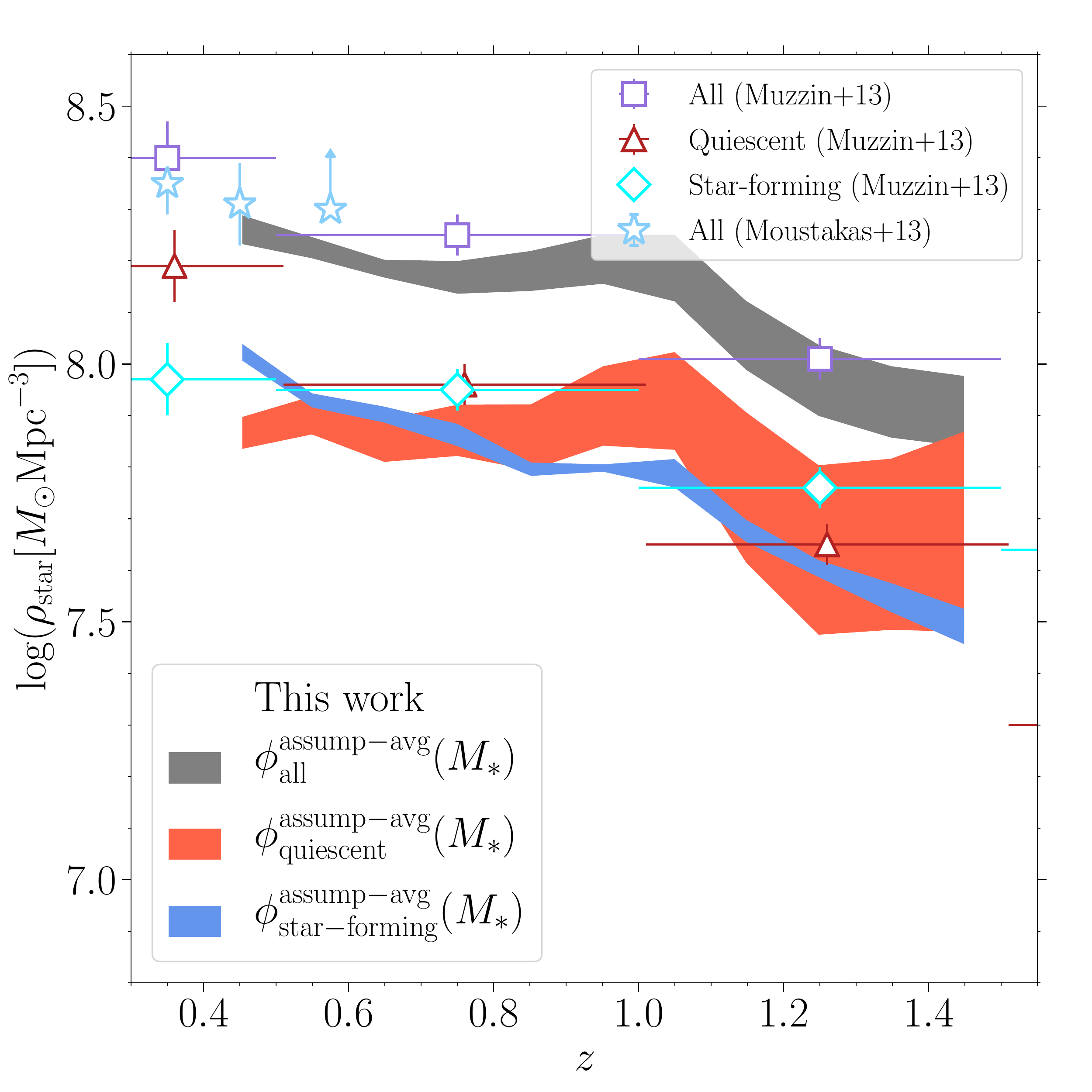}

\caption{The evolution of the stellar mass density of galaxies from $z=1.5$ to $z=0.4$ down to a limit of $\log M_{\ast} / M_{\odot}=9.0$. Plotted are the SHELA measurements and the 68\%-tile error range over all four stellar mass estimators. The results from the assumption-averaged SHELA SMF for all galaxies, quiescent galaxies, and star-forming galaxies are shown as grey, red, and blue shaded regions, respectively. The purple squares and light blue stars are from UltraVISTA \citep{Muzzin2013}, and PRIMUS \citep{Moustakas2013} surveys, respectively. The red triangles and cyan diamonds are from UltraVISTA for quiescent and star-forming galaxies. }
\label{fig:muzzinsfd}
\end{figure}

The stellar mass density we derive from SHELA shows an overall increase from $z=1.5$ to $z=0.4$.  However, most of this evolution occurs before $z \sim 1$.  This is consistent with our previous statements that the number density evolution at $ z < 1$ is dominated by quiescent galaxies.  Figure~\ref{fig:muzzinsfd} illustrates this: quiescent galaxies show no measurable growth in stellar mass density at $z < 1$. Most evolution occurs in star-forming galaxies, which show a continuous increase in the stellar mass density from $z = 1.5$ to $z=0.4$.   

\par We compare our stellar mass density with those from UltraVISTA \citep{Muzzin2013} and PRIMUS \citep{Moustakas2013}. The values from UltraVISTA are measured by integrating the maximum-likelihood Schechter function fits of the observed SMF, down to a limit of $10^{9}~M_{\odot}$; those from PRIMUS are derived by integrating the observed SMF and also the best-fit Schechter function down to a limit of $10^{9.5}~M_{\odot}$. Overall, the evolution of the stellar mass density for all galaxy populations are consistent, with $\sim0.2$ dex growth in the stellar mass density from  $z\sim1.5$ to $z\sim1$ (but our data provide the most detailed accounting of uncertainties). At lower redshifts, our result and those from UltraVISTA and PRIMUS are generally consistent with $\lesssim0.1$ dex growth $z\sim1$ to $z \sim 0.4$.  However, we note that our measurements encompass the largest area, and include in the error budget the effects from forward modeling, and uncertainties associated with the derivation of stellar masses.  

\subsection{On the Selection Quiescent and Star-forming Galaxies and their Number Density Evolution}
\label{sec:choiceofssfr}

\par \edittwo{Lastly, we consider how our results would be impacted using different thresholds to separate quiescent from star-forming galaxies.   In this study we have adopted the evolving threshold of $\log (\mathrm{sSFR} / \mathrm{yr}^{-1})$ that ranges from $-11$ at $z=$0.4 to  $-10.2$ at $z$=1.5 (see Section~\ref{sec:selectquisf} and Figure~\ref{fig:sfrmass}).  We tested how our results would change if we  used instead a non-evolving (fixed) threshold of $\log$ (sSFR / yr$^{-1}$) of $-11$ at all redshifts.  With the unevolving sSFR selection, we find no significant evolution ($\lesssim0.1$ dex) in the cumulative number density of massive quiescent galaxies (with $\log M_\ast / M_\odot > 11$) from $z=1.0$ to $z=0.4$, consistent with the behavior derived using the evolving threshold of sSFR (Figure~\ref{fig:muzzincumudensity}).  We do observe a difference when we consider the number density of quiescent galaxies down to more moderate masses of $\log (M_{\ast}/M_{\odot}) > 10$, which now exhibits a 0.2 dex increase from $z=1.0$ to $z=0.4$. Of course, this comes with a trade off in the evolution of star-forming galaxies, which now show  $\lesssim 0.1$ dex evolution in the cumulative number density from $z=1.0$ to $z=0.4$ in all stellar mass bins. }

\par \edittwo{The choice of sSFR threshold to differentiate between star-forming and quiescent galaxies clearly impacts our interpretation of evolution (and this is true for other studies in the literature).  We favor a redshift dependent sSFR cut because we wish to study galaxies that are star-forming (with a current SFR higher or equal to their past average) and quiescent (with a current SFR that is much less than their past average).  This is only achieved by using a redshift dependent sSFR for the reason that the SFR--stellar mass relation itself evolves (see Figure~\ref{fig:sfrmass}) and because there is less time available at higher redshift.  In contrast, using a fixed sSFR threshold would change the definition of quiescence, and would even include galaxies that are still on the star-forming main sequence at $z\sim 0.3$.   Our results specifically describe the evolution of quiescent and star-forming galaxies as defined here.   Using a different selection threshold to identify different populations (and describe the evolution of different populations of galaxies) would impact the interpretation.  It is therefore important to take into account the definition of quiescent and star-forming galaxies when comparing results in the literature. }

\section{Summary and Conclusions} \label{sec:conclusion}
We have exploited optical to mid-infrared photometric catalog of the 17.5 $\mathrm{deg}^2$  \textit{Spitzer}/HETDEX Exploratory Large-Area Survey (SHELA) to measure the galaxy SMF in 11 redshift bins from $z=0.4$ to $z=1.5$ down to $\log (M_{\ast} / M_{\odot})=10.3$. The large area and depth of SHELA drastically reduces the statistical uncertainties due to Poissonian errors and cosmic variance. The results can be summarized as follows.

\par We performed forward modeling to account for random and systematic errors in our stellar mass estimates and investigate their effects on the derived mass functions. We combined $M_{\ast}$ estimates that use a range of currently uncertain assumptions about star formation history and SPS models.
%
%
We find very little evidence for evolution in the SMF:  there is $\lesssim0.1$~dex evolution in both the characteristic stellar mass and the cumulative number density of massive galaxies ($>10^{11.0}~M_{\odot}$) between $0.4 < z < 1.0$ with \editthree{an uncertainty of only 13\%}.  We also present evidence for evolution in the cumulative number density of massive galaxies at higher redshift, which increases by $\lesssim0.4$~dex from $z=1.5$ to $z=1.0$.

\par  We discuss the contributions to the error budget, which are dominated by systematics.  This includes differences in SPS models and assumptions about the star formation history  used to derive the stellar mass. Among the effects considered here, the systematic uncertainties arising from the choice of star formation history and SPS models contribute $\lesssim0.1$~dex to the error budget in the growth of the characteristic stellar mass of massive galaxies at $z<1$ and increase to $0.2$~dex at $1.0 < z < 1.5$.

\par We discuss the evolution of the SMF, cumulative number density and stellar mass density.   We also consider the evolution of these as a function of galaxy star-formation activity (selected on the basis of their sSFR), using samples of quiescent and star-forming galaxies.   We find that quiescent galaxies dominate the evolution at the massive end of the SMF at all redshifts under consideration.    We do not detect evolution ($\lesssim0.1$~dex) in the number density of massive quiescent galaxies ($>10^{11.0}~M_{\odot}$) over $0.4 < z < 1.0$, even after accounting for the systematic and random uncertainties in the $M_{\ast}$ measurement.
We also find that quiescent galaxies dominate the massive end of the SMF by a ratio of 3:1 over star-forming galaxies.     Because we expect quiescent galaxies to experience stellar mass losses of 45\% over this redshift range ($0.4 < z < 1.0$), additional growth must occur to balance these effects.   Assuming this growth is dominated by (dry) mergers, we can derive an upper limit on the mass growth from these events.
%
%
Our observation suggests that the upper limit on mass growth by mergers over this redshift range is $\sim45\%$ ($\sim0.16$~dex) for quiescent galaxies more massive that $\log M_\ast/M_\odot > 11$.

\acknowledgements

We thank the anonymous referee for providing insightful comments and suggestions that improved the quality of this work. We are grateful for the support from the World Premier International Research Center Initiative (WPI Initiative), MEXT, Japan. We would like to thank to Darren L. DePoy, Robert C. Kennicutt, and Kim-Vy H. Tran for their helpful comments and suggestions. This work is supported by the National Science Foundation through grants AST 1413317 and 1614668, the NASA Astrophysics and Data Analysis Program through grant NNX16AN46G.  We acknowledge generous support from the George P. and Cynthia Woods Institute for Fundamental Physics and Astronomy at Texas A\&M University. LK and CP acknowledge the Texas A\&M University Brazos HPC cluster that contributed to the research reported here. The Institute for Gravitation and the Cosmos is supported by the Eberly College of Science and the Office of the Senior Vice President for Research at the Pennsylvania State University. This research made use of Astropy,\footnote{http://www.astropy.org} a community-developed core Python package for Astronomy \citep{astropy:2013, astropy:2018}.

\appendix

\begin{figure}
\centering
\includegraphics[width=1.0\textwidth]{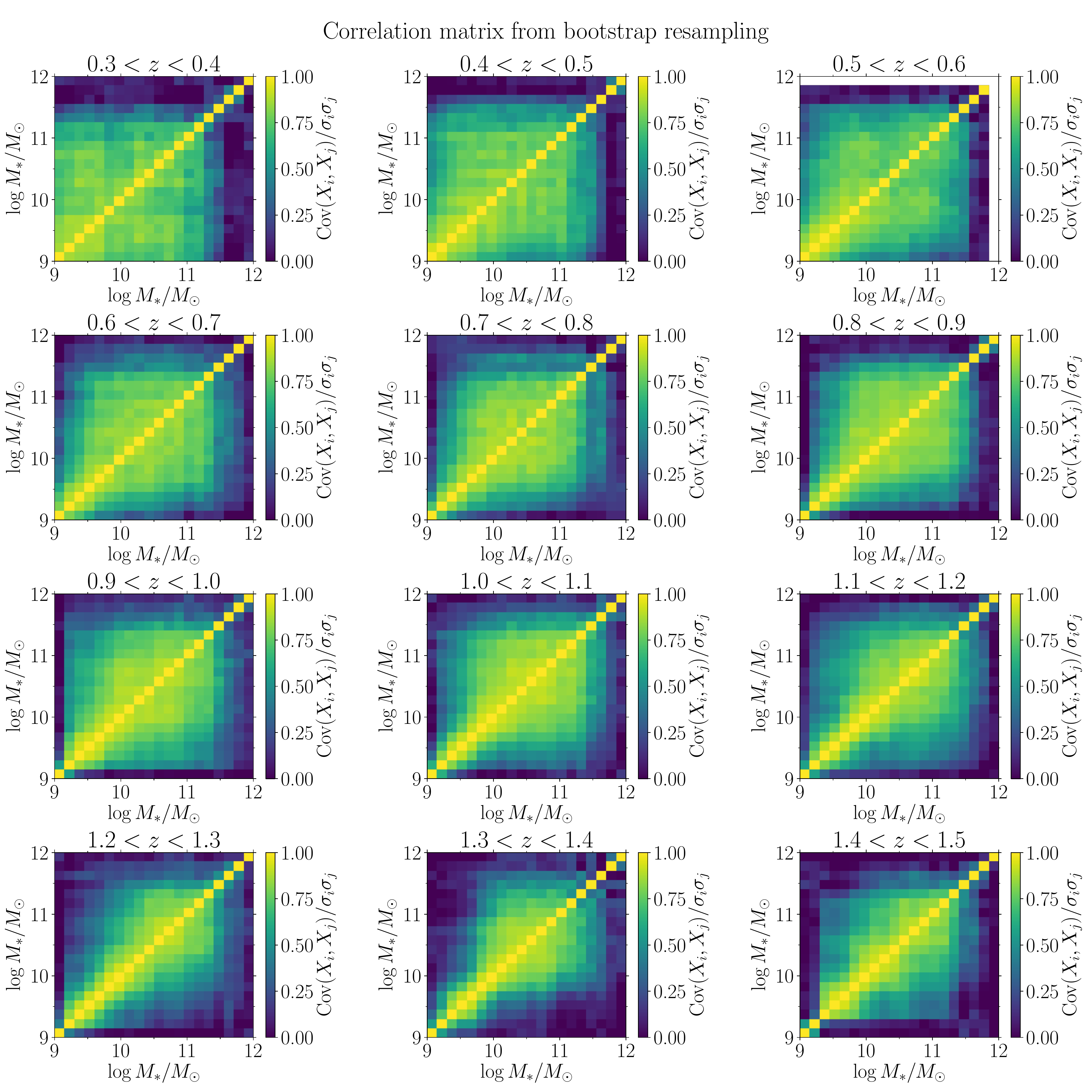}
\caption{Correlation matrices from the normalized covariance of the SHELA stellar mass functions as determined from gridding the survey footprint into 150 subregions and resampling with replacement. Colour indicates the strength of correlation between bins, according to the scale shown on the right.}
\label{fig:corrmatrices}
\end{figure}

\begin{figure}
\centering
\includegraphics[width=1.0\textwidth]{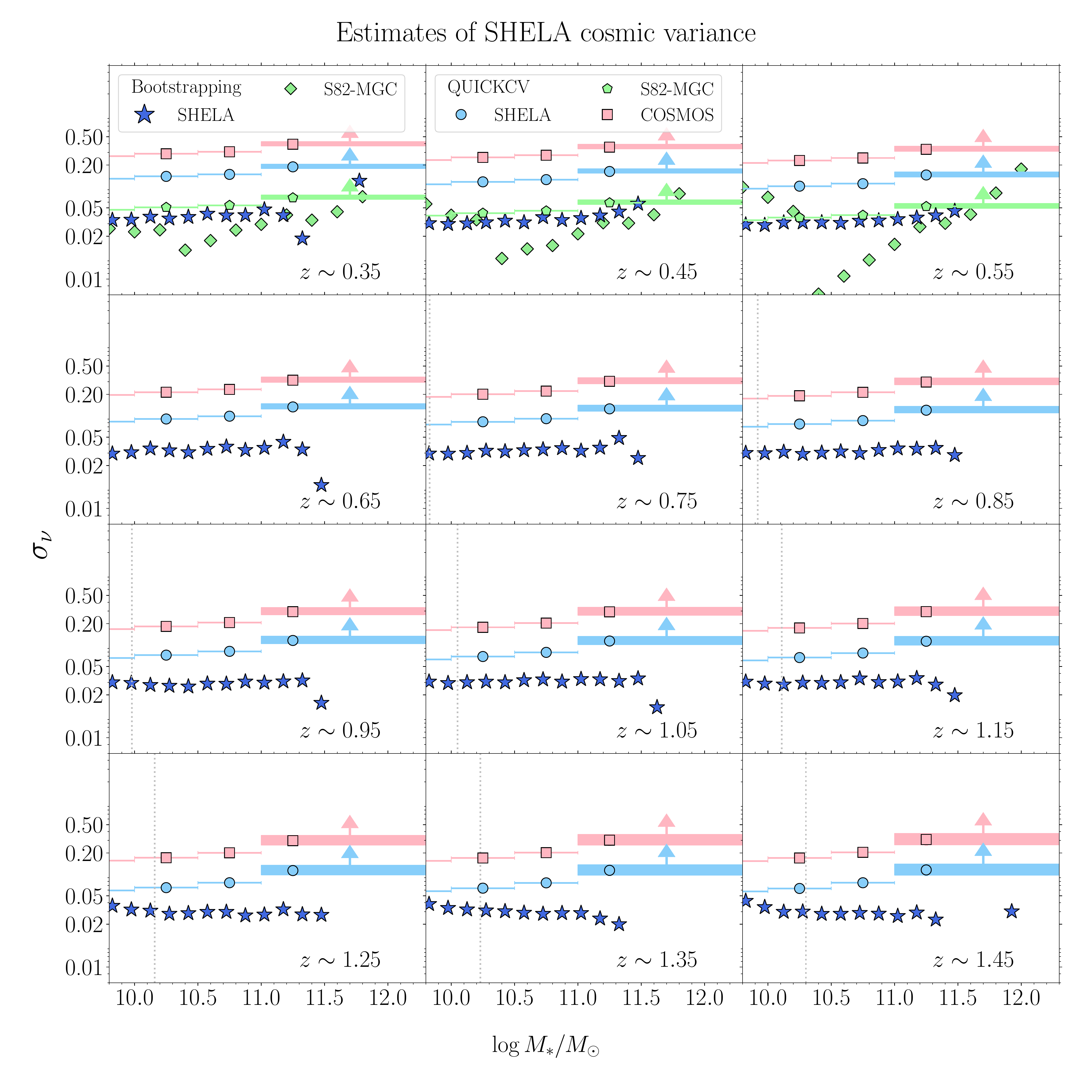}
\caption{Estimates of \edittwo{the relative cosmic variance ($\sigma_{v}$)} in our SHELA sample over 17.5 $\mathrm{deg}^2$ (blue stars) \edittwo{derived from bootstrap resampling} and taking the diagonal elements of the covariance matrix. For comparison, we show the cosmic variance of S82-MGC \edittwo{derived from bootstrap resampling} \citep[$140~\mathrm{deg}^2$]{Bundy2017} for galaxies at $0.3 < z < 0.65$ (green diamonds). \edittwo{We use the code QUICKCV \citep[$\sigma_{v,\mathrm{QUICKCV}}$;][]{Moster2011,Newman2014} to estimate the cosmic variance in SHELA (pink squares), S82-MGC \citep{Bundy2017} (green pentagons) and COSMOS (light blue circles, $1.4\times1.4~\mathrm{deg}^2$) samples in bin of $\mathrm{d}\log M_{\ast}/M_{\odot}=0.5$. The lower limits of $\sigma_{v,\mathrm{QUICKCV}}$ for SHELA, S82-MGC, and COSMOS for  galaxies more massive than $10^{11}~M_{\odot}$ are shown as the thick blue, green, and pink horizontal lines, respectively}. The estimated stellar mass completeness at a given redshift bin is indicated by the vertical dotted line in each panel.}

\label{fig:sigmacv}
\end{figure}

\section{Estimate of Cosmic Variance in SHELA}
\label{sec:appendix}
\par In Section~\ref{sec:cvestimate} we discuss a method based on \citet{Bundy2017} to estimate the cosmic variance on our SMFs.   As discussed in that section, we divide the footprint of SHELA survey into 150 sub-fields and measure the variance in the SMF. Figure~\ref{fig:corrmatrices} shows the correlation matrices between the SMFs derived in the sub-fields (resampling these sub-fields with replacement) as a function of stellar mass. \editone{The correlation matrix of the galaxy SMF can be decomposed into three terms: a diagonal term arising from Poisson noise (due to the finite number of galaxies in each bin); a large-scale structure term arising from clustering in the Universe on scales comparable to, and larger than, the survey volume; and an occupancy covariance term arising from the fact that galaxies of different stellar masses (luminosities) inhabit the same groups or clusters \citep{Smith2012,Benson2014}.  The correlation matrix of the galaxy SMF for our SHELA survey, which covers $\sim0.15~\mathrm{Gpc}^3$ comoving volume between $0.4 < z < 1.5$, shows that galaxies with stellar masses lower than the characteristic stellar mass $M^{*}$ are highly correlated, i.e., have a correlation coefficient $r>0.8$. This means that if there is an upward fluctuation of one bin with respect to the mean, then all other bins share the same upward fluctuation \citep{Smith2012}. Additionally, \citeauthor{Smith2012} demonstrated that for the case of a volume limited 2dFGRS-like survey with size $V=0.40~\mathrm{Gpc}^{3}$, the off-diagonal elements of the correlation matrix are entirely dominated by the cosmic variance term.  }

\edittwo{From the resampled SMFs, the diagonal elements of the covariance matrix  ($\sigma_{\mathrm{tot}}^2$; Figure~\ref{fig:corrmatrices})
would be the sum of the Poisson and cosmic variance uncertainties, $\sigma_{\mathrm{tot}}^2=\sigma_{\mathrm{Poisson}}^2 + \sigma_{\mathrm{CV}}^2$. We then first subtract the Poisson uncertainties from 
the diagonal elements of the covariance matrix. Finally, we derive the relative cosmic variance ($\sigma_{v}(\log M_{\ast},z)$) as the square-root of Poisson uncertainties-subtracted diagonal elements ($\sigma_{\mathrm{CV}}$) divided by the galaxy number density (SMF) at a given stellar mass bin ($\mathrm{d}N/\mathrm{d}\log M_{\ast}$).  For low mass galaxies  ($\log (M_{\ast}/M_{\odot})<11.5$), bootstrap resampling yields $\sigma_{v}$ of $2\%-5\%$ (corresponding to 1$\sigma$ error of 0.01-0.02 dex) at $0.3 < z <1.5$. At higher mass ($\log (M_{\ast}/M_{\odot})>11.5$), $\sigma_{v}$ rises to 6\%-12\% ($0.03-0.05$ dex).} These are illustrated in Figure~\ref{fig:sigmacv}.  In the redshift range that overlaps with 140 deg$^2$ S82-MGC survey \citep{Bundy2017}, we compare the results.  The cosmic variance in SHELA is roughly a factor of $\sim$2 larger than that measured by \citeauthor{Bundy2017}, which is to be expected assuming $\sigma_{v} \sim V^{-\gamma/0.3}$ for $\gamma=1.8$ within a volume $V$ \citep{Somerville2004}.


\par \edittwo{However, for massive galaxies with $\log (M_{\ast}/M_{\odot})>11.5$, the diagonal elements of the covariance matrix (Figure~\ref{fig:corrmatrices}) are largely dominated by the Poisson uncertainties. As a result, we cannot estimate the cosmic variance for the SHELA sample in the highest stellar mass bins, where $\sigma_{\mathrm{Poisson}} > \sigma_{\mathrm{CV}}$ using the bootstrap resampling method.
To get the estimate of $\sigma_{v}$ for these galaxies ($M_{\ast}/M_{\odot})>11.5$), we follow a method presented by \cite{Moster2011}. We use the code QUICKCV \citep{Newman2014} to compute the cosmic variance of dark matter ($\sigma_{\mathrm{dm}}(\bar{z})$) as a function of mean redshift for a given survey geometry. We then use the galaxy bias ($b(M_{\ast},\bar{z})$) predicted by \cite{Moster2010} for bin size of $\mathrm{d}\log M_{\ast}/M_{\odot}=0.5$ and $\log M_{\ast}/M_{\odot} > 11$. For example, the predicted bias for galaxies with $\log M_{\ast}/M_{\odot}=10-11.5$ at $z=0.35$ is $b=1.4-1.8$ with the uncertainty of $\sim0.1$ and $b=2-3.7$ with the uncertainty of $0.3-0.6$ at $z=1.45$. In the linear regime, the cosmic variance of the galaxy sample is the product of the galaxy bias and the dark matter cosmic variance, $\sigma_{v,\mathrm{QUICKCV}}=b(M_{\ast},\bar{z})\sigma_{\mathrm{dm}}(\bar{z})$. As expected, $\sigma_{v,\mathrm{QUICKCV}}$ increases with stellar mass and redshift due to the increasing bias with increasing stellar mass; massive galaxies are biased more strongly than galaxies at lower mass. The $\sigma_{v,\mathrm{QUICKCV}}$ is higher than that estimated using bootstrap resampling method
%
%
by a factor of 2-3.
%
%
This likely arises
%
%
from the assumed galaxy bias, with $b\sim2-3$. We therefore adopt the bootstrap method, which makes no assumptions about galaxy bias (and only depends on galaxy mass) to estimate $\sigma_{v}$ in our SHELA field. }



\par \edittwo{We also apply this method to estimate the relative cosmic variance in the S82-MGC and COSMOS survey. Figure~\ref{fig:sigmacv} shows that for massive galaxies ($\log M_{\ast}/M_{\odot}>11.0$), the cosmic variance in our SHELA sample is lower than that in COSMOS sample by a factor of 2 at $z=0.35$. On the other hand, over the same stellar mass and redshift, the cosmic variance in SHELA is roughly a factor of $\sim3$ larger than that that of the S82-MGC sample.} 


\par \editone{Finally, we note that we do not include the cosmic variance uncertainties in our forward modeling method because each galaxy in the SHELA survey  should be equally affected by the same large-scale fluctuation. As a result, the measurement of the galaxy SMF is mainly affected by the random errors in the stellar mass estimates rather than the cosmic variance. }
 
\section{Assumption-averaged stellar mass functions for quiescent and star forming galaxies}
\label{sec:appendix2}
We present our measurements of assumption-averaged stellar mass function for quiescent and star forming galaxies (Figure~\ref{fig:massfuncsqui} and Figure~\ref{fig:massfuncssf}) in Table~\ref{table:smfavgqui} and Table~\ref{table:smfavgsf}, respectively.
\begin{splitdeluxetable*}{lcccccBcccccc}
\tablecaption{Assumption-Averaged Stellar Mass Functions For Quiescent Galaxies\label{table:smfavgqui}}
\tablehead{
& \colhead{$0.4 < z < 0.5$} &  \colhead{$0.5 < z < 0.6$} & \colhead{$0.6 < z < 0.7$} & \colhead{$0.7 < z < 0.8$} & \colhead{$0.8 < z < 0.9$} &  \colhead{$0.9 < z < 1.0$} & \colhead{$1.0 < z < 1.1$} & \colhead{$1.1 < z < 1.2$} & \colhead{$1.2 < z < 1.3$} & \colhead{$1.3 < z < 1.4$} & \colhead{$1.4 < z < 1.5$}\\
\colhead{ $\log(M_{\ast}/M_{\odot})$} & \colhead{$\log(\phi/\mathrm{Mpc}^{-3}\mathrm{dex}^{-1}$)} &  \colhead{$\log(\phi/\mathrm{Mpc}^{-3}\mathrm{dex}^{-1}$)} & 
\colhead{$\log(\phi/\mathrm{Mpc}^{-3}\mathrm{dex}^{-1}$)} &
\colhead{$\log(\phi/\mathrm{Mpc}^{-3}\mathrm{dex}^{-1}$)} & 
\colhead{$\log(\phi/\mathrm{Mpc}^{-3}\mathrm{dex}^{-1}$)} &
\colhead{$\log(\phi/\mathrm{Mpc}^{-3}\mathrm{dex}^{-1}$)} &
\colhead{$\log(\phi/\mathrm{Mpc}^{-3}\mathrm{dex}^{-1}$)} &
\colhead{$\log(\phi/\mathrm{Mpc}^{-3}\mathrm{dex}^{-1}$)} &
\colhead{$\log(\phi/\mathrm{Mpc}^{-3}\mathrm{dex}^{-1}$)} &
\colhead{$\log(\phi/\mathrm{Mpc}^{-3}\mathrm{dex}^{-1}$)} &
\colhead{$\log(\phi/\mathrm{Mpc}^{-3}\mathrm{dex}^{-1}$)} \\
}  

\startdata
9.53 & $-3.53\pm0.03$ & ... & ... & ... & ... & ... & ... & ... & ... & ... & ... \\
9.68 & $-3.35\pm0.02$ & $-3.76\pm0.03$ & ... & ... & ... & ... & ... & ... & ... & ... & ... \\
9.83 & $-3.23\pm0.02$ & $-3.51\pm0.02$ & $-3.84\pm0.03$ & ... & ... & ... & ... & ... & ... & ... & ... \\
9.98 & $-3.10\pm0.02$ & $-3.34\pm0.02$ & $-3.56\pm0.02$ & $-3.67\pm0.02$ & $-3.76\pm0.02$ & ... & ... & ... & ... & ... & ... \\
10.13 & $-3.08\pm0.02$ & $-3.18\pm0.02$ & $-3.33\pm0.02$ & $-3.39\pm0.02$ & $-3.45\pm0.02$ & $-3.57\pm0.02$ & $-3.77\pm0.02$ & $-3.86\pm0.02$ & ... & ... & ... \\
10.28 & $-3.09\pm0.02$ & $-3.11\pm0.01$ & $-3.18\pm0.01$ & $-3.20\pm0.01$ & $-3.23\pm0.01$ & $-3.28\pm0.01$ & $-3.38\pm0.01$ & $-3.43\pm0.01$ & $-3.47\pm0.01$ & $-3.49\pm0.01$ & ... \\
10.43 & $-3.01\pm0.01$ & $-3.07\pm0.01$ & $-3.07\pm0.01$ & $-3.07\pm0.01$ & $-3.09\pm0.01$ & $-3.10\pm0.01$ & $-3.14\pm0.01$ & $-3.19\pm0.01$ & $-3.21\pm0.01$ & $-3.19\pm0.01$ & $-3.27\pm 0.01$ \\
10.58 & $-2.96\pm0.01$ & $-3.01\pm0.01$ & $-3.00\pm0.01$ & $-2.96\pm0.01$ & $-3.02\pm0.01$ & $-2.97\pm0.01$ & $-3.01\pm0.01$ & $-3.13\pm0.01$ & $-3.14\pm0.01$ & $-3.13\pm0.01$ & $-3.12\pm 0.01$ \\
10.73 & $-2.97\pm0.01$ & $-2.97\pm0.01$ & $-2.92\pm0.01$ & $-2.92\pm0.01$ & $-2.95\pm0.01$ & $-2.89\pm0.01$ & $-2.91\pm0.01$ & $-3.05\pm0.01$ & $-3.14\pm0.01$ & $-3.15\pm0.01$ & $-3.13\pm 0.01$ \\
10.88 & $-2.99\pm0.01$ & $-2.96\pm0.01$ & $-2.96\pm0.01$ & $-2.93\pm0.01$ & $-2.94\pm0.01$ & $-2.85\pm0.01$ & $-2.85\pm0.01$ & $-2.99\pm0.01$ & $-3.12\pm0.01$ & $-3.15\pm0.01$ & $-3.14\pm 0.01$ \\
11.03 & $-3.08\pm0.02$ & $-3.06\pm0.01$ & $-3.05\pm0.01$ & $-3.03\pm0.01$ & $-3.00\pm0.01$ & $-2.91\pm0.01$ & $-2.93\pm0.01$ & $-3.07\pm0.01$ & $-3.17\pm0.01$ & $-3.19\pm0.01$ & $-3.17\pm 0.01$ \\
11.18 & $-3.26\pm0.02$ & $-3.21\pm0.02$ & $-3.26\pm0.02$ & $-3.23\pm0.01$ & $-3.19\pm0.01$ & $-3.09\pm0.01$ & $-3.09\pm0.01$ & $-3.27\pm0.01$ & $-3.36\pm0.01$ & $-3.35\pm0.01$ & $-3.33\pm 0.01$ \\
11.33 & $-3.56\pm0.03$ & $-3.40\pm0.02$ & $-3.53\pm0.02$ & $-3.53\pm0.02$ & $-3.44\pm0.02$ & $-3.37\pm0.01$ & $-3.35\pm0.01$ & $-3.52\pm0.02$ & $-3.70\pm0.02$ & $-3.65\pm0.02$ & $-3.63\pm 0.02$ \\
11.48 & $-4.01\pm0.04$ & $-3.83\pm0.03$ & $-3.91\pm0.03$ & $-3.96\pm0.03$ & $-3.87\pm0.03$ & $-3.76\pm0.02$ & $-3.79\pm0.02$ & $-3.91\pm0.02$ & $-4.05\pm0.03$ & $-4.11\pm0.03$ & $-4.13\pm 0.03$ \\
11.63 & $-4.69\pm0.09$ & $-4.42\pm0.06$ & $-4.53\pm0.06$ & $-4.40\pm0.05$ & $-4.36\pm0.04$ & $-4.36\pm0.04$ & $-4.36\pm0.04$ & $-4.52\pm0.05$ & $-4.63\pm0.05$ & $-4.64\pm0.05$ & $-4.73\pm 0.06$ \\
11.78 & $-5.34\pm0.18$ & $-5.11\pm0.12$ & $-5.06\pm0.11$ & $-5.28\pm0.12$ & $-5.14\pm0.10$ & $-5.07\pm0.09$ & $-5.22\pm0.10$ & $-5.17\pm0.09$ & $-5.40\pm0.12$ & $-5.29\pm0.10$ & $-5.29\pm 0.10$ \\
11.93 & $-6.07\pm0.33$ & ... & $-5.98\pm0.26$ & $-5.93\pm0.23$ & $-5.99\pm0.23$ & $-6.16\pm0.26$ & $-6.28\pm0.28$ & $-6.31\pm0.28$ & $-6.26\pm0.26$ & $-5.91\pm0.19$ & $-6.23\pm 0.24$ \\
12.08 & $-6.54\pm0.48$ & $-6.07\pm0.30$ & $-6.76\pm0.48$ & $-6.14\pm0.28$ & $-6.59\pm0.38$ & ... & ... & ... & ... & $-6.58\pm0.33$ & $-6.77\pm 0.38$ \\
\enddata
\end{splitdeluxetable*}


\begin{splitdeluxetable*}{lcccccBcccccc}
\tablecaption{Assumption-Averaged Stellar Mass Functions For Star Forming Galaxies\label{table:smfavgsf}}
\tablehead{
& \colhead{$0.4 < z < 0.5$} &  \colhead{$0.5 < z < 0.6$} & \colhead{$0.6 < z < 0.7$} & \colhead{$0.7 < z < 0.8$} & \colhead{$0.8 < z < 0.9$} &  \colhead{$0.9 < z < 1.0$} & \colhead{$1.0 < z < 1.1$} & \colhead{$1.1 < z < 1.2$} & \colhead{$1.2 < z < 1.3$} & \colhead{$1.3 < z < 1.4$} & \colhead{$1.4 < z < 1.5$}\\
\colhead{ $\log(M_{\ast}/M_{\odot})$} & \colhead{$\log(\phi/\mathrm{Mpc}^{-3}\mathrm{dex}^{-1}$)} &  \colhead{$\log(\phi/\mathrm{Mpc}^{-3}\mathrm{dex}^{-1}$)} & 
\colhead{$\log(\phi/\mathrm{Mpc}^{-3}\mathrm{dex}^{-1}$)} &
\colhead{$\log(\phi/\mathrm{Mpc}^{-3}\mathrm{dex}^{-1}$)} & 
\colhead{$\log(\phi/\mathrm{Mpc}^{-3}\mathrm{dex}^{-1}$)} &
\colhead{$\log(\phi/\mathrm{Mpc}^{-3}\mathrm{dex}^{-1}$)} &
\colhead{$\log(\phi/\mathrm{Mpc}^{-3}\mathrm{dex}^{-1}$)} &
\colhead{$\log(\phi/\mathrm{Mpc}^{-3}\mathrm{dex}^{-1}$)} &
\colhead{$\log(\phi/\mathrm{Mpc}^{-3}\mathrm{dex}^{-1}$)} &
\colhead{$\log(\phi/\mathrm{Mpc}^{-3}\mathrm{dex}^{-1}$)} &
\colhead{$\log(\phi/\mathrm{Mpc}^{-3}\mathrm{dex}^{-1}$)} \\
}  

\startdata
9.53 & $-2.24\pm0.01$ & ... & ... & ... & ... & ... & ... & ... & ... & ... & ... \\
9.68 & $-2.31\pm0.01$ & $-2.32\pm0.01$ & ... & ... & ... & ... & ... & ... & ... & ... & ... \\
9.83 & $-2.38\pm0.01$ & $-2.40\pm0.01$ & $-2.51\pm0.01$ & ... & ... & ... & ... & ... & ... & ... & ... \\
9.98 & $-2.47\pm0.01$ & $-2.49\pm0.01$ & $-2.57\pm0.01$ & $-2.62\pm0.01$ & $-2.57\pm0.01$ & ... & ... & ... & ... & ... & ... \\
10.13 & $-2.57\pm0.01$ & $-2.59\pm0.01$ & $-2.64\pm0.01$ & $-2.69\pm0.01$ & $-2.63\pm0.01$ & $-2.56\pm0.01$ & $-2.55\pm0.01$ & $-2.59\pm0.01$ & ... & ... & ... \\
10.28 & $-2.63\pm0.01$ & $-2.72\pm0.01$ & $-2.71\pm0.01$ & $-2.76\pm0.01$ & $-2.70\pm0.01$ & $-2.64\pm0.01$ & $-2.64\pm0.01$ & $-2.70\pm0.01$ & $-2.74\pm0.01$ & $-2.78\pm0.01$ & ... \\
10.43 & $-2.69\pm0.01$ & $-2.81\pm0.01$ & $-2.77\pm0.01$ & $-2.81\pm0.01$ & $-2.78\pm0.01$ & $-2.74\pm0.01$ & $-2.77\pm0.01$ & $-2.87\pm0.01$ & $-2.92\pm0.01$ & $-2.93\pm0.01$ & $-2.88\pm 0.01$ \\
10.58 & $-2.73\pm0.01$ & $-2.88\pm0.01$ & $-2.82\pm0.01$ & $-2.87\pm0.01$ & $-2.87\pm0.01$ & $-2.83\pm0.01$ & $-2.91\pm0.01$ & $-3.02\pm0.01$ & $-3.10\pm0.01$ & $-3.13\pm0.01$ & $-3.10\pm 0.01$ \\
10.73 & $-2.81\pm0.01$ & $-2.93\pm0.01$ & $-2.89\pm0.01$ & $-2.96\pm0.01$ & $-2.95\pm0.01$ & $-2.93\pm0.01$ & $-2.99\pm0.01$ & $-3.10\pm0.01$ & $-3.22\pm0.01$ & $-3.28\pm0.01$ & $-3.28\pm 0.01$ \\
10.88 & $-2.94\pm0.01$ & $-3.01\pm0.01$ & $-3.01\pm0.01$ & $-3.06\pm0.01$ & $-3.09\pm0.01$ & $-3.09\pm0.01$ & $-3.14\pm0.01$ & $-3.22\pm0.01$ & $-3.34\pm0.01$ & $-3.45\pm0.01$ & $-3.42\pm 0.01$ \\
11.03 & $-3.08\pm0.02$ & $-3.17\pm0.02$ & $-3.19\pm0.01$ & $-3.22\pm0.01$ & $-3.25\pm0.01$ & $-3.29\pm0.01$ & $-3.39\pm0.01$ & $-3.43\pm0.01$ & $-3.54\pm0.02$ & $-3.66\pm0.02$ & $-3.65\pm 0.02$ \\
11.18 & $-3.31\pm0.02$ & $-3.39\pm0.02$ & $-3.43\pm0.02$ & $-3.46\pm0.02$ & $-3.51\pm0.02$ & $-3.59\pm0.02$ & $-3.67\pm0.02$ & $-3.73\pm0.02$ & $-3.80\pm0.02$ & $-3.93\pm0.02$ & $-3.93\pm 0.02$ \\
11.33 & $-3.69\pm0.03$ & $-3.77\pm0.03$ & $-3.82\pm0.03$ & $-3.81\pm0.03$ & $-3.88\pm0.03$ & $-3.96\pm0.03$ & $-4.08\pm0.03$ & $-4.04\pm0.03$ & $-4.15\pm0.03$ & $-4.30\pm0.03$ & $-4.33\pm 0.04$ \\
11.48 & $-4.21\pm0.06$ & $-4.43\pm0.06$ & $-4.32\pm0.05$ & $-4.39\pm0.05$ & $-4.38\pm0.05$ & $-4.54\pm0.05$ & $-4.61\pm0.05$ & $-4.58\pm0.05$ & $-4.57\pm0.05$ & $-4.68\pm0.05$ & $-4.63\pm 0.05$ \\
11.63 & $-4.78\pm0.10$ & $-5.27\pm0.15$ & $-5.02\pm0.10$ & $-4.92\pm0.09$ & $-5.06\pm0.09$ & $-5.36\pm0.12$ & $-5.28\pm0.11$ & $-5.20\pm0.10$ & $-5.09\pm0.08$ & $-5.18\pm0.09$ & $-5.07\pm 0.08$ \\
& $-5.20\pm0.10$ & $-5.09\pm0.08$ & $-5.18\pm0.09$ & $-5.07\pm 0.08$ \\
11.78 & $-5.40\pm0.19$ & $-6.19\pm0.33$ & $-5.86\pm0.23$ & $-5.76\pm0.20$ & $-5.64\pm0.17$ & $-5.86\pm0.20$ & $-6.08\pm0.23$ & $-6.11\pm0.23$ & $-5.59\pm0.14$ & $-5.51\pm0.13$ & $-5.87\pm 0.18$ \\
11.93 & $-5.76\pm0.26$ & ... & $-6.76\pm0.48$ & $-6.54\pm0.38$ & $-6.90\pm0.48$ & ... & $-6.98\pm0.48$ & $-6.23\pm0.26$ & $-6.13\pm0.23$ & $-6.10\pm0.22$ & $-5.96\pm 0.19$ \\
12.08 & $-6.07\pm0.33$ & ... & $-6.46\pm0.38$ & ... & $-6.42\pm0.33$ & $-6.16\pm0.26$ & ... & $-7.01\pm0.48$ & ... & $-7.06\pm0.48$ & $-6.38\pm 0.28$ \\
\enddata
\end{splitdeluxetable*}

\bibliography{references}

\begin{thebibliography}{}
\expandafter\ifx\csname natexlab\endcsname\relax\def\natexlab#1{#1}\fi
\providecommand{\url}[1]{\href{#1}{#1}}
\providecommand{\dodoi}[1]{doi:~\href{http://doi.org/#1}{\nolinkurl{#1}}}
\providecommand{\doeprint}[1]{\href{http://ascl.net/#1}{\nolinkurl{http://ascl.net/#1}}}
\providecommand{\doarXiv}[1]{\href{https://arxiv.org/abs/#1}{\nolinkurl{https://arxiv.org/abs/#1}}}

\bibitem[{{Albareti} {et~al.}(2017){Albareti}, {Allende Prieto}, {Almeida},
  {Anders}, {Anderson}, {Andrews}, {Arag{\'o}n-Salamanca},
  {Argudo-Fern{\'a}ndez}, {Armengaud}, {Aubourg}, \& et~al.}]{Albareti2017}
{Albareti}, F.~D., {Allende Prieto}, C., {Almeida}, A., {et~al.} 2017, \apjs,
  233, 25, \dodoi{10.3847/1538-4365/aa8992}

\bibitem[{{Arcila-Osejo} {et~al.}(2019){Arcila-Osejo}, {Sawicki}, {Arnouts},
  {Golob}, {Moutard}, \& {Sorba}}]{Arcila-Osejo2019}
{Arcila-Osejo}, L., {Sawicki}, M., {Arnouts}, S., {et~al.} 2019, \mnras, 486,
  4880, \dodoi{10.1093/mnras/stz1169}

\bibitem[{{Asplund} {et~al.}(2009){Asplund}, {Grevesse}, {Sauval}, \&
  {Scott}}]{Asplund2009}
{Asplund}, M., {Grevesse}, N., {Sauval}, A.~J., \& {Scott}, P. 2009, \araa, 47,
  481, \dodoi{10.1146/annurev.astro.46.060407.145222}

\bibitem[{{Astropy Collaboration} {et~al.}(2013){Astropy Collaboration},
  {Robitaille}, {Tollerud}, {Greenfield}, {Droettboom}, {Bray}, {Aldcroft},
  {Davis}, {Ginsburg}, {Price-Whelan}, {Kerzendorf}, {Conley}, {Crighton},
  {Barbary}, {Muna}, {Ferguson}, {Grollier}, {Parikh}, {Nair}, {Unther},
  {Deil}, {Woillez}, {Conseil}, {Kramer}, {Turner}, {Singer}, {Fox}, {Weaver},
  {Zabalza}, {Edwards}, {Azalee Bostroem}, {Burke}, {Casey}, {Crawford},
  {Dencheva}, {Ely}, {Jenness}, {Labrie}, {Lim}, {Pierfederici}, {Pontzen},
  {Ptak}, {Refsdal}, {Servillat}, \& {Streicher}}]{astropy:2013}
{Astropy Collaboration}, {Robitaille}, T.~P., {Tollerud}, E.~J., {et~al.} 2013,
  \aap, 558, A33, \dodoi{10.1051/0004-6361/201322068}

\bibitem[{{Baldry} {et~al.}(2004){Baldry}, {Glazebrook}, {Brinkmann},
  {Ivezi{\'c}}, {Lupton}, {Nichol}, \& {Szalay}}]{Baldry2004}
{Baldry}, I.~K., {Glazebrook}, K., {Brinkmann}, J., {et~al.} 2004, \apj, 600,
  681, \dodoi{10.1086/380092}

\bibitem[{{Baldry} {et~al.}(2008){Baldry}, {Glazebrook}, \&
  {Driver}}]{Baldry2008}
{Baldry}, I.~K., {Glazebrook}, K., \& {Driver}, S.~P. 2008, \mnras, 388, 945,
  \dodoi{10.1111/j.1365-2966.2008.13348.x}

\bibitem[{{Baldry} {et~al.}(2012){Baldry}, {Driver}, {Loveday}, {Taylor},
  {Kelvin}, {Liske}, {Norberg}, {Robotham}, {Brough}, {Hopkins}, {Bamford},
  {Peacock}, {Bland-Hawthorn}, {Conselice}, {Croom}, {Jones}, {Parkinson},
  {Popescu}, {Prescott}, {Sharp}, \& {Tuffs}}]{Baldry2012}
{Baldry}, I.~K., {Driver}, S.~P., {Loveday}, J., {et~al.} 2012, \mnras, 421,
  621, \dodoi{10.1111/j.1365-2966.2012.20340.x}

\bibitem[{{Beare} {et~al.}(2019){Beare}, {Brown}, {Pimbblet}, \&
  {Taylor}}]{Beare2019}
{Beare}, R., {Brown}, M. J.~I., {Pimbblet}, K., \& {Taylor}, E.~N. 2019, \apj,
  873, 78, \dodoi{10.3847/1538-4357/ab041a}

\bibitem[{{B{\'e}dorf} \& {Portegies Zwart}(2013)}]{Bedorf2013}
{B{\'e}dorf}, J., \& {Portegies Zwart}, S. 2013, \mnras, 431, 767,
  \dodoi{10.1093/mnras/stt208}

\bibitem[{{Behroozi} {et~al.}(2013{\natexlab{a}}){Behroozi}, {Marchesini},
  {Wechsler}, {Muzzin}, {Papovich}, \& {Stefanon}}]{Behroozi2013nov}
{Behroozi}, P.~S., {Marchesini}, D., {Wechsler}, R.~H., {et~al.}
  2013{\natexlab{a}}, \apjl, 777, L10, \dodoi{10.1088/2041-8205/777/1/L10}

\bibitem[{{Behroozi} {et~al.}(2013{\natexlab{b}}){Behroozi}, {Wechsler}, \&
  {Conroy}}]{Behroozi2013}
{Behroozi}, P.~S., {Wechsler}, R.~H., \& {Conroy}, C. 2013{\natexlab{b}}, \apj,
  770, 57, \dodoi{10.1088/0004-637X/770/1/57}

\bibitem[{{Behroozi} {et~al.}(2013{\natexlab{c}}){Behroozi}, {Wechsler}, \&
  {Conroy}}]{Behroozi2013jan}
---. 2013{\natexlab{c}}, \apjl, 762, L31, \dodoi{10.1088/2041-8205/762/2/L31}

\bibitem[{{Belli} {et~al.}(2019){Belli}, {Newman}, \& {Ellis}}]{Belli2019}
{Belli}, S., {Newman}, A.~B., \& {Ellis}, R.~S. 2019, \apj, 874, 17,
  \dodoi{10.3847/1538-4357/ab07af}

\bibitem[{{Benson}(2014)}]{Benson2014}
{Benson}, A.~J. 2014, \mnras, 444, 2599, \dodoi{10.1093/mnras/stu1630}

\bibitem[{{Bernardi} {et~al.}(2016){Bernardi}, {Meert}, {Sheth},
  {Huertas-Company}, {Maraston}, {Shankar}, \& {Vikram}}]{Bernardi2016}
{Bernardi}, M., {Meert}, A., {Sheth}, R.~K., {et~al.} 2016, \mnras, 455, 4122,
  \dodoi{10.1093/mnras/stv2487}

\bibitem[{{Blumenthal} {et~al.}(1984){Blumenthal}, {Faber}, {Primack}, \&
  {Rees}}]{Blumenthal1984}
{Blumenthal}, G.~R., {Faber}, S.~M., {Primack}, J.~R., \& {Rees}, M.~J. 1984,
  \nat, 311, 517, \dodoi{10.1038/311517a0}

\bibitem[{{Brammer} {et~al.}(2008){Brammer}, {van Dokkum}, \&
  {Coppi}}]{Brammer2008}
{Brammer}, G.~B., {van Dokkum}, P.~G., \& {Coppi}, P. 2008, \apj, 686, 1503,
  \dodoi{10.1086/591786}

\bibitem[{{Brammer} {et~al.}(2011){Brammer}, {Whitaker}, {van Dokkum},
  {Marchesini}, {Franx}, {Kriek}, {Labb{\'e}}, {Lee}, {Muzzin}, {Quadri},
  {Rudnick}, \& {Williams}}]{Brammer2011}
{Brammer}, G.~B., {Whitaker}, K.~E., {van Dokkum}, P.~G., {et~al.} 2011, \apj,
  739, 24, \dodoi{10.1088/0004-637X/739/1/24}

\bibitem[{{Bruzual} \& {Charlot}(2003)}]{Bruzual2003}
{Bruzual}, G., \& {Charlot}, S. 2003, \mnras, 344, 1000,
  \dodoi{10.1046/j.1365-8711.2003.06897.x}

\bibitem[{{Bundy} {et~al.}(2017){Bundy}, {Leauthaud}, {Saito}, {Maraston},
  {Wake}, \& {Thomas}}]{Bundy2017}
{Bundy}, K., {Leauthaud}, A., {Saito}, S., {et~al.} 2017, \apj, 851, 34,
  \dodoi{10.3847/1538-4357/aa9896}

\bibitem[{{Capozzi} {et~al.}(2016){Capozzi}, {Maraston}, {Daddi}, {Renzini},
  {Strazzullo}, \& {Gobat}}]{Capozzi2016}
{Capozzi}, D., {Maraston}, C., {Daddi}, E., {et~al.} 2016, \mnras, 456, 790,
  \dodoi{10.1093/mnras/stv2692}

\bibitem[{{Capozzi} {et~al.}(2017){Capozzi}, {Etherington}, {Thomas},
  {Maraston}, {Rykoff}, {Sevilla-Noarbe}, {Bechtol}, {Carrasco Kind},
  {Drlica-Wagner}, \& {Pforr}}]{Capozzi2017}
{Capozzi}, D., {Etherington}, J., {Thomas}, D., {et~al.} 2017, arXiv e-prints,
  arXiv:1707.09066.
\newblock \doarXiv{1707.09066}

\bibitem[{{Caputi} {et~al.}(2015){Caputi}, {Ilbert}, {Laigle}, {McCracken}, {Le
  F{\`e}vre}, {Fynbo}, {Milvang-Jensen}, {Capak}, {Salvato}, \&
  {Taniguchi}}]{Caputi2015}
{Caputi}, K.~I., {Ilbert}, O., {Laigle}, C., {et~al.} 2015, \apj, 810, 73,
  \dodoi{10.1088/0004-637X/810/1/73}

\bibitem[{{Chabrier}(2003)}]{Chabrier2003}
{Chabrier}, G. 2003, \pasp, 115, 763, \dodoi{10.1086/376392}

\bibitem[{{Charlot} \& {Fall}(2000)}]{Charlot2000}
{Charlot}, S., \& {Fall}, S.~M. 2000, \apj, 539, 718, \dodoi{10.1086/309250}

\bibitem[{{Coil} {et~al.}(2011){Coil}, {Blanton}, {Burles}, {Cool},
  {Eisenstein}, {Moustakas}, {Wong}, {Zhu}, {Aird}, {Bernstein}, {Bolton}, \&
  {Hogg}}]{Coil2011}
{Coil}, A.~L., {Blanton}, M.~R., {Burles}, S.~M., {et~al.} 2011, \apj, 741, 8,
  \dodoi{10.1088/0004-637X/741/1/8}

\bibitem[{{Conroy} \& {Gunn}(2010{\natexlab{a}})}]{Conroy2010}
{Conroy}, C., \& {Gunn}, J.~E. 2010{\natexlab{a}}, \apj, 712, 833,
  \dodoi{10.1088/0004-637X/712/2/833}

\bibitem[{{Conroy} \& {Gunn}(2010{\natexlab{b}})}]{Conroy2010ascl}
---. 2010{\natexlab{b}}, {FSPS: Flexible Stellar Population Synthesis},
  Astrophysics Source Code Library.
\newblock \doeprint{1010.043}

\bibitem[{{Conroy} \& {Wechsler}(2009)}]{Conroy2009}
{Conroy}, C., \& {Wechsler}, R.~H. 2009, \apj, 696, 620,
  \dodoi{10.1088/0004-637X/696/1/620}

\bibitem[{{Conselice} {et~al.}(2007){Conselice}, {Bundy}, {Trujillo}, {Coil},
  {Eisenhardt}, {Ellis}, {Georgakakis}, {Huang}, {Lotz}, {Nandra}, {Newman},
  {Papovich}, {Weiner}, \& {Willmer}}]{Conselice2007}
{Conselice}, C.~J., {Bundy}, K., {Trujillo}, I., {et~al.} 2007, \mnras, 381,
  962, \dodoi{10.1111/j.1365-2966.2007.12316.x}

\bibitem[{{Cooper} {et~al.}(2013){Cooper}, {D'Souza}, {Kauffmann}, {Wang},
  {Boylan-Kolchin}, {Guo}, {Frenk}, \& {White}}]{Cooper2013}
{Cooper}, A.~P., {D'Souza}, R., {Kauffmann}, G., {et~al.} 2013, \mnras, 434,
  3348, \dodoi{10.1093/mnras/stt1245}

\bibitem[{{Davidzon} {et~al.}(2017){Davidzon}, {Ilbert}, {Laigle}, {Coupon},
  {McCracken}, {Delvecchio}, {Masters}, {Capak}, {Hsieh}, \& {Le
  F{\`e}vre}}]{Davidzon2017}
{Davidzon}, I., {Ilbert}, O., {Laigle}, C., {et~al.} 2017, \aap, 605, A70,
  \dodoi{10.1051/0004-6361/201730419}

\bibitem[{{Dekel} {et~al.}(2009){Dekel}, {Sari}, \& {Ceverino}}]{Dekel2009apj}
{Dekel}, A., {Sari}, R., \& {Ceverino}, D. 2009, \apj, 703, 785,
  \dodoi{10.1088/0004-637X/703/1/785}

\bibitem[{{Drory} {et~al.}(2009){Drory}, {Bundy}, {Leauthaud}, {Scoville},
  {Capak}, {Ilbert}, {Kartaltepe}, {Kneib}, {McCracken}, {Salvato}, {Sanders},
  {Thompson}, \& {Willott}}]{Drory2009}
{Drory}, N., {Bundy}, K., {Leauthaud}, A., {et~al.} 2009, \apj, 707, 1595,
  \dodoi{10.1088/0004-637X/707/2/1595}

\bibitem[{{D'Souza} {et~al.}(2015){D'Souza}, {Vegetti}, \&
  {Kauffmann}}]{Souza2015}
{D'Souza}, R., {Vegetti}, S., \& {Kauffmann}, G. 2015, \mnras, 454, 4027,
  \dodoi{10.1093/mnras/stv2234}

\bibitem[{{Eddington}(1913)}]{Eddington1913}
{Eddington}, A.~S. 1913, \mnras, 73, 359, \dodoi{10.1093/mnras/73.5.359}

\bibitem[{{Fontana} {et~al.}(2006){Fontana}, {Salimbeni}, {Grazian},
  {Giallongo}, {Pentericci}, {Nonino}, {Fontanot}, {Menci}, {Monaco},
  {Cristiani}, {Vanzella}, {de Santis}, \& {Gallozzi}}]{Fontana2006}
{Fontana}, A., {Salimbeni}, S., {Grazian}, A., {et~al.} 2006, \aap, 459, 745,
  \dodoi{10.1051/0004-6361:20065475}

\bibitem[{{Fontanot} {et~al.}(2009){Fontanot}, {De Lucia}, {Monaco},
  {Somerville}, \& {Santini}}]{Fontanot2009}
{Fontanot}, F., {De Lucia}, G., {Monaco}, P., {Somerville}, R.~S., \&
  {Santini}, P. 2009, \mnras, 397, 1776,
  \dodoi{10.1111/j.1365-2966.2009.15058.x}

\bibitem[{{Girardi} {et~al.}(2000){Girardi}, {Bressan}, {Bertelli}, \&
  {Chiosi}}]{Girardi2000}
{Girardi}, L., {Bressan}, A., {Bertelli}, G., \& {Chiosi}, C. 2000, \aaps, 141,
  371, \dodoi{10.1051/aas:2000126}

\bibitem[{{Grazian} {et~al.}(2015){Grazian}, {Fontana}, {Santini}, {Dunlop},
  {Ferguson}, {Castellano}, {Amorin}, {Ashby}, {Barro}, \&
  {Behroozi}}]{Grazian2015}
{Grazian}, A., {Fontana}, A., {Santini}, P., {et~al.} 2015, \aap, 575, A96,
  \dodoi{10.1051/0004-6361/201424750}

\bibitem[{{Hilz} {et~al.}(2013){Hilz}, {Naab}, \& {Ostriker}}]{Hilz2013}
{Hilz}, M., {Naab}, T., \& {Ostriker}, J.~P. 2013, \mnras, 429, 2924,
  \dodoi{10.1093/mnras/sts501}

\bibitem[{{Huang} {et~al.}(2018){Huang}, {Leauthaud}, {Greene}, {Bundy}, {Lin},
  {Tanaka}, {Miyazaki}, \& {Komiyama}}]{Huang2018}
{Huang}, S., {Leauthaud}, A., {Greene}, J.~E., {et~al.} 2018, \mnras, 475,
  3348, \dodoi{10.1093/mnras/stx3200}

\bibitem[{{Ilbert} {et~al.}(2013){Ilbert}, {McCracken}, {Le F{\`e}vre},
  {Capak}, {Dunlop}, {Karim}, {Renzini}, {Caputi}, {Boissier}, \&
  {Arnouts}}]{Ilbert2013}
{Ilbert}, O., {McCracken}, H.~J., {Le F{\`e}vre}, O., {et~al.} 2013, \aap, 556,
  A55, \dodoi{10.1051/0004-6361/201321100}

\bibitem[{{Johansson} {et~al.}(2012){Johansson}, {Naab}, \&
  {Ostriker}}]{Johansson2012}
{Johansson}, P.~H., {Naab}, T., \& {Ostriker}, J.~P. 2012, \apj, 754, 115,
  \dodoi{10.1088/0004-637X/754/2/115}

\bibitem[{Kawinwanichakij(2018)}]{Kawinwanichakij2018}
Kawinwanichakij, L. 2018, PhD thesis.
\newblock
  \url{http://proxy.library.tamu.edu/login?url=https://search.proquest.com/docview/2197682266?accountid=7082}

\bibitem[{{Kawinwanichakij} {et~al.}(2017){Kawinwanichakij}, {Papovich},
  {Quadri}, {Glazebrook}, {Kacprzak}, {Allen}, {Bell}, {Croton}, {Dekel}, \&
  {Ferguson}}]{Kawinwanichakij2017}
{Kawinwanichakij}, L., {Papovich}, C., {Quadri}, R.~F., {et~al.} 2017, \apj,
  847, 134, \dodoi{10.3847/1538-4357/aa8b75}

\bibitem[{{Kere{\v s}} {et~al.}(2005){Kere{\v s}}, {Katz}, {Weinberg}, \&
  {Dav{\'e}}}]{Keres2005}
{Kere{\v s}}, D., {Katz}, N., {Weinberg}, D.~H., \& {Dav{\'e}}, R. 2005,
  \mnras, 363, 2, \dodoi{10.1111/j.1365-2966.2005.09451.x}

\bibitem[{{Khochfar} \& {Silk}(2006)}]{Khochfar2006}
{Khochfar}, S., \& {Silk}, J. 2006, \apjl, 648, L21, \dodoi{10.1086/507768}

\bibitem[{Kriek {et~al.}(2010)Kriek, Labb{\'{e}}, Conroy, Whitaker, van Dokkum,
  Brammer, Franx, Illingworth, Marchesini, Muzzin, Quadri, \&
  Rudnick}]{Kriek2010}
Kriek, M., Labb{\'{e}}, I., Conroy, C., {et~al.} 2010, The Astrophysical
  Journal, 722, L64, \dodoi{10.1088/2041-8205/722/1/l64}

\bibitem[{{Kroupa}(2001)}]{Kroupa2001}
{Kroupa}, P. 2001, \mnras, 322, 231, \dodoi{10.1046/j.1365-8711.2001.04022.x}

\bibitem[{{Lacey} \& {Cole}(1993)}]{Lacey1993}
{Lacey}, C., \& {Cole}, S. 1993, \mnras, 262, 627,
  \dodoi{10.1093/mnras/262.3.627}

\bibitem[{{Lackner} {et~al.}(2012){Lackner}, {Cen}, {Ostriker}, \&
  {Joung}}]{Lackner2012}
{Lackner}, C.~N., {Cen}, R., {Ostriker}, J.~P., \& {Joung}, M.~R. 2012, \mnras,
  425, 641, \dodoi{10.1111/j.1365-2966.2012.21525.x}

\bibitem[{{Lamassa} {et~al.}(2016){Lamassa}, {Urry}, {Cappelluti}, {Bohringer},
  {Comastri}, {Glikman}, {Richards}, {Ananna}, {Brusa}, {Cardamone}, {Chon},
  {Civano}, {Farrah}, {Gilfanov}, {Green}, {Komossa}, {Lira}, {Makler},
  {Marchesi}, {Pecoraro}, {Ranalli}, {Salvato}, {Schawinski}, {Stern},
  {Treister}, \& {Viero}}]{Lamassa2016viz}
{Lamassa}, S.~M., {Urry}, C.~M., {Cappelluti}, N., {et~al.} 2016, VizieR Online
  Data Catalog, 181

\bibitem[{{LaMassa} {et~al.}(2016){LaMassa}, {Urry}, {Cappelluti},
  {B{\"o}hringer}, {Comastri}, {Glikman}, {Richards}, {Ananna}, {Brusa},
  {Cardamone}, {Chon}, {Civano}, {Farrah}, {Gilfanov}, {Green}, {Komossa},
  {Lira}, {Makler}, {Marchesi}, {Pecoraro}, {Ranalli}, {Salvato}, {Schawinski},
  {Stern}, {Treister}, \& {Viero}}]{Lamassa2016}
{LaMassa}, S.~M., {Urry}, C.~M., {Cappelluti}, N., {et~al.} 2016, \apj, 817,
  172, \dodoi{10.3847/0004-637X/817/2/172}

\bibitem[{{Lang} {et~al.}(2016{\natexlab{a}}){Lang}, {Hogg}, \&
  {Mykytyn}}]{Lang2016ascl}
{Lang}, D., {Hogg}, D.~W., \& {Mykytyn}, D. 2016{\natexlab{a}}, {The Tractor:
  Probabilistic astronomical source detection and measurement}, Astrophysics
  Source Code Library.
\newblock \doeprint{1604.008}

\bibitem[{{Lang} {et~al.}(2016{\natexlab{b}}){Lang}, {Hogg}, \&
  {Schlegel}}]{Lang2016aj}
{Lang}, D., {Hogg}, D.~W., \& {Schlegel}, D.~J. 2016{\natexlab{b}}, \aj, 151,
  36, \dodoi{10.3847/0004-6256/151/2/36}

\bibitem[{{Laporte} {et~al.}(2013){Laporte}, {White}, {Naab}, \&
  {Gao}}]{Laporte2013}
{Laporte}, C.~F.~P., {White}, S.~D.~M., {Naab}, T., \& {Gao}, L. 2013, \mnras,
  435, 901, \dodoi{10.1093/mnras/stt912}

\bibitem[{{Lee} \& {Yi}(2013)}]{Lee2013}
{Lee}, J., \& {Yi}, S.~K. 2013, \apj, 766, 38,
  \dodoi{10.1088/0004-637X/766/1/38}

\bibitem[{{Lee} \& {Yi}(2017)}]{Lee2017}
---. 2017, \apj, 836, 161, \dodoi{10.3847/1538-4357/aa5b87}

\bibitem[{{Mancone} \& {Gonzalez}(2012)}]{Mancone2012}
{Mancone}, C., \& {Gonzalez}, A. 2012, {EzGal: A Flexible Interface for Stellar
  Population Synthesis Models}, Astrophysics Source Code Library.
\newblock \doeprint{1208.021}

\bibitem[{{Maraston}(2005)}]{Maraston2005}
{Maraston}, C. 2005, \mnras, 362, 799, \dodoi{10.1111/j.1365-2966.2005.09270.x}

\bibitem[{{Maraston} {et~al.}(2006){Maraston}, {Daddi}, {Renzini}, {Cimatti},
  {Dickinson}, {Papovich}, {Pasquali}, \& {Pirzkal}}]{Maraston2006}
{Maraston}, C., {Daddi}, E., {Renzini}, A., {et~al.} 2006, \apj, 652, 85,
  \dodoi{10.1086/508143}

\bibitem[{{Maraston} {et~al.}(2013){Maraston}, {Pforr}, {Henriques}, {Thomas},
  {Wake}, {Brownstein}, {Capozzi}, {Tinker}, {Bundy}, {Skibba}, {Beifiori},
  {Nichol}, {Edmondson}, {Schneider}, {Chen}, {Masters}, {Steele}, {Bolton},
  {York}, {Weaver}, {Higgs}, {Bizyaev}, {Brewington}, {Malanushenko},
  {Malanushenko}, {Snedden}, {Oravetz}, {Pan}, {Shelden}, \&
  {Simmons}}]{Maraston2013}
{Maraston}, C., {Pforr}, J., {Henriques}, B.~M., {et~al.} 2013, \mnras, 435,
  2764, \dodoi{10.1093/mnras/stt1424}

\bibitem[{{Marchesini} {et~al.}(2009){Marchesini}, {van Dokkum}, {F{\"o}rster
  Schreiber}, {Franx}, {Labb{\'e}}, \& {Wuyts}}]{Marchesini2009}
{Marchesini}, D., {van Dokkum}, P.~G., {F{\"o}rster Schreiber}, N.~M., {et~al.}
  2009, \apj, 701, 1765, \dodoi{10.1088/0004-637X/701/2/1765}

\bibitem[{{Marchesini} {et~al.}(2014){Marchesini}, {Muzzin}, {Stefanon},
  {Franx}, {Brammer}, {Marsan}, {Vulcani}, {Fynbo}, {Milvang-Jensen}, {Dunlop},
  \& {Buitrago}}]{Marchesini2014}
{Marchesini}, D., {Muzzin}, A., {Stefanon}, M., {et~al.} 2014, \apj, 794, 65,
  \dodoi{10.1088/0004-637X/794/1/65}

\bibitem[{{Marigo} \& {Girardi}(2007)}]{Marigo2007}
{Marigo}, P., \& {Girardi}, L. 2007, \aap, 469, 239,
  \dodoi{10.1051/0004-6361:20066772}

\bibitem[{{Marigo} {et~al.}(2008){Marigo}, {Girardi}, {Bressan}, {Groenewegen},
  {Silva}, \& {Granato}}]{Marigo2008}
{Marigo}, P., {Girardi}, L., {Bressan}, A., {et~al.} 2008, \aap, 482, 883,
  \dodoi{10.1051/0004-6361:20078467}

\bibitem[{{Matsuoka} \& {Kawara}(2010)}]{Matsuoka2010}
{Matsuoka}, Y., \& {Kawara}, K. 2010, \mnras, 405, 100,
  \dodoi{10.1111/j.1365-2966.2010.16456.x}

\bibitem[{{Mortlock} {et~al.}(2015){Mortlock}, {Conselice}, {Hartley},
  {Duncan}, {Lani}, {Ownsworth}, {Almaini}, {Wel}, {Huang}, {Ashby}, {Willner},
  {Fontana}, {Dekel}, {Koekemoer}, {Ferguson}, {Faber}, {Grogin}, \&
  {Kocevski}}]{Mortlock2015}
{Mortlock}, A., {Conselice}, C.~J., {Hartley}, W.~G., {et~al.} 2015, \mnras,
  447, 2, \dodoi{10.1093/mnras/stu2403}

\bibitem[{{Moster} {et~al.}(2013){Moster}, {Naab}, \& {White}}]{Moster2013}
{Moster}, B.~P., {Naab}, T., \& {White}, S.~D.~M. 2013, \mnras, 428, 3121,
  \dodoi{10.1093/mnras/sts261}

\bibitem[{{Moster} {et~al.}(2018){Moster}, {Naab}, \& {White}}]{Moster2018}
{Moster}, B.~P., {Naab}, T., \& {White}, S. D.~M. 2018, \mnras, 477, 1822,
  \dodoi{10.1093/mnras/sty655}

\bibitem[{{Moster} {et~al.}(2010){Moster}, {Somerville}, {Maulbetsch}, {van den
  Bosch}, {Macci{\`o}}, {Naab}, \& {Oser}}]{Moster2010}
{Moster}, B.~P., {Somerville}, R.~S., {Maulbetsch}, C., {et~al.} 2010, \apj,
  710, 903, \dodoi{10.1088/0004-637X/710/2/903}

\bibitem[{{Moster} {et~al.}(2011){Moster}, {Somerville}, {Newman}, \&
  {Rix}}]{Moster2011}
{Moster}, B.~P., {Somerville}, R.~S., {Newman}, J.~A., \& {Rix}, H.-W. 2011,
  \apj, 731, 113, \dodoi{10.1088/0004-637X/731/2/113}

\bibitem[{{Moustakas}(2017)}]{Moustakas2017}
{Moustakas}, J. 2017, {iSEDfit: Bayesian spectral energy distribution modeling
  of galaxies}, Astrophysics Source Code Library.
\newblock \doeprint{1708.029}

\bibitem[{{Moustakas} {et~al.}(2013){Moustakas}, {Coil}, {Aird}, {Blanton},
  {Cool}, {Eisenstein}, {Mendez}, {Wong}, {Zhu}, \& {Arnouts}}]{Moustakas2013}
{Moustakas}, J., {Coil}, A.~L., {Aird}, J., {et~al.} 2013, \apj, 767, 50,
  \dodoi{10.1088/0004-637X/767/1/50}

\bibitem[{{Moutard} {et~al.}(2016){Moutard}, {Arnouts}, {Ilbert}, {Coupon},
  {Davidzon}, {Guzzo}, {Hudelot}, {McCracken}, {Van Werbaeke}, {Morrison}, {Le
  F{\`e}vre}, {Comte}, {Bolzonella}, {Fritz}, {Garilli}, \&
  {Scodeggio}}]{Moutard2016}
{Moutard}, T., {Arnouts}, S., {Ilbert}, O., {et~al.} 2016, \aap, 590, A103,
  \dodoi{10.1051/0004-6361/201527294}

\bibitem[{{Mutch} {et~al.}(2013){Mutch}, {Croton}, \& {Poole}}]{Mutch2013}
{Mutch}, S.~J., {Croton}, D.~J., \& {Poole}, G.~B. 2013, \mnras, 435, 2445,
  \dodoi{10.1093/mnras/stt1453}

\bibitem[{{Muzzin} {et~al.}(2013){Muzzin}, {Marchesini}, {Stefanon}, {Franx},
  {Milvang-Jensen}, {Dunlop}, {Fynbo}, {Brammer}, {Labb{\'e}}, \& {van
  Dokkum}}]{Muzzin2013}
{Muzzin}, A., {Marchesini}, D., {Stefanon}, M., {et~al.} 2013, \apjs, 206, 8,
  \dodoi{10.1088/0067-0049/206/1/8}

\bibitem[{{Naab} {et~al.}(2006){Naab}, {Khochfar}, \& {Burkert}}]{Naab2006}
{Naab}, T., {Khochfar}, S., \& {Burkert}, A. 2006, \apjl, 636, L81,
  \dodoi{10.1086/500205}

\bibitem[{{Newman} {et~al.}(2012){Newman}, {Ellis}, {Bundy}, \&
  {Treu}}]{Newman2012}
{Newman}, A.~B., {Ellis}, R.~S., {Bundy}, K., \& {Treu}, T. 2012, \apj, 746,
  162, \dodoi{10.1088/0004-637X/746/2/162}

\bibitem[{{Newman} \& {Moster}(2014)}]{Newman2014}
{Newman}, J.~A., \& {Moster}, B.~P. 2014, {QUICKCV: Cosmic variance
  calculator}.
\newblock \doeprint{1402.012}

\bibitem[{{Noeske} {et~al.}(2007){Noeske}, {Weiner}, {Faber}, {Papovich},
  {Koo}, {Somerville}, {Bundy}, {Conselice}, {Newman}, {Schiminovich}, {Le
  Floc'h}, {Coil}, {Rieke}, {Lotz}, {Primack}, {Barmby}, {Cooper}, {Davis},
  {Ellis}, {Fazio}, {Guhathakurta}, {Huang}, {Kassin}, {Martin}, {Phillips},
  {Rich}, {Small}, {Willmer}, \& {Wilson}}]{Noeske2007}
{Noeske}, K.~G., {Weiner}, B.~J., {Faber}, S.~M., {et~al.} 2007, \apj, 660,
  L43, \dodoi{10.1086/517926}

\bibitem[{{Oogi} \& {Habe}(2013)}]{Oogi2013}
{Oogi}, T., \& {Habe}, A. 2013, \mnras, 428, 641, \dodoi{10.1093/mnras/sts047}

\bibitem[{{Oser} {et~al.}(2012){Oser}, {Naab}, {Ostriker}, \&
  {Johansson}}]{Oser2012}
{Oser}, L., {Naab}, T., {Ostriker}, J.~P., \& {Johansson}, P.~H. 2012, \apj,
  744, 63, \dodoi{10.1088/0004-637X/744/1/63}

\bibitem[{{Oser} {et~al.}(2010){Oser}, {Ostriker}, {Naab}, {Johansson}, \&
  {Burkert}}]{Oser2010}
{Oser}, L., {Ostriker}, J.~P., {Naab}, T., {Johansson}, P.~H., \& {Burkert}, A.
  2010, \apj, 725, 2312, \dodoi{10.1088/0004-637X/725/2/2312}

\bibitem[{{Ownsworth} {et~al.}(2014){Ownsworth}, {Conselice}, {Mortlock},
  {Hartley}, {Almaini}, {Duncan}, \& {Mundy}}]{Ownsworth2014}
{Ownsworth}, J.~R., {Conselice}, C.~J., {Mortlock}, A., {et~al.} 2014, \mnras,
  445, 2198, \dodoi{10.1093/mnras/stu1802}

\bibitem[{{Papovich} {et~al.}(2016){Papovich}, {Shipley}, {Mehrtens}, {Lanham},
  {Lacy}, {Ciardullo}, {Finkelstein}, {Bassett}, {Behroozi}, {Blanc}, {de
  Jong}, {DePoy}, {Drory}, {Gawiser}, {Gebhardt}, {Gronwall}, {Hill}, {Hopp},
  {Jogee}, {Kawinwanichakij}, {Marshall}, {McLinden}, {Mentuch Cooper},
  {Somerville}, {Steinmetz}, {Tran}, {Tuttle}, {Viero}, {Wechsler}, \&
  {Zeimann}}]{Papovich2016}
{Papovich}, C., {Shipley}, H.~V., {Mehrtens}, N., {et~al.} 2016, \apjs, 224,
  28, \dodoi{10.3847/0067-0049/224/2/28}

\bibitem[{{Papovich} {et~al.}(2018){Papovich}, {Kawinwanichakij}, {Quadri},
  {Glazebrook}, {Labb{\'e}}, {Tran}, {Forrest}, {Kacprzak}, {Spitler}, \&
  {Straatman}}]{Papovich2018}
{Papovich}, C., {Kawinwanichakij}, L., {Quadri}, R.~F., {et~al.} 2018, \apj,
  854, 30, \dodoi{10.3847/1538-4357/aaa766}

\bibitem[{{Patel} {et~al.}(2013){Patel}, {van Dokkum}, {Franx}, {Quadri},
  {Muzzin}, {Marchesini}, {Williams}, {Holden}, \& {Stefanon}}]{Patel2013}
{Patel}, S.~G., {van Dokkum}, P.~G., {Franx}, M., {et~al.} 2013, \apj, 766, 15,
  \dodoi{10.1088/0004-637X/766/1/15}

\bibitem[{{Pillepich} {et~al.}(2018){Pillepich}, {Nelson}, {Hernquist},
  {Springel}, {Pakmor}, {Torrey}, {Weinberger}, {Genel}, {Naiman}, {Marinacci},
  \& {Vogelsberger}}]{Pillepich2018}
{Pillepich}, A., {Nelson}, D., {Hernquist}, L., {et~al.} 2018, \mnras, 475,
  648, \dodoi{10.1093/mnras/stx3112}

\bibitem[{{Planck Collaboration} {et~al.}(2018){Planck Collaboration},
  {Aghanim}, {Akrami}, {Ashdown}, {Aumont}, {Baccigalupi}, {Ballardini},
  {Banday}, {Barreiro}, {Bartolo}, {Basak}, {Battye}, {Benabed}, {Bernard},
  {Bersanelli}, {Bielewicz}, {Bock}, {Bond}, {Borrill}, {Bouchet}, {Boulanger},
  {Bucher}, {Burigana}, {Butler}, {Calabrese}, {Cardoso}, {Carron},
  {Challinor}, {Chiang}, {Chluba}, {Colombo}, {Combet}, {Contreras}, {Crill},
  {Cuttaia}, {de Bernardis}, {de Zotti}, {Delabrouille}, {Delouis}, {Di
  Valentino}, {Diego}, {Dor{\'e}}, {Douspis}, {Ducout}, {Dupac}, {Dusini},
  {Efstathiou}, {Elsner}, {En{\ss}lin}, {Eriksen}, {Fantaye}, {Farhang},
  {Fergusson}, {Fernandez-Cobos}, {Finelli}, {Forastieri}, {Frailis},
  {Franceschi}, {Frolov}, {Galeotta}, {Galli}, {Ganga}, {G{\'e}nova-Santos},
  {Gerbino}, {Ghosh}, {Gonz{\'a}lez-Nuevo}, {G{\'o}rski}, {Gratton},
  {Gruppuso}, {Gudmundsson}, {Hamann}, {Handley}, {Herranz}, {Hivon}, {Huang},
  {Jaffe}, {Jones}, {Karakci}, {Keih{\"a}nen}, {Keskitalo}, {Kiiveri}, {Kim},
  {Kisner}, {Knox}, {Krachmalnicoff}, {Kunz}, {Kurki-Suonio}, {Lagache},
  {Lamarre}, {Lasenby}, {Lattanzi}, {Lawrence}, {Le Jeune}, {Lemos},
  {Lesgourgues}, {Levrier}, {Lewis}, {Liguori}, {Lilje}, {Lilley}, {Lindholm},
  {L{\'o}pez-Caniego}, {Lubin}, {Ma}, {Mac{\'{\i}}as-P{\'e}rez}, {Maggio},
  {Maino}, {Mandolesi}, {Mangilli}, {Marcos-Caballero}, {Maris}, {Martin},
  {Martinelli}, {Mart{\'{\i}}nez-Gonz{\'a}lez}, {Matarrese}, {Mauri}, {McEwen},
  {Meinhold}, {Melchiorri}, {Mennella}, {Migliaccio}, {Millea}, {Mitra},
  {Miville-Desch{\^e}nes}, {Molinari}, {Montier}, {Morgante}, {Moss}, {Natoli},
  {N{\o}rgaard-Nielsen}, {Pagano}, {Paoletti}, {Partridge}, {Patanchon},
  {Peiris}, {Perrotta}, {Pettorino}, {Piacentini}, {Polastri}, {Polenta},
  {Puget}, {Rachen}, {Reinecke}, {Remazeilles}, {Renzi}, {Rocha}, {Rosset},
  {Roudier}, {Rubi{\~n}o-Mart{\'{\i}}n}, {Ruiz-Granados}, {Salvati}, {Sandri},
  {Savelainen}, {Scott}, {Shellard}, {Sirignano}, {Sirri}, {Spencer},
  {Sunyaev}, {Suur-Uski}, {Tauber}, {Tavagnacco}, {Tenti}, {Toffolatti},
  {Tomasi}, {Trombetti}, {Valenziano}, {Valiviita}, {Van Tent}, {Vibert},
  {Vielva}, {Villa}, {Vittorio}, {Wandelt}, {Wehus}, {White}, {White},
  {Zacchei}, \& {Zonca}}]{Planck2018}
{Planck Collaboration}, {Aghanim}, N., {Akrami}, Y., {et~al.} 2018, ArXiv
  e-prints.
\newblock \doarXiv{1807.06209}

\bibitem[{{Pozzetti} {et~al.}(2007){Pozzetti}, {Bolzonella}, {Lamareille},
  {Zamorani}, {Franzetti}, {Le F{\`e}vre}, {Iovino}, {Temporin}, {Ilbert},
  {Arnouts}, {Charlot}, {Brinchmann}, {Zucca}, {Tresse}, {Scodeggio}, {Guzzo},
  {Bottini}, {Garilli}, {Le Brun}, {Maccagni}, {Picat}, {Scaramella},
  {Vettolani}, {Zanichelli}, {Adami}, {Bardelli}, {Cappi}, {Ciliegi},
  {Contini}, {Foucaud}, {Gavignaud}, {McCracken}, {Marano}, {Marinoni},
  {Mazure}, {Meneux}, {Merighi}, {Paltani}, {Pell{\`o}}, {Pollo}, {Radovich},
  {Bondi}, {Bongiorno}, {Cucciati}, {de la Torre}, {Gregorini}, {Mellier},
  {Merluzzi}, {Vergani}, \& {Walcher}}]{Pozzetti2007}
{Pozzetti}, L., {Bolzonella}, M., {Lamareille}, F., {et~al.} 2007, \aap, 474,
  443, \dodoi{10.1051/0004-6361:20077609}

\bibitem[{{Price-Whelan} {et~al.}(2018){Price-Whelan}, {Sip{\H{o}}cz},
  {G{\"u}nther}, {Lim}, {Crawford}, {Conseil}, {Shupe}, {Craig}, {Dencheva},
  {Ginsburg}, {VanderPlas}, {Bradley}, {P{\'e}rez-Su{\'a}rez}, {de Val-Borro},
  {Paper Contributors}, {Aldcroft}, {Cruz}, {Robitaille}, {Tollerud},
  {Coordination Committee}, {Ardelean}, {Babej}, {Bach}, {Bachetti}, {Bakanov},
  {Bamford}, {Barentsen}, {Barmby}, {Baumbach}, {Berry}, {Biscani}, {Boquien},
  {Bostroem}, {Bouma}, {Brammer}, {Bray}, {Breytenbach}, {Buddelmeijer},
  {Burke}, {Calderone}, {Cano Rodr{\'\i}guez}, {Cara}, {Cardoso}, {Cheedella},
  {Copin}, {Corrales}, {Crichton}, {D{\textquoteright}Avella}, {Deil},
  {Depagne}, {Dietrich}, {Donath}, {Droettboom}, {Earl}, {Erben}, {Fabbro},
  {Ferreira}, {Finethy}, {Fox}, {Garrison}, {Gibbons}, {Goldstein}, {Gommers},
  {Greco}, {Greenfield}, {Groener}, {Grollier}, {Hagen}, {Hirst}, {Homeier},
  {Horton}, {Hosseinzadeh}, {Hu}, {Hunkeler}, {Ivezi{\'c}}, {Jain}, {Jenness},
  {Kanarek}, {Kendrew}, {Kern}, {Kerzendorf}, {Khvalko}, {King}, {Kirkby},
  {Kulkarni}, {Kumar}, {Lee}, {Lenz}, {Littlefair}, {Ma}, {Macleod},
  {Mastropietro}, {McCully}, {Montagnac}, {Morris}, {Mueller}, {Mumford},
  {Muna}, {Murphy}, {Nelson}, {Nguyen}, {Ninan}, {N{\"o}the}, {Ogaz}, {Oh},
  {Parejko}, {Parley}, {Pascual}, {Patil}, {Patil}, {Plunkett}, {Prochaska},
  {Rastogi}, {Reddy Janga}, {Sabater}, {Sakurikar}, {Seifert}, {Sherbert},
  {Sherwood-Taylor}, {Shih}, {Sick}, {Silbiger}, {Singanamalla}, {Singer},
  {Sladen}, {Sooley}, {Sornarajah}, {Streicher}, {Teuben}, {Thomas},
  {Tremblay}, {Turner}, {Terr{\'o}n}, {van Kerkwijk}, {de la Vega}, {Watkins},
  {Weaver}, {Whitmore}, {Woillez}, {Zabalza}, \& {Contributors}}]{astropy:2018}
{Price-Whelan}, A.~M., {Sip{\H{o}}cz}, B.~M., {G{\"u}nther}, H.~M., {et~al.}
  2018, \aj, 156, 123, \dodoi{10.3847/1538-3881/aabc4f}

\bibitem[{{Qu} {et~al.}(2017){Qu}, {Helly}, {Bower}, {Theuns}, {Crain},
  {Frenk}, {Furlong}, {McAlpine}, {Schaller}, {Schaye}, \& {White}}]{Qu2017}
{Qu}, Y., {Helly}, J.~C., {Bower}, R.~G., {et~al.} 2017, \mnras, 464, 1659,
  \dodoi{10.1093/mnras/stw2437}

\bibitem[{{Riess} {et~al.}(2019){Riess}, {Casertano}, {Yuan}, {Macri}, \&
  {Scolnic}}]{Riess2019}
{Riess}, A.~G., {Casertano}, S., {Yuan}, W., {Macri}, L.~M., \& {Scolnic}, D.
  2019, \apj, 876, 85, \dodoi{10.3847/1538-4357/ab1422}

\bibitem[{{Rodighiero} {et~al.}(2011){Rodighiero}, {Daddi}, {Baronchelli},
  {Cimatti}, {Renzini}, {Aussel}, {Popesso}, {Lutz}, {Andreani}, {Berta},
  {Cava}, {Elbaz}, {Feltre}, {Fontana}, {F{\"o}rster Schreiber},
  {Franceschini}, {Genzel}, {Grazian}, {Gruppioni}, {Ilbert}, {Le Floch},
  {Magdis}, {Magliocchetti}, {Magnelli}, {Maiolino}, {McCracken}, {Nordon},
  {Poglitsch}, {Santini}, {Pozzi}, {Riguccini}, {Tacconi}, {Wuyts}, \&
  {Zamorani}}]{Rodighiero2011}
{Rodighiero}, G., {Daddi}, E., {Baronchelli}, I., {et~al.} 2011, \apj, 739,
  L40, \dodoi{10.1088/2041-8205/739/2/L40}

\bibitem[{{Rodriguez-Gomez} {et~al.}(2016){Rodriguez-Gomez}, {Pillepich},
  {Sales}, {Genel}, {Vogelsberger}, {Zhu}, {Wellons}, {Nelson}, {Torrey},
  {Springel}, {Ma}, \& {Hernquist}}]{Rodriguez2016}
{Rodriguez-Gomez}, V., {Pillepich}, A., {Sales}, L.~V., {et~al.} 2016, \mnras,
  458, 2371, \dodoi{10.1093/mnras/stw456}

\bibitem[{{Rodr{\'{\i}}guez-Puebla} {et~al.}(2017){Rodr{\'{\i}}guez-Puebla},
  {Primack}, {Avila-Reese}, \& {Faber}}]{Rodriguez2017}
{Rodr{\'{\i}}guez-Puebla}, A., {Primack}, J.~R., {Avila-Reese}, V., \& {Faber},
  S.~M. 2017, \mnras, 470, 651, \dodoi{10.1093/mnras/stx1172}

\bibitem[{{Salim} {et~al.}(2007){Salim}, {Rich}, {Charlot}, {Brinchmann},
  {Johnson}, {Schiminovich}, {Seibert}, {Mallery}, {Heckman}, \&
  {Forster}}]{Salim2007}
{Salim}, S., {Rich}, R.~M., {Charlot}, S., {et~al.} 2007, \apjs, 173, 267,
  \dodoi{10.1086/519218}

\bibitem[{{S{\'a}nchez-Bl{\'a}zquez} {et~al.}(2006){S{\'a}nchez-Bl{\'a}zquez},
  {Peletier}, {Jim{\'e}nez-Vicente}, {Cardiel}, {Cenarro},
  {Falc{\'o}n-Barroso}, {Gorgas}, {Selam}, \& {Vazdekis}}]{Sanchez2006}
{S{\'a}nchez-Bl{\'a}zquez}, P., {Peletier}, R.~F., {Jim{\'e}nez-Vicente}, J.,
  {et~al.} 2006, \mnras, 371, 703, \dodoi{10.1111/j.1365-2966.2006.10699.x}

\bibitem[{{Schreiber} {et~al.}(2016){Schreiber}, {Elbaz}, {Pannella}, {Ciesla},
  {Wang}, {Koekemoer}, {Rafelski}, \& {Daddi}}]{Schreiber2016}
{Schreiber}, C., {Elbaz}, D., {Pannella}, M., {et~al.} 2016, \aap, 589, A35,
  \dodoi{10.1051/0004-6361/201527200}

\bibitem[{{Shimizu} \& {Inoue}(2013)}]{Shimizu2013}
{Shimizu}, I., \& {Inoue}, A.~K. 2013, \pasj, 65, 96,
  \dodoi{10.1093/pasj/65.5.96}

\bibitem[{{Shivaei} {et~al.}(2015){Shivaei}, {Reddy}, {Shapley}, {Kriek},
  {Siana}, {Mobasher}, {Coil}, {Freeman}, {Sanders}, {Price}, {de Groot}, \&
  {Azadi}}]{Shivaei2015}
{Shivaei}, I., {Reddy}, N.~A., {Shapley}, A.~E., {et~al.} 2015, \apj, 815, 98,
  \dodoi{10.1088/0004-637X/815/2/98}

\bibitem[{{Smith}(2012)}]{Smith2012}
{Smith}, R.~E. 2012, \mnras, 426, 531, \dodoi{10.1111/j.1365-2966.2012.21745.x}

\bibitem[{{Somerville} {et~al.}(2004){Somerville}, {Lee}, {Ferguson},
  {Gardner}, {Moustakas}, \& {Giavalisco}}]{Somerville2004}
{Somerville}, R.~S., {Lee}, K., {Ferguson}, H.~C., {et~al.} 2004, \apjl, 600,
  L171, \dodoi{10.1086/378628}

\bibitem[{{Speagle} {et~al.}(2014){Speagle}, {Steinhardt}, {Capak}, \&
  {Silverman}}]{Speagle2014}
{Speagle}, J.~S., {Steinhardt}, C.~L., {Capak}, P.~L., \& {Silverman}, J.~D.
  2014, \apjs, 214, 15, \dodoi{10.1088/0067-0049/214/2/15}

\bibitem[{{Springel} {et~al.}(2005){Springel}, {Di Matteo}, \&
  {Hernquist}}]{Springel2005}
{Springel}, V., {Di Matteo}, T., \& {Hernquist}, L. 2005, \mnras, 361, 776,
  \dodoi{10.1111/j.1365-2966.2005.09238.x}

\bibitem[{{Strateva} {et~al.}(2001){Strateva}, {Ivezi{\'c}}, {Knapp},
  {Narayanan}, {Strauss}, {Gunn}, {Lupton}, {Schlegel}, {Bahcall}, {Brinkmann},
  {Brunner}, {Budav{\'a}ri}, {Csabai}, {Castander}, {Doi}, {Fukugita}, {Gy{\H
  o}ry}, {Hamabe}, {Hennessy}, {Ichikawa}, {Kunszt}, {Lamb}, {McKay},
  {Okamura}, {Racusin}, {Sekiguchi}, {Schneider}, {Shimasaku}, \&
  {York}}]{Strateva2001}
{Strateva}, I., {Ivezi{\'c}}, {\v Z}., {Knapp}, G.~R., {et~al.} 2001, \aj, 122,
  1861, \dodoi{10.1086/323301}

\bibitem[{{Szomoru} {et~al.}(2012){Szomoru}, {Franx}, \& {van
  Dokkum}}]{Szomoru2012}
{Szomoru}, D., {Franx}, M., \& {van Dokkum}, P.~G. 2012, \apj, 749, 121,
  \dodoi{10.1088/0004-637X/749/2/121}

\bibitem[{{Tomczak} {et~al.}(2014){Tomczak}, {Quadri}, {Tran}, {Labb{\'e}},
  {Straatman}, {Papovich}, {Glazebrook}, {Allen}, {Brammer}, {Kacprzak},
  {Kawinwanichakij}, {Kelson}, {McCarthy}, {Mehrtens}, {Monson}, {Persson},
  {Spitler}, {Tilvi}, \& {van Dokkum}}]{Tomczak2014}
{Tomczak}, A.~R., {Quadri}, R.~F., {Tran}, K.-V.~H., {et~al.} 2014, \apj, 783,
  85, \dodoi{10.1088/0004-637X/783/2/85}

\bibitem[{{van der Wel} {et~al.}(2014){van der Wel}, {Franx}, {van Dokkum},
  {Skelton}, {Momcheva}, {Whitaker}, {Brammer}, {Bell}, {Rix}, {Wuyts},
  {Ferguson}, {Holden}, {Barro}, {Koekemoer}, {Chang}, {McGrath},
  {H{\"a}ussler}, {Dekel}, {Behroozi}, {Fumagalli}, {Leja}, {Lundgren},
  {Maseda}, {Nelson}, {Wake}, {Patel}, {Labb{\'e}}, {Faber}, {Grogin}, \&
  {Kocevski}}]{VanderWel2014}
{van der Wel}, A., {Franx}, M., {van Dokkum}, P.~G., {et~al.} 2014, \apj, 788,
  28, \dodoi{10.1088/0004-637X/788/1/28}

\bibitem[{{van Dokkum} {et~al.}(2010){van Dokkum}, {Whitaker}, {Brammer},
  {Franx}, {Kriek}, {Labb{\'e}}, {Marchesini}, {Quadri}, {Bezanson},
  {Illingworth}, {Muzzin}, {Rudnick}, {Tal}, \& {Wake}}]{vanDokkum2010}
{van Dokkum}, P.~G., {Whitaker}, K.~E., {Brammer}, G., {et~al.} 2010, \apj,
  709, 1018, \dodoi{10.1088/0004-637X/709/2/1018}

\bibitem[{{Vassiliadis} \& {Wood}(1994)}]{Vassiliadis1994}
{Vassiliadis}, E., \& {Wood}, P.~R. 1994, \apjs, 92, 125,
  \dodoi{10.1086/191962}

\bibitem[{{Weinzirl} {et~al.}(2011){Weinzirl}, {Jogee}, {Conselice},
  {Papovich}, {Chary}, {Bluck}, {Gr{\"u}tzbauch}, {Buitrago}, {Mobasher},
  {Lucas}, {Dickinson}, \& {Bauer}}]{Weinzirl2011}
{Weinzirl}, T., {Jogee}, S., {Conselice}, C.~J., {et~al.} 2011, \apj, 743, 87,
  \dodoi{10.1088/0004-637X/743/1/87}

\bibitem[{{Wellons} {et~al.}(2015){Wellons}, {Torrey}, {Ma}, {Rodriguez-Gomez},
  {Vogelsberger}, {Kriek}, {van Dokkum}, {Nelson}, {Genel}, {Pillepich},
  {Springel}, {Sijacki}, {Snyder}, {Nelson}, {Sales}, \&
  {Hernquist}}]{Wellons2015}
{Wellons}, S., {Torrey}, P., {Ma}, C.-P., {et~al.} 2015, \mnras, 449, 361,
  \dodoi{10.1093/mnras/stv303}

\bibitem[{{Whitaker} {et~al.}(2012){Whitaker}, {van Dokkum}, {Brammer}, \&
  {Franx}}]{Whitaker2012}
{Whitaker}, K.~E., {van Dokkum}, P.~G., {Brammer}, G., \& {Franx}, M. 2012,
  \apj, 754, L29, \dodoi{10.1088/2041-8205/754/2/L29}

\bibitem[{{White} \& {Frenk}(1991)}]{White1991}
{White}, S.~D.~M., \& {Frenk}, C.~S. 1991, \apj, 379, 52,
  \dodoi{10.1086/170483}

\bibitem[{{White} \& {Rees}(1978)}]{White1978}
{White}, S.~D.~M., \& {Rees}, M.~J. 1978, \mnras, 183, 341,
  \dodoi{10.1093/mnras/183.3.341}

\bibitem[{{Wild} {et~al.}(2009){Wild}, {Walcher}, {Johansson}, {Tresse},
  {Charlot}, {Pollo}, {Le F{\`e}vre}, \& {de Ravel}}]{Wild2009}
{Wild}, V., {Walcher}, C.~J., {Johansson}, P.~H., {et~al.} 2009, \mnras, 395,
  144, \dodoi{10.1111/j.1365-2966.2009.14537.x}

\bibitem[{{Wold} {et~al.}(2019){Wold}, {Kawinwanichakij}, {Stevans},
  {Finkelstein}, {Papovich}, {Devarakonda}, {Ciardullo}, {Feldmeier}, {Florez},
  {Gawiser}, {Gronwall}, {Jogee}, {Marshall}, {Sherman}, {Shipley},
  {Somerville}, {Valdes}, \& {Zeimann}}]{Wold2019}
{Wold}, I.~G.~B., {Kawinwanichakij}, L., {Stevans}, M.~L., {et~al.} 2019,
  \apjs, 240, 5, \dodoi{10.3847/1538-4365/aaee85}

\bibitem[{{Wright} {et~al.}(2018){Wright}, {Driver}, \&
  {Robotham}}]{Wright2018}
{Wright}, A.~H., {Driver}, S.~P., \& {Robotham}, A.~S.~G. 2018, \mnras, 480,
  3491, \dodoi{10.1093/mnras/sty2136}

\bibitem[{{Wuyts} {et~al.}(2011){Wuyts}, {F{\"o}rster Schreiber}, {van der
  Wel}, {Magnelli}, {Guo}, {Genzel}, {Lutz}, {Aussel}, {Barro}, {Berta},
  {Cava}, {Graci{\'a}-Carpio}, {Hathi}, {Huang}, {Kocevski}, {Koekemoer},
  {Lee}, {Le Floc'h}, {McGrath}, {Nordon}, {Popesso}, {Pozzi}, {Riguccini},
  {Rodighiero}, {Saintonge}, \& {Tacconi}}]{Wuyts2011}
{Wuyts}, S., {F{\"o}rster Schreiber}, N.~M., {van der Wel}, A., {et~al.} 2011,
  \apj, 742, 96, \dodoi{10.1088/0004-637X/742/2/96}

\end{thebibliography}
\end{document}